\documentclass[10pt]{article}
\usepackage[utf8]{inputenc}
\usepackage[T1]{fontenc}
\usepackage{amsmath}
\usepackage{amsfonts}
\usepackage{amssymb}
\usepackage[version=4]{mhchem}
\usepackage{hyperref}
\hypersetup{colorlinks=true, linkcolor=blue, filecolor=magenta, urlcolor=cyan}
\usepackage{graphicx}
\usepackage[export]{adjustbox}
\graphicspath{ {./images/} }
\newcommand{\bra}[1]{\langle #1 |}
\newcommand{\ket}[1]{| #1 \rangle}
\usepackage{enumitem}
\usepackage{placeins}
\usepackage[english]{babel}
\usepackage{microtype}
\usepackage{url}
\usepackage{amsthm}

\usepackage[numbers]{natbib}  
\usepackage{setspace}
\usepackage{float}
\usepackage{subcaption}

\newtheoremstyle{nodotstyle} 
  {12pt} 
  {12pt} 
  {} 
  {} 
  {} 
  {} 
  { } 
  {} 

\theoremstyle{nodotstyle}
\newtheorem*{nodotdef}{} 

\title{A new indeterminacy-based quantum theory}

\author{Francisco Pipa\thanks{franciscosapipa@gmail.com}\\[0.5cm]
Department of Philosophy, University of Kansas}

\date{}

\begin{document}

\maketitle

\begin{abstract}
\noindent I propose a novel interpretation of quantum theory, which I will call Environmental Determinacy-based (EnDQT). In contrast to the well-known interpretations of quantum theory, EnDQT has the benefit of not adding non-local, superdeterministic, or retrocausal hidden variables. Also, it is not in tension with relativistic causality by providing a local causal explanation of quantum correlations. Furthermore, measurement outcomes don't vary according to, for example, systems or worlds. It is a conservative QT in the sense that, unlike theories such as spontaneous collapse theories, no modifications of the fundamental equations of quantum theory are required to establish when determinate values arise. Moreover, in principle, arbitrary systems can be in a coherent superposition for an arbitrary amount of time. According to EnDQT, at a certain stage of the evolution of the universe, some systems acquire the capacity to have and give rise to other systems having determinate values through an indeterministic process. Furthermore, this capacity propagates via local interactions between systems. When systems are isolated from others that have this capacity, they can, in principle, evolve unitarily indefinitely. EnDQT may provide payoffs to other areas of physics and their foundations, such as cosmology, via the features of the systems that start the chains of interactions.
\end{abstract}






\section{The measurement problem and EnDQT's desiderata}

The measurement problem\footnote{See, e.g., \cite{Maudlin1995ThreeProblems, sep-qt-issues}.} can be seen as arising from interactions in quantum theory (QT), which, without introducing additional assumptions, lead to the quantum state of a macroscopic system being in a superposition. This superposition does not correspond to a physical magnitude with determinate values. However, we know from classical physics and experimental evidence that this cannot be the case at macroscopic scales.

One way to see this problem more concretely is by first assuming the so-called Eigenstate-Eigenvalue Link (EEL). The EEL gives a clear way of making sense of how the mathematical framework of QT represents physical magnitudes and it played an important role in interpreting QT, especially
within the more orthodox interpretations.\footnote{See \cite{2016SHPMP..55...92G}
  for a historical overview of the importance of this link.} According
to this link:\\

\noindent \emph{A system S has a determinate value q of an observable O if and
only if the quantum state of S is in eigenstate of} \(O\) \emph{with an
eigenvalue} \(q\)\emph{.}\\

This link often leads to the assumption that if the quantum state of $S$ is not an eigenstate of some observable, $S$ has, in a sense, an \emph{indeterminate value} of that observable.
 
 Now, let's consider a quantum system \( S \) and a measurement device \( M \), where \( M \) will measure a certain property of \( S \). The system \( S \) is initially in a superposition of quantum states:
 
\begin{equation}
   |\psi\rangle_S = \sum_i c_i |\psi_i\rangle_S   
\end{equation}
where \( \{|\psi_i\rangle_S\} \) are the eigenstates of the observable \( \hat{O} \) of the system, with corresponding eigenvalues \( \{o_i\} \). Furthermore, the measurement device \( M \) is initially in a ready state \( |m_0\rangle_M \). States \( \{|m_i\rangle_M\} \) 
of the device $M$ are eigenstates of the observable \( \hat{M} \) that represents the different measurement records of $M$ when it interacts with $S$.

During the measurement process, the system \( S \) interacts with the device \( M \). According to the standard rules for interactions in unitary QT, this interaction results in the quantum state of the measurement device becoming correlated with the eigenstates of the observable \( \hat{O} \), leading to the following entangled state with $c_i \neq 0$:
   \begin{equation}
   |\Psi\rangle_{SM} = \sum_i c_i |\psi_i\rangle_S |m_i\rangle_M.
   \end{equation}
 Given the EEL, this superposition implies that the system \( S \) doesn't have a determinate value of observable \( \hat{O} \), as well as the observable \( \hat{M} \). Now, the problem is that this is contrary to what our measurement records indicate, as well as to what classical physics successfully predicts at these scales. We might choose to deal with this issue by assuming the so-called collapse postulate:\\

\noindent \emph{When the system S is measured, the state of S transitions
stochastically to one the eigenstates of} \(\hat{O}\)\emph{, where the
probability for such transition is given by the Born rule.}\\

However, it is unclear precisely when to apply this postulate and how to ground it because it is unclear what constitutes a measurement. This issue, in a nutshell, forms the measurement problem.

Diverse quantum theories (QTs) were developed to deal with the measurement problem. Standard QT, whose basic axioms were initially developed by Dirac and von Neumann,\footnote{\cite{Dirac1930TheMechanics, VonNeumann1932MathematischeQuantenmechanik}. See Appendix A, and e.g., Chapter 4 in \cite{Barrett2019TheMechanics} for an introduction.}, has enjoyed incredible empirical success. Given the success of this theory, a plausible desideratum for a satisfactory quantum theory that solves the measurement problem is to be as conservative as possible, staying as close as possible to standard QT. In the search for a solution to this problem, given the empirical success of QT, I think we should adopt a conservative approach. This conservative approach should seek to fulfill the following desideratum:\\

*) A precise criterion for when determinate values arise that doesn't modify the fundamental equations of QT like spontaneous collapse theories or postulates a special force that causes such collapse like gravitational collapse theories.\footnote{See, e.g., \cite{sep-qm-collapse, Oppenheim2023, Diosi1995QuantumLimit}.}\\   

This desideratum is conservative because we currently have no clear evidence that we need to modify the equations of QT or posit that a force like gravity causes the collapse. Similarly, current evidence doesn't point towards limits regarding the scales in which we can place systems in a coherent superposition so that they give rise to interference phenomena if they are sufficiently isolated from their environment. In other words, it is plausible to consider that any system or systems in a superposition of quantum states could, in principle, evolve unitarily indefinitely regarding any dynamical physical magnitude. So, this conservative solution should also fulfill the desideratum of\\

 **) Allowing for any system to, in principle, be in a coherent superposition of quantum states associated with at least any dynamical physical magnitude for an arbitrary amount of time.\\

I will consider that an approach that fulfills *) and **) fulfills what I will call the desideratum UT) for unitarity.

Decoherence theory suggests that interactions between a $target$ system $S$ with a certain kind of system $E$, which includes measurement devices, lead $S$ to have determinate values.\footnote{See, e.g., \cite{Schlosshauer2007DecoherenceTransition} and references therein.} Furthermore, we have evidence that interactions between quantum systems involving decoherence play some role in giving rise to determinate outcomes.\footnote{Ibid.} Decoherence doesn't modify the fundamental equations of standard QT and thus allows for a conservative approach that fulfills *). Therefore, a conservative strategy fulfilling *) should consider that only interactions with systems $E$ lead a target system $S$ to have determinate values, where these interactions are described quantum mechanically via decoherence.

 But interactions with which systems $E$, which kinds of interactions? We don't want just to vaguely claim that these would just be interactions with systems that constitute the ``large environment'' of a system $S$ that undergoes decoherence (as it is traditionally claimed). First, it is unclear which systems we are talking precisely about since the environment of a system can involve many systems. Second, considering the desideratum **), we aim to ensure that any systems, including those interacting with their environment, can, under certain suitable conditions, be put into a superposition of quantum states that result in interference. Specifically, as we make progress in being able to observe interference phenomena with larger systems made up of multiple systems (which could form a ``large'' environment for each other), a conservative approach should not impose any limitations on this ability.

In a conservative approach that relies solely on decoherence, we can assume that system $S$ will only have definite values if it interacts with systems $E$ that have been decohered shortly before the interaction with $S$ begins. These systems $E$ would have been decohered by other systems that were previously decohered by some other systems, and so on. This assumption is plausible because it is reasonable to consider that the constituents of our measurement devices were previously subject to decoherence due to the large environment they interact with, and so on. However, this idea seems to start giving rise to an infinite regress and some vagueness regarding how to think about these interactions since it is unclear what decoherence even represents. Thus, even if we decide to maintain a conservative strategy, there are some issues to deal with and choices to make when interpreting QT.

EnDQT will adopt the above conservative strategy but also deal with these issues by appealing to some plausible special systems that establish when these interactions began, a more precise structure that represents such interactions and simple and conservative rules that establish how determinate values arise from them. This strategy will also advance a clear view of what decoherence is about. The structure that EnDQT will introduce is a network structure whose edges represent interactions between certain systems represented via decoherence, and that roughly establishes when these interactions give rise to them having determinate values. These interactions form what I will call Stable Determination Chains (SDCs), where what I mean by ``stable'' will be clarified below.

Furthermore, to circumvent an infinite regress, SDCs started somewhere in spacetime. As I will argue, the first systems with determinate values arose in the past through some systems. These systems started chains of local interactions over time and space, i.e., the SDCs. More concretely, by interacting with these systems, a system $S'$ acquires the capacity to have a determinate value of an observable during these interactions and to give rise to other systems having determinate values in interactions with $S'$, which allows these later systems to lead other systems to have determinate values, and so on. I will call this capacity, the determination capacity. So, these chains allow the determination capacity to propagate between systems, where it is indeterministic which values will arise under interactions among the possible ones. 

These interactions are modeled via decoherence; thus, as I have said, they don't lead to any modification of fundamental equations of QT. Moreover, note that the systems that never were part of this chain of interactions can, in principle, continue their deterministic evolution represented unitarily, and without any indeterministic process interrupting this evolution. This allows EnDQT to fulfill the desideratum **).

I will argue that one possible systems that initiate SDCs can be understood through inflation, a dominant paradigm in modern cosmology. So, EnDQT will assign a new role to the systems that start inflation. I will also argue that this assumption regarding the special systems that give rise to SDCs isn't problematic because every QT has to postulate some kind of initial conditions in the early universe to explain diverse physical phenomena (e.g., the temporal asymmetries through the so-called Past Hypothesis\footnote{\cite{Albert2000TimeChance, sep-time-thermo}.}) and the phenomena that the big bang and inflation aim to explain). However, EnDQT has the benefit of potentially grounding these special features of the early universe in fundamental aspects of QT.

Like QT, the theory of relativity has enjoyed numerous empirical successes. Thus, a conservative approach should make QT compatible with relativity and not in conflict with it. Therefore, another important desideratum for a conservative approach to QT is to make it compatible with relativity at the basic level. What I mean by at basic level is setting aside issues related with the possible need of a theory of quantum gravity (more on this below).  However, as is well known and will be further explained, already at the basic level of non-relativistic QT, Bell's theorem threatens such compatibility. Thus, another desideratum for a quantum theory is that \\

\noindent LC) In domains where the application of QT is established, ensure that the theory: does not conflict with relativity by favoring any specific reference frame, thus avoiding action-at-a-distance phenomena as seen in Bohmian mechanics,\footnote{See, e.g., \cite{Goldstein2021BohmianMechanics}.}; does not introduce hidden variables that result in retrocausality or superdeterminism.\footnote{See, e.g., \cite{Hossenfelder2020RethinkingSuperdeterminism, sep-qm-retrocausality}.}\\

For EnDQT to achieve LC), first, I will argue that it can deal with Bell's theorem by providing a local explanation of quantum correlations via Quantum Causal Models (Section 3).\footnote{See, e.g., \cite{Costa_2016, Allen2017QuantumModels, Allen2017QuantumModels}.} Second, EnDQT will adopt a perspective on quantum states where they don't literally and directly represent some physical entity; instead, together with other elements of the theory, such as observables and networks representing SDCs, they help make inferences, gain knowledge about and indirectly represent how systems evolve and affect each other, how SDCs evolve, when systems acquire or not determinate values, how systems evolve outside interactions, etc. The fulfillment of LC) will be mainly argued on the basis of non-relativistic QT by showing that quantum correlations can be explained without being in tension with relativity. It turns out that the tension between QT and relativity already arises in the non-relativistic case. I will not focus on the locality of a theory in the sense that it respects relativistic symmetries (i.e., it allows for Lorentz covariance and, more generally, the general covariance of the physical laws). Nevertheless, I will also provide arguments for why EnDQT can be local in this sense.

You may call this theory a ``collapse theory,'' but I resist the use of that terminology because we run the risk of running together very different interpretations of QT with very different consequences and ontological commitments. This view doesn't reify the wavefunction. In agreement with the above desideratum, contrary to spontaneous collapse theories, which clearly reify the quantum state, there is no literal physical collapse of the wavefunction/quantum states in a superposition. Instead, there is an epistemic local state update of the original state of the target system that can be implemented upon decoherence of this system by its environmental systems that belong to SDCs under their local interactions. These interactions give rise indeterministically to these systems having a determinate value. Therefore, given that EnDQT doesn't reify quantum states, in Bell-type scenarios, the measurements of an agent Alice on her system don't non-locally affect the space-like separated system of agent Bob and vice versa. There isn't such non-locality. On top of that, this view doesn't put any limits on the superposition principle, i.e., the ability to place objects in a coherent superposition, except whatever limits decoherence models already provide us. There isn't also a special force causing this indeterministic process to occur.

Note that this view on quantum states also considers that decoherence shouldn't be interpreted as representing a process of branching of the wave-function/quantum states but rather as a process in which, under local interactions, an environmental system that belongs to an SDC gives rise indeterministically to another system having determinate values.

A common way to fulfill UT) and LC) is by adopting a relationalist interpretation. On this view, outcomes are relative to and can vary in the same situation according to, for example, worlds, any systems, private perspectives, environments, simultaneity hyperplanes, etc.\footnote{See, e.g., \cite{Wallace2012TheInterpretation, Rovelli1996RelationalMechanics, DiBiagio2021StableFacts, Healey2017ThePhilosophy, Dieks2019QuantumCovariance, ormrod2024quantum}.} One example of a relationalist strategy is the Many-Worlds Interpretation (MWI), in which roughly each physically possible outcome corresponds to a $world.$ However, while maintaining our conservative approach, we should aim to circumvent such a counterintuitive strategy. Thus, another desiderata is to\\

\noindent NR) Not adopt a relationalist interpretation of QT.\\

 Furthermore, given the well-known issue of probabilities of the MWI,\footnote{See, e.g., \cite{Albert2010ProbabilityPicture, Price2010DecisionsProbability}.} fulfilling this desideratum presents another advantage. Also, it is unclear if relationalism in a single world is desirable.\footnote{See, e.g., \cite{Riedel2024RelationalRelativity}.} As it will become clear, EnDQT will fulfill this desideratum since outcomes will be absolute/non-relationalist.

Note that MWI-like views consider that decoherence in large enough regions of spacetime establishes criteria for systems to have determinate values of an observable. However, as we will see in the next section, EnDQT, in a sense, considers that such criteria form necessary but not sufficient conditions for determinate values to arise. This is because it matters if the environmental systems involved in this process belong to an SDC.

Moreover, as I will argue (section 3), to my knowledge, EnDQT is currently the only QT that doesn't modify the fundamental equations of standard QT,\footnote{It might be objected that quantum causal models modify the fundamental equation of QT. Note that I don't regard these models as a modification of equations of standard QT, but as a generalization or variation.} that is able to provide a local, non-hidden variable, non-relationalist common cause explanation of quantum correlations like the ones in Bell scenarios and the so-called extended Wigner's friend scenarios.\footnote{I will focus on the scenarios from \cite{Bong2020AParadox, Brukner_2018}.}\footnote{Contrary to the suggestions by others (e.g., \cite{e23080925, schmid2023review, ying2023relating}), we don't need to adapt QCMs to a relationalist approach, which might be considered as an advantage of this view.} 
Thus, by fulfilling all of the above plausible desiderata, EnDQT should be considered a conservative approach, which, given the empirical success of QT, I think is a virtue.

I will start by explaining the basics of EnDQT, argue that it provides UT) and start building the case that it provides LC) and NR) (Section 2). I will also provide a toy model to illustrate how EnDQT represents physical interactions in Appendix A. In Section 3, I will argue that EnDQT provides LC) and NR) by showing that it provides a non-relational, local, and non-superdeterministic/non-retrocausal explanation of quantum/Bell-type correlations. Throughout the text, I will suggest future developments.

Therefore, I will argue that in contrast to the well-known quantum theories, EnDQT has the benefit of addressing the measurement problem without adding non-local, superdeterministic, or retrocausal hidden variables. Also, it addresses that problem without being in tension with relativistic causality by providing a local causal explanation of quantum correlations. Furthermore, measurement outcomes don't vary according to, for example, systems or worlds. It is a conservative QT in the sense that, unlike theories such as spontaneous collapse theories, no modifications of the fundamental equations of quantum theory are required to establish when determinate values arise. Moreover, in principle, arbitrary systems can be in a superposition for an arbitrary amount of time. To simplify, I will mostly assume non-relativistic QT and the Schrödinger picture Hilbert space-based finite dimensional QT.

\section{Main features: four conditions and two hypotheses}
I will start by presenting the main features of EnDQT and show why it fulfills the desideratum UT), which was explained in the previous section. Also, I will begin building the argument for why it allows for LC) and NR). The main features of EnDQT presented here\footnote{We will see that there are other versions of EnDQT if we vary these hypotheses and conditions, but we will see that this version so far is enough to fulfill the desiderata presented in Section 1} involve four conditions and two hypotheses about how the capacity to have determinate values, and to provide other systems that capacity, spreads through interactions. After some key assumptions, I will start by explaining the four conditions, which will require that I give an account of the role of quantum states (which was already mostly given in the previous section), systems, interactions, decoherence, and how having a determinate value and allowing other systems to have determinate values propagate via interactions inferred via decoherence, where specific chains of interactions are formed. Afterward, I will present the two hypotheses that support EnDQT, where the second hypothesis involves an account of what the systems that start SDCs could be. I will mention some empirical predictions that EnDQT provides throughout the next sections.

\subsection{The four conditions}
Before presenting the four conditions concerning how the capacity to have determinate values, and to provide other systems that capacity, spreads through interactions, I will explain some background key assumptions that I will make. To simplify, throughout this article, I will employ the familiar view that what exists are systems; a system is characterized by a collection of observables, and an observable of a system sometimes has a determinate value, where its eigenvalues represent the latter. This leads systems to ``have a determinate value.'' Or its observables sometimes have indeterminate values, leading systems to ``have an indeterminate value.'' Interactions are represented via QT, and some of them (which are represented via decoherence) lead systems to have a determinate value of an observable.

Different ontologies can make the above view more precise and allow EnDQT to adopt a more robust realism. One may understand determinate values of systems as referring to flashes that arise or are produced under interactions, i.e., an ontology of local events in spacetime (but differently from spontaneous collapse theories and with a different interpretation of the quantum state), but there are other ways.\footnote{The flash ontology was first proposed by \cite{Bell2004AreJumps} and named by \cite{Tumulka2006AModel}.} We could also view observables as representing determinables (e.g., position, energy, etc.) and determinate values as representing determinates of those determinables. Interactions give rise to a determinable with a determinate. Systems are collections of determinables, which at different moments of time, have determinates or not (e.g., having a spin-z with or without a determinate of spin-z) depending on their interactions like in the gappy version of quantum indeterminacy presented in \cite{Calosi2018QuantumIndeterminacy}. Quantum indeterminacy arises when we have a state of affairs constituted by a system lacking a determinate of a determinable.\footnote{Alternatively, we could have an ontology of quantum properties, and this is the one I prefer. See Appendix C and \cite{Pipa2024AnTheory}.} To be clear, according to EnDQT, determinacy and indeterminacy are objective features of the world. EnDQT postulates that (like other theories postulate other things). Determinacy is also associated with classicality (i.e., phenomena associated with classical physics) and measurement records, and our evidence for determinacy will involve these phenomena, as we will see.

I will consider a (quantum) system as occupying local regions of spacetime and being represented at a moment in time by (an equivalence class of) quantum states and observables that act on the quantum states that belong to the Hilbert space of the system. Given the aim of not being in tension with relativistic causality, I will be interested in an ontology constituted fundamentally by local systems and their local interactions,\footnote{This assumption can be made more adequate under a quantum field theoretic treatment.} and hence on systems whose observables act on quantum states concerning a single region of space.\footnote{For example, the larger system that forms a Bell pair would be a system localized in multiple regions of space. The quantum state of this system is an eigenstate of non-local observables.} I will be very liberal about what constitutes a system. For example, an atom's internal degrees of freedom could constitute one.

Concerning the observables of a system $S$, for the sake of parsimony and for the purposes of allowing for a local theory (more on this in Section 3), I will assume that:\\ 

\noindent Any observable $O$ of $S$, including the non-dynamical ones, outside of specific interactions of $S$ involving $O$, cannot have determinate values but rather have indeterminate values.\\

The assumption that systems have indeterminate values of any observable by default except during certain interactions comes from the perspective that the so-called links that connect state assignments to assignments of determinate values are idealizations that never occur in practice. For instance, \cite{Wallace2019WhatMechanics} provides good arguments for why the so-called Eigenstate-Eigenvalue Link is not a very useful assumption. As we have seen in the introduction, this link says that a system has a determinate value of an observable $O$ if and only if the system is in a state that is an eigenstate of $O$. A system rarely is in an eigenstate of such observable. When it is and if it is (since this occurs upon typically idealized projective measurements), it is, for a very brief moment of time, quickly evolving out of those states. Decoherence and decohering interactions provide a much less problematic way of assigning determinate values to the observables of systems because it is less grounded in idealized assumptions.\footnote{Furthermore, as argued by \cite{GIULINI1995291}, the so-called environmentally-induced superselection can be used to justify why systems are never found in a superposition of the eigenstates of certain observables, always having determinate values of that observable (e.g., mass or electric charge) upon interactions. More concretely, the eigenstates of the so-called non-dynamical observables, which are observables that are never observed in a superposition, are typically considered to be subject to superselection rules (see, e.g., \cite{Bartlett2007ReferenceInformation}). These rules can be regarded as prohibiting the preparation of quantum states in a superposition, which are eigenstates of some observable (e.g., electric charge) and assume a coherent behavior. Rather than postulating these rules, decoherence in a widespread environment in spacetime might be used to explain this superselection (see, e.g., \cite{Earman2008SuperselectionPhilosophers, GIULINI1995291}). EnDQT will assume this perspective here (not that if systems evolve back into a superposition of such initially superselected states, if they are always subject to those interactions, they will very likely maintain the original eigenstate). However, the validity of EnDQT is not tied to this assumption. Instead, for example, we can assume that there are certain non-dynamical observables that always have determinate values. Or, we can rather assume that these non-dynamical observables should be regarded more like constants and considered to be less fundamental than the dynamical ones. It is rather the behavior of systems due to dynamical observables (which are subject to decoherence) that is more fundamental and explains the features of non-dynamical observables, where this behavior is described/governed by laws.}
An important feature that EnDQT introduces is, as I have mentioned, the Determination Capacity (DC). The DC will have the great benefit of allowing for the formulation of a set of conditions that establish when systems have determinate values, but without modifying fundamental equations of QT and allowing for the fulfillment of the desiderata explained in Section 1.

Below, I will explain the conditions that will establish how the DC is transmitted between systems, where the DC is the capacity to give rise to other systems having a determinate value via interactions and to transmit the DC to other systems in the non-relativistic regime. Also, these conditions establish what it takes for a system to have a determinate value in this regime.\footnote{In a relativistic quantum field theoretic of these conditions, some further precisifications will be needed, but they won't affect their core features \cite{PipaToyEnDQT}.} Due to their conservativeness, I will call them Conservative Determination Conditions (CDCs). CDCs are laws or law-like features that describe/govern how determinate values arise and the capacity to have determinate values is transmitted via interactions. There are alternative determination conditions, but I found these to be the simplest and most conservative ones, as well as the ones that more appropriately describe a chain of decohering local interactions in the non-relativistic regime.\\ 

\noindent CDC1) For EnDQT, the world is fundamentally constituted by systems with indeterminate values of any observable. The determination capacity (DC) of  system $X$ concerning system $Y$ (DC-Y) is the capacity that $X$ has while interacting locally with $Y$,\vspace{5mm}

\noindent i) to allow $Y$ to have a determinate value under the interaction with $X$ that also leads $X$ to have a determinate value, where $X$ and $Y$ have a determinate value in the same spacetime region,\vspace{5mm}\\

\noindent ii) to provide the DC to $Y$ concerning another system $Z$ (DC-Z) if and only if a) $Z$ starts interacting locally with $Y$ while $Y$ is already interacting with $X$, and b) $Y$ has a determinate value due to $X$ and $Z$ doesn't disturb this process.\\

So, the DC propagates between systems via interactions because $Z$ can then have the DC concerning a system $K$ (DC-K) 
if and only if a) $K$ starts interacting with $Z$ while $Z$ is already interacting with $Y$, and b) $Z$ has a determinate value due to $Y$, 
and so on for a system $L$ that interacts with $K$ while $K$\footnote{Note that more systems may be needed so that $L$ has a determinate value.} interacts with $Z$, etc. Note that $X$ having a determinate value and 
$Y$ having a determinate value in the interaction in i) is the same event (i.e., it occurs in the same spacetime region. This is why, for EnDQT, interactions give rise to determinate values.

As we can see, CDC1) doesn't invoke the EEL to assign determinate values to observables of quantum systems. Rather, it considers that systems only have determinate values of observables during interactions with certain other systems. Otherwise, systems have indeterminate values of any observable. As we will see, the motivations for these features of CDC1) are connected with the aim of EnDQT of fulfilling the desideratum LC) of maintaining locality. Note that the assumption that systems only have determinate values under interactions doesn't mean that these values are less objective. The analogy here is with the binding energy of atoms. Binding energy arises under interactions, but we tend to consider binding energy as existing out there. 

I consider the assumption that the target system and environment have a determinate value in the same spacetime region, i.e., while interacting, as a conservative one. As foreshadowed in the previous section, I will use decoherence to establish when systems have determinate values. Since decoherence involves entanglement between the system and its environment, this assumption agrees with how systems become entangled at a time in local interaction. If system $A$ gets entangled with system $B$, system $B$ gets entangled with $A$. The quantum framework doesn't distinguish which one happens first. This assumption is also important in order for EnDQT to fulfill its aims. If these systems had determinate values at the same time in different spatial regions, the related events would be space-like separated, and according to special relativity, we would have relativistic reference frames where one event happens first, then the other, and thus we would fall into relationalism since which system has a determinate value would vary with the reference frame unless we favored a preferred reference frame. Furthermore, if this were the case, it would be more problematic to establish that the environment gave rise to the target systems having a determinate value. 

Now, why not consider instead that the events involving elementary systems of the environment having determinate values are time-like separated (i.e., occurring at different times in the same spatial region) rather than occurring in the same spacetime region? The reason why is more obvious but worth mentioning. This alternative assumption would be unsatisfactory since it could give rise to a certain future dependency in terms of when determinate values arise. To see this, let’s assume that each system of the environment has the DC and interacts with the target system, having a determinate value in this interaction before the target system, and thus the target system doesn’t yet have a determinate value (so the events where they have determinate values are time-like separated). Furthermore, let’s consider that we have a process that doesn’t give rise to decoherence because the environmental overlap terms oscillate between zero and non-zero. So, the entanglement between the system and the target system would oscillate over time. Since the environmental systems would already have determinate values, due to their entanglement with the target system (at least for some time), this would allow for an observer to gain knowledge about the determinate value of the target system by looking at the environmental systems that are highly entangled with the target system at least for a brief moment in time. Hence, we would have a contradiction with the assumption that the target system doesn’t yet have a determinate value. Although these values would oscillate quickly, this would still be a contradiction. To deal with this situation, one could argue that systems of the environment would only have determinate values in the present if there was decoherence due to these systems in the future. However, this would give rise to future dependency and likely either superdeterminism or retrocausality. Hence, I have considered that the target system and the environmental systems have a determinate value in the same spacetime region while they interact.\footnote{One may object to the assumption that decohering systems of the environment and the target system have determinate values in the same spacetime region by mentioning the collisional models of decoherence (see \cite{Zeh2003BasicTheory, Schlosshauer2007DecoherenceTransition} and references therein). In these models, one might be tempted to consider that the collisions that give rise to decoherence happen over time, and hence, the events that give rise to decoherence should be time-like separated. However, note that if collisions involving many systems happened at the same time, there wouldn’t exist a significant change in these models. Also, the fundamental theory of the world will likely be one of quantum fields, which doesn’t use the variable position, and ultimately, these models are only effective. The assumption regarding systems having determinate values in the same spacetime region shouldn't be strange. In physics, we are used to reify features that arise due to interactions, such as the bonding energy of atoms.}

Given CDC1), the DC is transmitted between systems via local interactions over spacetime,\\

\noindent For a system $X$ to interact with system $Y$ from time $t$ to $t^{\prime}$, the quantum states of $X$ and $Y$ must at least evolve from $t$ to $t'$ under the Hamiltonian of interaction representing the local interaction between them.\footnote{So, the Hamiltonian of interaction should represent interactions in the same spacetime region. This assumption can be made more precise via quantum field theory.}\\

As I have mentioned in the previous section, the reasons for posing CDC1-ii) are that, first, I want to provide clear criteria for systems to give rise to other systems having determinate values and to transmit that capacity to these systems. Without a criterion like CDC1), tracking when systems have the DC would be hard.
Second, I also want clear criteria that establish that they can lose that capacity since, given the above desiderata, I want to allow for the possibility of arbitrary systems to be in a superposition for an arbitrary amount of time, even if they are interacting with other systems. So, we want these latter systems to lose that capacity. Third, I want to appeal $only$ to (local) interactions represented by QT for systems to have the DC and not some other criteria (except in the case of some special and plausible systems; more on this below). Hence, I have also assumed CDC-ii-a).\footnote{Note that establishing that interactions are local would add some complications in the quantum field theoretic case because, in that domain, we will want the quantum field systems that transmit the DC to be localized in a single bounded spacetime region.} Fourth, I want such criteria to be plausible in the sense that it is in agreement with what seems to occur in decoherence models and measurement-like situations where system $E$ (such as a measurement device) that gives rise to a system $S$ having a determinate value, before interacting with $S$, has a determinate value of some observable like we seem to have in the \textit{ready state} of a measurement device (i.e. before the measurement device does its measurement). Hence, I have assumed CDC1-ii-b). 
Now, which interactions give rise to determinate values? As I have explained in the previous section, since my aim here is to be conservative, I will use decoherence to represent those interactions because physicists standardly use it to represent measurement-like interactions.

I will now explain briefly decoherence and some of the assumptions I will make when interpreting it. I will also highlight with numbers $1)$ and $2)$  two of the assumptions made in decoherence models, which EnDQT will later justify. Let's consider a system $S$ in the following state with complex non-zero coefficients $\alpha_{i}$,

\begin{equation}
\ket{\psi}_{S}=\sum_{i=1}^{N} \alpha_{i}\left|s_{i}\right\rangle_{S} 
\end{equation}

\noindent and an environmental system $E$ of $S$, constituted by many subsystems, interacting strongly with system $S$. For instance, $|\psi\rangle_{S}$ could be a superposition of spin-z eigenstates. Furthermore, $S$ will be interacting strongly with the many subsystems with a spin in a specific direction that constitutes $E$, i.e., the Hamiltonian of interaction $H_{int}$ dominates the systems' evolution.\footnote{I.e., the spectral frequencies available in $H_{int}$ are much higher than the ones of the self-Hamiltonian of the systems. See, e.g., \cite{Cucchietti2005DecoherenceEnvironments}.} So, $1)$ the dynamics will be driven by an interaction Hamiltonian, which governs or describes the interactions between systems that can affect specific observables. For simplicity, throughout this article, I will assume this kind of evolution of systems under interactions with their environment.\footnote{This is the so-called quantum-measurement limit and is typically successful in describing many measurement-like interactions (\cite{Schlosshauer2007DecoherenceTransition}) In this case, the energy scales of the system-environment interaction are much larger than the energy scales associated with the self-Hamiltonians of the system and environment. More complex models of decoherence (see, e.g., \cite{Zurek1993CoherentDecoherence}) where the system doesn't interact strongly with the environment, and self-Hamiltonian also has some weight in the evolution of the system may give rise to different observables with determinate values depending on the initial quantum states. For simplicity, I will not talk about these more complex cases here or analyze how, in these cases, SDCs could be formed.} Now, let's assume that $S$ locally interacts with $E$, where their interaction is represented via the standard von Neumann interaction at least approximately by

\begin{equation}  
\left|s_{i}\right\rangle_{S}\left|E_{0}\right\rangle_{E} \rightarrow_{\widehat{U}}\left|s_{i}\right\rangle_{S}\left|E_{i}(t)\right\rangle_{E}
\label{eqdecoherence0}
\end{equation}
for all $i$ or

\begin{equation}
\left(\sum_{i=1}^{N} \alpha_{i}\left|s_{i}\right\rangle_{S}\right)\left|E_{0}\right\rangle_{E} \rightarrow_{\widehat{U}} \sum_{i=1}^{N} \alpha_{i}\left|s_{i}\right\rangle_{S}\left|E_{i}(t)\right\rangle_{E}=|\Psi\rangle_{S+E}.
\label{eqdecoherence}
\end{equation}

The distinguishability between the different states of $E$ concerning its interactions with $S$ can be quantified via the overlap between quantum states $\left\langle E_{i}(t) \mid E_{l}(t)\right\rangle_{E}$. The impact of this distinguishability of the states of $E$ on $S$ can be analyzed via the reduced density operator of $S$, obtained from tracing over the degrees of freedom of $E$ in $|\Psi\rangle_{S+E}$,

\begin{equation}
\begin{aligned}
\hat{\rho}_{S}(t)=\sum_{i=1}^{N}\left|\alpha_{i}\right|^{2}\left|s_{i}\right\rangle_{S}\bra{s_{i}} +\sum_{i, l=1, i \neq l}^{N} \alpha_{i}^{*} \alpha_{l}\ket{s_{i}}_{S}\left\langle s_{l}\right|\left\langle E_{i}(t) \mid E_{l}(t)\right\rangle_{E}+\\
\alpha_{l}^{*} \alpha_{i}\left|s_{l}\right\rangle_{S}\left\langle s_{i}\right|\left\langle E_{l}(t) \mid E_{i}(t)\right\rangle_{E}.
\label{eqdecoherence2}
\end{aligned}
\end{equation}

Under the Hamiltonian of interaction describing the interactions between the target system and many systems, and $2)$ systems having randomly distributed initial states and coupling constants,\footnote{A nonrandom initial distribution of the quantum states of the environment can lead to not having so many phases that would cancel each other out in the off-diagonal terms of the density matrix, not, therefore, leading to decoherence.} in decoherence models we obtain that $\left\langle E_{i}(t) \mid E_{l}(t)\right\rangle_{E}$ quickly decreases over time until $\left\langle E_{i}(t) \mid E_{l}(t)\right\rangle_{E} \approx 0$ when has $E$ is constituted by many systems. The recurrence time of this term (back to not being significantly small in comparison with the other terms) in this case tends to be so large that it can exceed the universe's age, giving rise to a quasi-irreversible process, mathematically speaking. Note that I will provide a distinct and ``more pragmatic'' sense of irreversibility below concerning other features of the environment that should be distinguished from this sense of irreversibility. 

So, when either eq. (\ref{eqdecoherence0}) or (\ref{eqdecoherence}), together with (\ref{eqdecoherence2}) hold to describe a system $S$ interacting with $E$, I will consider that $S$ was decohered by $E$ or that the superposition of the states of $S$ (also often called pointer states) were decohered by the states $\left|E_{i}(t)\right\rangle_{E}$ of $E$. The observable of $S$ formed by these pointer states is the pointer observable.\footnote{Although the processes represented by eq. (\ref{eqdecoherence0}) are not what is often called decoherence, I will call them that since (more realistically) they may hold only approximately. I consider these processes rare since, typically, systems aren't in an eigenstate of some observable. The latter should rather be seen as an idealization. See, e.g., \cite{Wallace2019WhatMechanics}.}

Importantly, note that by decoherence here, it does not necessarily mean the process of destruction of interference (like it is often assumed to mean) but rather whatever is represented via these models. As it will be clearer, it will only refer to the process of destruction of interference when $E$ has the DC. I will call the process that is modeled via decoherence that involves an environmental system having the DC, \textit{fundamental decoherence}. More concretely, I will assume that when $E$ has the DC, the reduced density operator $\hat{\rho}_{S}$ that is obtained from eqs. (\ref{eqdecoherence0}) or (\ref{eqdecoherence}) can be used to predict the determinate values and resultant statistics of the consequences of this interaction (i.e., the determinate values of $S$ and $E$), as well as the timescale in which we can update the state of $S$ to one of the $\left|s_{i}\right\rangle_{S}$ under decoherence. Relatedly, as it will become clearer below when $E$ has the DC, this model can directly account for the disappearance of interference effects due to $S$ in situations where it interacts with $E$. So, in this case, when states of the environment become extremely distinguishable under interactions between $S$ and $E$ over time, we have,

\begin{equation}
\hat{\rho}_{S} \approx \sum_{i=1}^{N}\left|\alpha_{i}\right|^{2}\left|s_{i}\right\rangle_{S}\left\langle s_{i}\right| . 
\end{equation}

 From now on, I will call the states $\left|E_{i}(t)\right\rangle_{E}$ and $\left|E_{l}(t)\right\rangle_{E}$ for all $i, j$ with $i \neq j$ when they are distinguishable, i.e., $\left\langle E_{i}(t) \mid E_{l}(t)\right\rangle_{E} \approx 0$, simply eigenstates of an observable $O^{\prime}$ of $E$ because the projectors onto these states will approximately commute with the observable $O^\prime$ of $E$.

 So, given the reasons above, I will assume that\\

\noindent CDC2) Interactions between system $X$ and a set of systems that form a system $Y$ that may be larger, which has the DC, lead system $X$ to have a certain determinate value, where the distinguishability of the physical state of $Y$ concerning the possible determinate values of $X$ allows us to infer if $X$ will have a determinate value among the possible ones and when that happens. Such distinguishability is inferred via the (fundamental) decoherence of $X$ by $Y$, and where it is indeterministic which one of the values will arise among the possible ones.\\

Thus, decoherence for EnDQT is a tool to infer how and when determinate values arise, where such values arise via an indeterministic process. More on this below.

The determinate value of $X$ could be a measurement outcome and the one of $Y$ could be a measurement device. Given the above assumptions and CDC2), we have that\\

\noindent\textit{In the simple situations that we will be concerned with here where the Hamiltonian of interaction dominates the evolution of the system, in order for system $Y$ to have a determinate value $v$ of $O_Y$, i) the observable $O_Y$ of $Y$ that is monitored by system $X$ that has the DC, and whose eigenstates are decohered by $X$ in the sense above, has to at least approximately commute with the Hamiltonian of interaction $H_{XY}$ representing the interaction between $X$ and $Y$, i.e., $[H_{XY},O_Y]\approx 0$ (this is the so-called \textit{commutativity criterion}),\footnote{See \cite{Schlosshauer2007DecoherenceTransition} and references therein. Note that $[H_{XY},O_Y]\approx 0 \Rightarrow [H_{XY},O_Y \otimes \mathbb{I}_Y]\approx 0 $. Furthermore, this criterion implies that all terms in a Hamiltonian of interaction $H_{int}$ will individually satisfy this criterion. In more complex models of decoherence (see previous footnote), note that this monitoring may be indirect, such as the decoherence of momentum in more complex models of decoherence than the ones mentioned here \cite{Zurek1993CoherentDecoherence}, where there is direct monitoring of the position. The latter is contained in the Hamiltonian of interaction of the system (but not the former), and that's why it is considered that the decoherence of the momentum is indirect.} and ii) where the eigenvalues of $O_Y$ include $v$.}\\

Thus, the determinate values that arise are the ones that are dynamically robust under interactions with certain systems that have the DC. Note that the times where a system has a determinate value due to interactions with environmental systems will be represented and inferred via the time that the overlap terms above go quasi-irreversibly to zero under decoherence due to a system having the DC (i.e., the decoherence timescale). So, the above overlap terms going quasi-irreversibly to zero will allow us to infer if the local interactions between $S$ and the environmental systems that have the DC succeeded in giving give to $S$ having a determinate value of $\mathrm{O}$. Furthermore, it is indeterministic which value $S$ will have among the possible ones, where the latter is given by the eigenvalues of $O$.\footnote{For simplicity, I will not address here the case where we don't have maximum distinguishability of the states of the environment $E$, concerning the states of the target system. Roughly, this case can be inferred by the overlap terms of the environment not being zero or one stably over time. Furthermore, in these situations, the target system won't have a determinate value due to $E$. See Appendix C and \cite{Pipa2024AnTheory} for an ontology of quantum properties that addresses this case.}

The reader familiar with decoherence might have found the above description of fundamental decoherence as missing an often-cited ingredient of decoherence, which is that decoherence is associated with the openness of the environment or is associated with ``the entanglement of the degrees of freedom of the system'' with inaccessible or uncontrollable environments that make the state of the whole larger system hard/impossible to reverse. I will now relate this feature to fundamental decoherence models. First, let's call the models of decoherence where we don't know if the environmental systems have the DC or that is not specified, \textit{pragmatic decoherence models}. In contrast with fundamental decoherence models, pragmatic decoherence models involve other considerations that go beyond the model itself, such as whether the environment is open or not. \textit{Pragmatic irreversible decoherence models} are models that involve situations where it’s assumed that it is very difficult or impossible to reverse the process represented by them because they concern the interaction with many systems that are difficult to control and which often involve open environments. Importantly, these are the assumptions that go beyond the models of decoherence themselves, being rather extra assumptions postulated by the modeler. The processes represented by the pragmatic irreversible decoherence models are the processes that we normally call decoherence. However, along with my explanation of the other CDCs, I will explain below that the processes involving fundamental decoherence resemble, in important ways, the processes represented by the pragmatic irreversible decoherence models. Afterward, I will put the relation between these models in a clearer foundation.

Returning to the CDCs, we can now use CDC2) to spell out CDC1) in terms of fundamental decoherence.\\

\noindent CDC1*)The DC-Y of $X$ is the capacity that $X$ has while interacting with $Y$,\vspace{5mm}

\noindent i*) to decohere $Y$, which leads both systems to have a determinate value of an observable O during their interaction, where, in the absence of these kinds of interactions, systems have indeterminate values of any observable. Let's suppose that system $S$ in eq.\eqref{eqdecoherence} is an instance of $X$, and system $E$ is an instance of $Y$. The possible values of $X$ are represented by the eigenvalues of the observable that the quantum states $\left|s_{i}\right\rangle_{S}$ of $S$ in eq.\eqref{eqdecoherence}  are eigenstates of. The possible values of $Y$ are represented by the eigenvalues of the observable that the quantum states $\left|E_{i}(t)\right\rangle_{E}$ in eq.\eqref{eqdecoherence} are eigenstates of and \vspace{5mm}

\noindent ii*) to provide the DC-Z to $Y$ if and only if ii-a) $Z$ starts interacting with $Y$ while $Y$ is interacting with $X$ and ii-b) $Y$ is decohered by $X$, and $Z$ doesn't disturb this process, i.e., driving away to other states, the states of $Y$ that are being decohered by $X$.\\

We could suppose that while $S$ in the example above interacts with $E$, it starts interacting with another system $S^\prime$, which would be an instance of $Z$. Then, when $S$ is finally decohered by $E$, it could decohere $S^\prime$ and give rise to both $S$ and $E^\prime$ having a determinate value. Such values would be represented like in the case of eq.\eqref{eqdecoherence}, but now the target system would be $S^\prime$, and the environment would be $S$. So, $S$ would have a determinate value when it interacts with $E$ and another when it interacts with $S^\prime$.

I want to emphasize that the CDSs are just one possible condition, and there are other possible ones. As we can see, via the CDC1*) and CDC1), I postulate that in order for systems to have the DC and transmit it to other systems, they need to form a tight chain of interaction with some rules. As we will see further below, the idea is that I want to explain how sometimes we manage to isolate systems and give rise to interference phenomena via them. Imagine that this tight chain wasn't postulated, and I had more loose determination conditions. For instance, imagine that these conditions allowed a system to simply retain the DC forever when it interacts with another system that has the DC. We would have $stray$ systems with the DC, and it becomes hard to justify why we sometimes can isolate systems from an SDC and, thus, from systems with the DC.

I also want to emphasize that these criteria and a few others will allow EnDQT to provide the great benefit of giving conditions for when determinate values arise, but without modifying the fundamental equations of QT like spontaneous collapse theories or adopting a relationalist view or hidden variables. As we can see, systems with the DC and determinate values propagate through interactions. I will call a chain of interactions between systems that propagate the DC or just give rise to systems having determinate values a \textit{stable determination chain} (SDC). It is called SDC because the process that gives rise to systems having determinate values can be seen as a process that, in order to occur, needs to be stable in the sense that it stably obeys CDC1*). More concretely, in order to infer that systems have determinate values of some observable, it is necessary that the overlap terms of the quantum states of the environment go $stably$ to zero. Also, given CDC1*) again, it is also necessary that systems that interact with systems that are going over this process, are $stably$ not disturbed from going over this process.

One may object that contrary to what is claimed, EnDQT is really modifying the equations because it’s posing a time where an indeterministic process occurs and stops a process represented deterministically and unitarily. Note, however, that the point is that we won’t have to modify the fundamental equations of QT to infer this process contrary to, for example, as I have said, spontaneous collapse theories.

Returning to our conditions, given an SDC, we run the risk of an infinite regress because it is unclear where and when it starts. To circumvent these issues, I will consider that\\

\noindent CDC3) There are two kinds of systems that constitute an SDC, or two interrelated kinds of roles they have \vspace{5mm}

\noindent -Initiator systems or initiators, which are systems that a) have the DC concerning any system by default (i.e., they always have the DC-X for any system X), i.e., independently of their interactions with other systems. Or b) they are the first systems that have the DC or the ones that we assign the DC initially in our models. Because of this, initiators are the systems that start SDCs.\vspace{5mm}

\noindent -Non-initiator systems are systems that don’t have the DC concerning a system by default but have it due to their interactions with other systems that have the DC.\\

So, the (fundamental) decoherence of some system $S$ by an initiator is necessary and sufficient to allow that later system to have a determinate value of some observable $O$ of $S$. Also, $S$ can acquire the DC concerning some other system $S^{\prime}$ if it interacts with this system while it interacts with the initiator. We will see further below a system that is a plausible candidate to be an initiator, and which is widely accepted in cosmology. I will argue that although initiators arose to address the measurement problem, they have the advantage of potentially addressing other problems in the foundations of physics, which is a good sign. Note that initiators of the kind b) are identical to non-initiators. They just end up being the first systems with the DC in the universe. Or if we conceive that the universe existed forever, as well as the DC, they are the systems that we assign the DC to in our models at first. On the other hand, initiators of the kind a) are special systems because they always have the DC concerning any system. It is less parsimonious to postulate initiators of the kind a) than b).

SDCs are represented by directed graphs, which represent the propagation of the DCs, or its potential propagation, which gives rise to systems that belong to it having determinate values. I will represent this interaction between $X$ that leads $Y$ to have a determinate value (together with $X$) and potentially leads $Y$ to have the DC concerning some other system as $X \rightarrow Y$. When $Y$ has the DC-Z, which leads some other system $Z$ to have a determinate value, I will represent it as $X \rightarrow Y \rightarrow Z$.\footnote{For simplicity, here I will mostly not care about the distinction between a token network, which represents concrete interactions between systems, and type networks, which represent interactions between types of systems that exist in specific regions of spacetime.} In some graphs that aim to depict the whole situation, the systems with only directed arrows towards them represent systems that have the DC but won't end up transmitting it to other systems. An SDC ends when it reaches these systems. The nodes with no directed arrows towards them represent the initiators.

Let's turn to the last CDC). In decoherence models, the environment of a system is typically composed of many subsystems. So, it is more realistic and plausible to assume that\\

\noindent CDC4) For a system S to have the DC concerning some system S’, its subsystems must have the DC concerning S’ or its subsystems (value-mereology assumption).\\

\begin{figure}[hbt]
    \centering
    \includegraphics[scale=0.4]{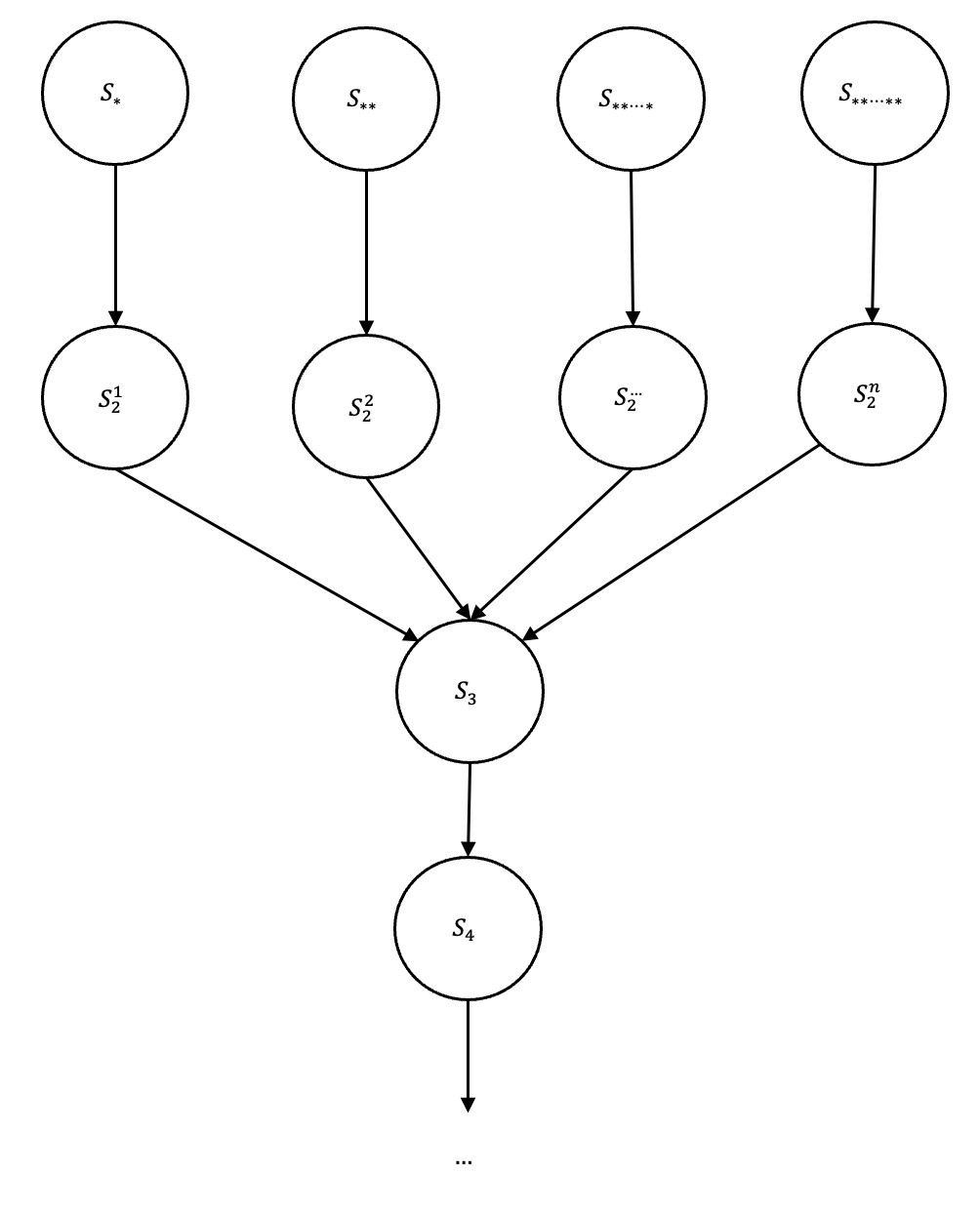}
    \caption{Directed graph that represents the transmission of the DC between systems.}
    \label{fig:example1}
\end{figure}

\begin{figure}[hbt]
    \centering
    \includegraphics[scale=0.4]{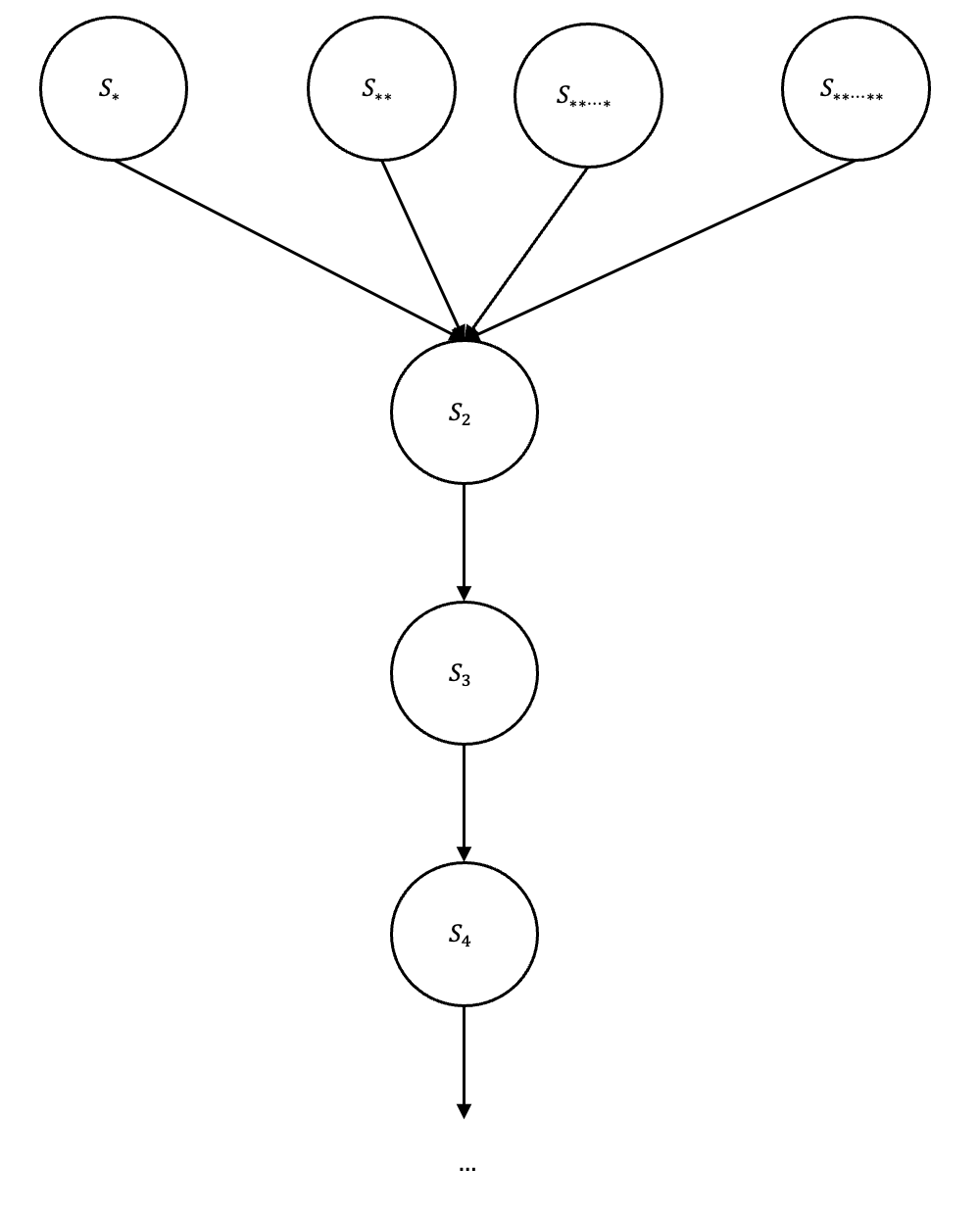}
    \caption{Directed graph involving common effects like the previous one, but just represents the whole system $S_2$.}
    \label{fig:example2}
\end{figure}

For instance, let's consider that instead of $S_{2}$ above, we have a subsystem $S_{2}^{i}$ of $S_{2}$ for some $i$ where $S_{2}^{i}$ is not able to decohere $S_{3}$ alone, but $S_{2}$ is. $S_{2}$ would just be able to give rise to $S_{3}$ having a determinate value, having the DC, if its subsystems $S_{2}^{i}$ for all $i$ interacted with subsystems of $S_{1}$, acquiring the DC, where $S_{1}$ and its subsystems have the DC. So, subsystems of a system, such as $S_{2}^{i}$ for all $i$ are considered to form a ``cause'' for the ``common effect,'' which is a system $S_{3}$ having a determinate value of one of its observables. Each subsystem of $S_{2}$ would also have another determinate value when $S_{3}$ has a determinate value. These interactions are represented by a directed graph with ``colliders'' (i.e., a series of arrows that point to a single system, see Figure \ref{fig:example1}). We can also simplify the structure of the above graph by treating $S_{2}$ as a whole, neglecting its subsystems (Figure \ref{fig:example2}). 

CDC4) further constrains the structure and persistence of SDCs, whose elements already have to obey the other CDCs. It is plausible to consider that typically, a system will interact with many systems in spatiotemporal regions, which form a larger system $E$. In order for a system $E$ with its subsystems to decohere another $S$, its subsystems will have to be interacting with other systems that have the DC concerning them, and those latter systems will have to be interacting with other systems, and so on. So, we will need many systems for a system interacting with a larger system $E$ to have a determinate value due to $E$. Also, the more macroscopic $S$ is, the more subsystems in principle has, and so more systems belonging to SDCs need to be interacting with $S$ in order for it to have a determinate value and the DC concerning other systems. This process resembles a process represented by a pragmatic irreversible decoherence model because we will have many systems involved in the interaction that leads $S$ to have a determinate value and perhaps the DC, and these systems will be hard to control. Also, the SDCs give rise to indeterministic processes, which make these interactions impossible to reverse unitarily.\footnote{Note that it’s plausible to consider that systems that belong to an SDC having determinate values due to a target system $A$ will depend on which other systems are interacting with $A$ at that time (all of them having a determinate value, constituting single events).  Via their influence on $A$ having a determinate value, they also influence each other to have determinate values. Because of this, decoherence of a target system due to a collective of systems allows us to infer the process that gives rise to a collective of systems having determinate values (such as in the case of central spin models, i.e., a model with a system with a spin interacting with multiple systems during a period, see \cite{Cucchietti2005DecoherenceEnvironments} and Appendix A).}

CDC1)-CDC4) constitute the CDCs. As we can begin to see more clearly, EnDQT has the benefit of providing a no non-local, retrocausal, or superdeterministic hidden-variable criterion for systems to have determinate values without modifying the fundamental equations of QT or appealing to relationalism.

In principle, the fundamental decoherence models will be grounded in interactions between the fundamental systems that have the DC.\footnote{I am setting aside here exotic views that consider that such systems don’t exist.} Since we are working in a non-relativistic regime, I have not provided such a model here, and I will be neutral about it. Also, given that we are also working in this regime, I haven’t provided a detailed relation between these different levels.

In Appendix A, I will also briefly provide a toy model to illustrate how EnDQT represents physical interactions. Readers are encouraged to read for more thorough details. That model will involve a series of interactions modeled via the spin-spin decoherence models.\footnote{\cite{Zurek1982Environment-inducedRules, Cucchietti2005DecoherenceEnvironments}.} These are the simplest decoherence models, and the example in the appendix is the simplest SDC model that one can conceive. I will also briefly argue that in a quantum field theoretic model, Lorentz and general covariance of the laws in principle will be respected. Furthermore, we will see that, in principle, any model of decoherence fulfilling CDC1)-CDC4) can be incorporated into this kind of SDC model. Here, in the main text, I will rather consider a simple and very schematic example. This example will involve systems $A$, $B$, and $C$, where $A$ is an initiator, in a toy mini-universe where the SDC that will be formed has the following structure, $A \rightarrow B \rightarrow C$. We assume that these will be quick interactions occurring at timescales where we can neglect the intrinsic evolution of the systems. For example, $A$ could be a system well approximated by a large set of quantum harmonic oscillators\footnote{Or more precisely, bosonic modes. These interactions can be represented via the spin-boson mode. See, e.g., \cite{Leggett1987DynamicsSystem, Schlosshauer2007DecoherenceTransition}.} or a set of spin-1/2 systems or two-level systems that can be approximated as spin-1/2 systems (see toy model in Appendix A). Each set of quantum systems interacts with a single spin-1/2 system. The collection of these systems constitutes system $B$. Then, system $C$ could be another two-level system that will interact with $B$.\footnote{The interactions between the systems that constitute $B$ and system $C$ would be modeled via the so-called spin-spin decoherence models developed in \cite{Zurek1982Environment-inducedRules, Cucchietti2005DecoherenceEnvironments}.} Let's assume that $C$ starts interacting with the systems that constitute $B$ while $B$ are interacting with $A$ so that $B$ has the DC-$C$ (in agreement with the value mereology assumption), and $B$ can end up transmitting the DC to $C$ concerning some other system that $C$ might end up interacting with.

 Moreover, in order for this interaction to fulfill CDC-ii) and to simplify, let's assume that when $C$ starts interacting with $B$, it doesn't change significantly the initial state that $B$ had when it started interacting with $A$. So, when $B$ and $C$ begin interacting, let’s neglect the evolution of the quantum states of $B$ while $A$ and $B$ interact, such that we can idealize that $B$ and $C$ start interacting only when the interaction between $A$ and $B$ ends.\footnote{Another way to pose this assumption is to consider that while $A$ interacts with $B$, $C$ starts interacting with B in such a way that it doesn’t drive the states of $B$ out of being states that $A$ decoheres in the following sense: the Hamiltonian of interaction of $A$ and $B$ would still at least approximately commute with the (pointer) observable that these states are eigenstates of.} Thus, we can just analyze the evolution of the quantum states of $A$ while $A$ and $B$ are interacting, where this interaction ends approximately at $t^{\prime}$ (in a certain relativistic reference frame). Let's put a subscript SDC on the quantum states of a system if that system is an initiator or has the DC relative to some system belonging to an SDC. We then have the following interaction between $A$ and $B$,
\begin{gather}
\begin{aligned}
\left|E_{\text {ready }}\right\rangle_{\text{A SDC}}\left(\alpha^{\prime}\left|E_{0}^{\prime}\right\rangle_{B}+\beta^{\prime}\left|E_{1}^{\prime}\right\rangle_{B}\right)\left(\alpha|\uparrow\rangle_{C}+\beta|\downarrow\rangle_{C}\right) \rightarrow_{t^{\prime}} \\
\left(\left|E_{0}\left(t^{\prime}\right)\right\rangle_{\text{A SDC}}\left|E_{0}^{\prime}\right\rangle_{B}+\left|E_{1}\left(t^{\prime}\right)\right\rangle_{\text{A SDC}}\left|E_{1}^{\prime}\right\rangle_{B}\right)\left(\alpha|\uparrow\rangle_{C}+\beta|\downarrow\rangle_{C}\right) . 
\end{aligned}
\label{exampleSDC1}
\end{gather}
 If $\left\langle E_{0}\left(t^{\prime}\right) \mid E_{1}\left(t^{\prime}\right)\right\rangle_{\text{A SDC}}\approx 0$ and $\left\langle E_{1}\left(t^{\prime}\right) \mid E_{0}\left(t^{\prime}\right)\right\rangle_{\text{A SDC}} \approx 0$ quasi-irreversibly when $A$ and $B$ end their interaction, we infer that $B$ has a determinate value of the (pointer) observable $monitored$ by $A$ at $t^{\prime}$ that arises from their interaction (i.e., let's assume that is either 0 or 1) and acquires the DC-C. Let's assume that $B$ has a determinate value 0. Now, let's consider the interaction between $B$ and $C$, and assume that, given their interaction Hamiltonian, it ends at $t^{\prime \prime}$,

\begin{equation}
\left|E_{0}\left(t^{\prime}\right)\right\rangle_{\text{A SDC}} \ket{E_{0}^{\prime \uparrow}\left(t^{\prime \prime}\right)}_{B}|\uparrow\rangle_{C}+
\left|E_{0}\left(t^{\prime}\right)\right\rangle_{\text{A SDC}} \ket{E_{0}^{\prime \downarrow}\left(t^{\prime \prime}\right)}_{B}|\downarrow\rangle_{C}.
\label{exampleSDC2}
\end{equation}

The evolution of the interaction between $B$ and $C$ could be analyzed via the reduced density operator $\rho_{C}(t)$. Their interaction will allow $C$ to have a determinate value ($\uparrow$ or $\downarrow$) at $t^{\prime \prime}$ if $\bra{E_{0}^{\prime \uparrow}\left(t^{\prime \prime}\right)} E_{0}^{\downarrow}\left(t^{\prime \prime}\right)\rangle_{B} \approx 0$
 and $\bra{E_{0}^{\prime \downarrow}\left(t^{\prime \prime}\right)} E_{0}^{\uparrow}\left(t^{\prime \prime}\right)\rangle_{B} \approx 0$ quasi-irreversibly when $B$ and $C$ end their interaction. $B$ will have a determinate value at $t^{\prime \prime}$ that arises from its interaction with $C$ where the possible values that it can have are represented via the eigenvalues of the observable that $\ket{E_{0}^{\prime \and \uparrow}\left(t^{\prime \prime}\right)}_{B}$ and $\ket{E_{0}^{\prime \and \downarrow}\left(t^{\prime \prime}\right)}_{B}$ are eigenstates of. Furthermore, $C$ can have the DC concerning some other system $D$ if it interacts with it before the interaction with $B$ ends. Note that since system $A$ is an initiator, if it is of the kind a), it has the DC concerning any system. So, if it is of this kind, its ability to give rise to other systems having determinate values and allowing them to have the DC doesn't depend on its interactions with other systems.

Now, given the CDCs we can see how EnDQT justifies the assumptions $1)$ and $2)$ of decoherence models. Regarding $1)$, it is important that the Hamiltonian of interaction assumes a specific form to account for decoherence (given by the above commutativity criterion), which depends on the observables of the environment and the system because we can regard it as representing the law-like dynamics of SDCs, which leads to determinate values (more on this below when we assume a hypothesis that connects the pragmatic irreversible decoherence models with the fundamental decoherence models).

Regarding $2)$, the initial states of the systems (and associated coupling constants) are randomly distributed (which gives rise to the overlap terms going quasi-irreversibly to zero in decoherence models) because the subsystems of $E$ also need to belong to SDCs that indeterministically gave rise to the subsystems of $E$ having determinate values before interacting with $S$ and, hence, to a random distribution of their quantum states (and associated coupling constants). What about the distribution of the quantum states of the initiators? There are various possibilities, which will depend on what we regard as initiators, which are speculative. For example, that can be regarded as a brute fact of ``the initial conditions of the universe'' where initiators could be located. Or, given enough time, the quantum states of the initiators can achieve this overlap that triggers the process of propagation of the DC (more on this below).\footnote{See also Appendix A.} 

As we will see, the explanation for why systems with the DC give rise to determinate values in this law-like way can be elegantly (as we will see) traced back to the initiators. So, one more benefit of adopting EnDQT with its CDCs is that it addresses some ad hoc or, in my view (at least), strange features of decoherence by seeing it as a tool to represent the dynamics of SDCs. In Appendix D, I go over a simple example of how EnDQT can model interference phenomena.

\subsection{Two hypotheses}
I will now explain some natural hypotheses about initiators and the structure of SDCs that I will make to satisfactorily fulfill the goal of achieving UT), LC), and NR). These hypotheses will clarify some of the claims above and address some of the vagueness associated with models of decoherence.

I have been assuming that we can rely on fundamental decoherence to infer and represent how determinate values arise via SDCs. Fundamental decoherence represents the interactions between systems, starting with the initiators. The example above and CDC1)-CDC4) involve these models. So, fundamental decoherence represents interactions that render systems with determinate values in a certain situation involving an environment with a DC, given an appropriate and local Hamiltonian. Importantly, in order for a system $S$ to have a DC and lead other systems to have determinate values, it is plausible that it will typically have to belong to an environment with many systems that propagate the DC, which seems to lead to a process that is hard to reverse/control due to these many systems that have determinate values indeterministically. Furthermore, the more macroscopic is $S$, the harder this seems to be. I also explained that fundamental decoherence models seem to represent phenomena that resemble those represented by pragmatic irreversible decoherence models. Remember that pragmatic irreversible decoherence models are models that involve situations where it’s impossible to reverse the process they represent because they concern open/inaccessible environments or environments with many systems (but there is no reference to systems having the DC).  It seems that we can use pragmatic irreversible decoherence models to infer when the processes represented by fundamental decoherence occur where the environmental systems have the DC. But, as we will see, more needs to be said about the behavior of SDCs and fundamental decoherence to ground the typically used pragmatic irreversible decoherence models as a proper inferential tool. More concretely, as an inferential tool to infer when systems have determinate values due to systems $E$ that have the DC in situations where we have not followed the interactions between systems that lead $E$ to have the DC.

As a reminder, I have called the models of decoherence that don’t necessarily track the interactions involving systems with the DC pragmatic decoherence models. Another kind of pragmatic decoherence model is what I will call the \textit{pragmatic reversible decoherence models}. These are models that represent a process that apparently involves decoherence in the sense that the overlap terms of the environment go quasi-irreversibly to zero. However, someone in some privileged position could reverse this process via operations on the system and environment, which is sometimes called recoherence. Or, to put it less pragmatically, they don’t involve enough degrees of freedom to be considered irreversible. This reversible process often occurs inside isolated environments or situations where the environmental degrees of freedom don’t become inaccessible to be reversed due to their in-practice isolation. Thus, the processes represented by these models aren't what we typically consider to be decoherence.

If we aren't careful, the distinction between a pragmatic reversible model and an irreversible one may be ambiguous in some situations. The Wigner's friend scenario\footnote{\cite{Wigner1961RemarksQuestion}.} is an example of a situation. Suppose an isolated lab occupies an arbitrarily large spatiotemporal region with a human agent inside (a ``friend''). The lab is isolated in such a way that Wigner outside the lab can unitarily manipulate the state of the friend plus their target system that the friend interacts with, treating both as being in an entangled state. So, in this case, we could have an enormous lab with many systems getting entangled with the friend and their target system S for a long time. However, if we were Wigner, we wouldn't consider that there was decoherence of S by the friend because Wigner could still unitarily reverse the state of the friend and their system. He would rather consider treating the friend plus target system interaction via a reversible decoherence pragmatic model. However, if the lab were open, he would treat their interaction via an irreversible decoherence pragmatic model. So, how do we exactly distinguish a reversible decoherence process from an irreversible one, since the reversible could also involve many systems? Let's call this problem the \textit{Wigner's friend ambiguity problem}. As the discussion in the previous paragraphs suggests, if we want to address the measurement problem without changing the fundamental equations of QT, adding hidden variables to it, or adopting a relationalist view (more on this below), it seems that a way to deal with this problem is via paying attention to what constitutes an open environment and how it relates with SDCs.

I have pointed out how fundamental decoherence involving members of an SDC seems to lead to processes represented by the pragmatic irreversible decoherence models. The first hypothesis aims to deal with the above ambiguity problem and ground the success of the pragmatic irreversible decoherence models in helping account for determinate values arising via SDCs in open environment situations, as opposed to the reversible models. These pragmatic irreversible decoherence models need to obey two plausible further constraints (given the local features of EnDQT): they need to represent local interactions (i.e., local), and they need to be empirically successful. Thus, I will hypothesize that\\

\noindent The SDCs in our world are widespread in such a way that the pragmatic empirically successful local irreversible decoherence models in open environments track the interactions between systems that belong to SDCs serving as an environment for a target system that doesn't belong to an SDC, but end up belonging to it. However, the SDCs in our world are also such that there can also exist processes represented via local empirically successful reversible decoherence pragmatic models, where the latter are tracking the interactions between systems that don't belong to SDCs (\textit{SDCs-decoherence hypothesis}).\\

 In other words, the SDCs-decoherence hypothesis states that the SDCs in our world are such that open environments will give rise to enough interactions between systems that have the DC and ones that may end up having it, allowing us to reliably approximate fundamental decoherence models using pragmatic irreversible decoherence models that concern open environments. From now on, I will just assume that processes represented by pragmatic irreversible decoherence models occur in open environments. 
 
 Under the plausible assumption that irreversible pragmatic decoherence models represent how the classical domain arises from quantum systems and that this classical domain is widespread, notice that the SDCs-decoherence hypothesis hypothesizes that systems tend to develop interactions that obey CDC1)-CDC4).
 
 On the other hand, it also hypothesizes that SDCs in our world are such that in isolated/controlled situations, we are able to isolate systems $E$ (which serve as ``environments'' for target systems in pragmatic decoherence models) from interacting with members of SDCs. They only interact with systems that don't have the DC. $E$ will then interact with another system $S$, and this interaction is modeled via reversible pragmatic decoherence models.

 So, given this hypothesis, if a pragmatic reversible decoherence model models specific situations $SI$ with predictive success, it allows us to infer the situations $SI$ where a system $S$ is interacting with environmental systems $E$ that don't have the DC. Thus, we infer that in situations $SI$, no process that gives rise to determinate values occurs.

 So, note that isolated regions with macroscopic systems inside might isolate a macroscopic system $S$ from the influence of SDCs (like in the Wigner's friend situation). Suppose this is done properly so that we can unitarily manipulate the contents of that region. In that case, we have a reversible process inside that region instead of an irreversible one. Thus, if we have some situation that even involves interactions between macroscopic systems (like inside the friend's lab) but that is modeled by reversible decoherence with predictive success, we can infer that we have managed to isolate the systems from the influence of SDCs (more on this in Section 3). 
 Whether we are able to achieve this isolation depends fundamentally on the history of the SDCs (modeled via fundamental decoherence models) that could interact with $S$ in those situations and on our ingenuity in shielding it from interacting with SDCs (more on this below). 
 
 This view held by EnDQT contrasts with the one often assumed by relationalist views, such as the MWI,\footnote{\cite{Wallace2012TheInterpretation}.} which would consider that determinacy arises within a large enough isolated spatiotemporal region with systems decohering each other inside of it. It is in this sense that, as mentioned in the introduction, for MWI, decoherence is necessary and sufficient for determinacy, and for EnDQT, the interaction with SDCs matters.

 So, to be clearer, according to EnDQT, the proper isolation of the friend's lab amounts to not simply the isolation of the lab but the isolation of the friend (and their system) from interacting with elements of SDCs. As I have mentioned, whether this isolation can be done in practice will depend on the particulars of the SDCs inside that lab and their history, represented by fundamental decoherence, going back to the initiators, and whether they will interact appropriately with the friend or not, giving rise to their decoherence. If there are enough members of an SDC inside an isolated lab, and assuming we know who the members are, the local processes represented by fundamental decoherence would be enough to model the process that leads the systems inside the isolated lab to have determinate values. 

As I have mentioned above, a friend could be isolated by simply not allowing SDCs to interact with them. For example, suppose that the target system of the friend is system $C$ in eqs. \eqref{exampleSDC1} and \eqref{exampleSDC2}, and that the friend is system $B$. System $B$ is just for a slight moment in a superposition. Furthermore, system $A$ is now some system that previously interacted with other members of an SDC. We could isolate $B$ from interacting with $A$. If this occurs, the friend would be unable to give rise to their target system $C$ having a determinate value. The friend and their target system would just be in an entangled superposition with their target system, where both systems would have indeterminate values. 

Relatedly,  a friend could also be shielded from interacting with members of SDCs in the following way. Let's return to the original example involving systems $A$, $B$, and $C$. As we can see via this example, in order for a system (like $C$) to continue having determinate values of an observable and giving rise to other systems having the DC and having determinate values, interactions of the above kind should proceed at other times, i.e., $C$ has to interact with other systems while interacting with $B$. This leads EnDQT to predict a phenomenon that I will call the \textit{dissolution of an SDC}. If, during the evolution of an SDC, no system interacts with the system like $C$ that is leading the expansion of that SDC, that SDC will disappear, not being able to give rise to further determinate values and the local destruction of superpositions.\footnote{Note that system $C$ may also continue having determinate values if its states are decohered by other systems that belong to other SDCs that are expanding.} Now, to isolate the friend from the influence of SDCs, we would $just$ need to dissolve the SDCs that could interact with the friend. Note that this is different from isolating systems from interacting with target systems. It is instead not allowing the target systems to interact with systems with the DC. The phenomenon of dissolution of SDCs is a distinct prediction of EnDQT.

So, what the SDCs-decoherence hypothesis is doing is connecting the pragmatic decoherence models with the fundamental decoherence models, grounding the empirical success of the first on the phenomena represented by the latter. Accepting EnDQT and the CDC1)-CDC4), we have good evidence that the SDCs-decoherence hypothesis holds, given the success of models of irreversible decoherence in accounting for measurement-like phenomena, and the success of reversible ones in not accounting for non-measurement-like phenomena.\footnote{See, e.g., virtual/reversible decoherence in \cite{Schlosshauer2007DecoherenceTransition}.} Also, we have seen above how this hypothesis is plausible given that the CDCs lead to phenomena that resemble those represented by pragmatic irreversible decoherence models.

Contrary to what one may worry about, there is no future dependency or retrocausality for EnDQT in the sense that something giving rise to determinate values in the present depends on how some events turn up to arise in the future, i.e., the interactions turn out to give rise in the future to irreversible decoherence because they become uncontrollable (pragmatically speaking). Instead, given the above SDCs-decoherence hypothesis, it is hypothesized that our world is such that the empirically successful local pragmatic irreversible decoherence models (in open environments) are a reliable indicator to infer that determinate values arise via local interactions because the environmental systems that participate in this process are such that they have the DC. 

Now, it is time to address the elephant in the room, which is to specify what kind of physical systems initiators are and when we expect the SDCs to start, i.e., when initiators begun to act. It seems to me that the hypothesis to establish when SDCs started should fulfill the following two desiderata: i') initiators should have a privileged position that allows them to give rise to widespread SDCs so that they can explain the standardly assumed widespread existence of systems with determinate values across spacetime, including in the early universe. In other words, it should explain the widespread classicality that we observe or assume to exist. Furthermore, ii') it should support the SDCs-decoherence hypothesis in the sense that the systems that start SDCs should not be starting SDCs in arbitrary spacetime regions so that it becomes plausible that sometimes we are able to isolate systems from the influence of SDCs (so that we are able, for example, to recohere the quantum states of a system). It is also undesirable that, if we manage to isolate a spacetime region to maintain the quantum systems inside it in a superposition, an initiator could likely manifest arbitrarily and destroy those superpositions. This desideratum is particularly important in the case of initiators of the kind a). These initiators always have the DC concerning any system.

It is plausible that initiators should manifest themselves mainly in the early universe since this is the earliest stage when we apply classical physics. Furthermore, it is plausible to consider that in the early universe, something special happened (more on this below) and that systems in the early universe could be more influential than in the latter stages (more on this below, too). So, given i') and ii'), it is plausible to assume the following general empirical hypothesis as one possibility for when SDCs started,\\

\noindent At least most current SDCs in our universe started in what we consider the early universe, and initiators had a privileged role in this stage in terms of interacting with other quantum systems, which led to the formation of these SDCs (\textit{SDCs-starting hypothesis}).\\

As we will now see, the SDCs-starting hypothesis should be seen as a placeholder for more concrete hypotheses that obey the desiderata i') and ii') as the field of cosmology develops. Given the above desiderata, we can develop heuristics to establish which concrete systems started SDCs and seek a more specific hypothesis. Probably the simplest and most conservative one that is in agreement with these desiderata and EnDQT use of decoherence involves looking at the earliest occurring phenomena where it was necessary to postulate a (pragmatic irreversible) decoherence process.  

 Inflation is typically considered to have been driven by a scalar field $\phi$, called the inflaton.\footnote{Some models postulate multiple inflaton fields; for simplicity, here I will just consider that there is one field.} It is hypothesized that the zero-point fluctuations of the quantized inflaton scalar field in some regions (i.e., fluctuations of the field in the vacuum state) and the associated energy-momentum fluctuations and gravitational field, amplified by the rapid expansion of inflation, attracted more matter than in other regions. Then, it is hypothesized that this phenomenon gave rise to the (non-evenly distributed) cosmic structure in our universe (e.g., galaxy, galaxy clusters, etc.).\footnote{\cite{Liddle2009TheStructure}.} The inflaton field is often treated classically, and the effects of these fluctuations are observed via slight temperature anisotropies in the Cosmic Microwave Background. The problem is to explain how these quantum fluctuations became classical during the early stages of the evolution of the universe. To my knowledge, the earliest reasonably accepted decoherence processes involved the decoherence of these fluctuations and aimed to address this problem. Many decoherence models were formulated to describe this process.\footnote{See, e.g., \cite{Boyanovsky2015EffectiveEquation, Burgess2008DecoherenceFluctuations, Burgess2003EffectiveInflation, Kiefer2007PointerCosmology, Martin2012CosmologicalProblem, Burgess2022MinimalInflation, Raveendran2022EnhancedModels}.}
 
 In this article, I am not going to enter into details concerning how decoherence helps address this problem; however, we can briefly see how that can be done. In a toy model, each quantum state $\left|s_{i}\right\rangle_{S}$ in  eq.\eqref{eqdecoherence} could be the field amplitude/fock state with momentum $\textbf{k}$ of the inflaton field fluctuations $S$ in a spacetime region, and thus $i$ would be the particle occupation number (i.e., the number of particles occupying a given mode $\textbf{k}$ ). $E$ would be the environment (more on this below).\footnote{See, e.g., \cite{Kiefer:2008kub} and references therein for more details on these decoherence models.} We could invoke decoherence plus the MWI to perhaps make these models more satisfactory. Alternative proposals appealed to spontaneous collapse theories to explain how quantum fluctuations become classical.\footnote{See, e.g., \cite{Perez_2006}.} However, we can instead appeal to the more conservative approach proposed here.

Given the above heuristic, we can assume that currently the inflaton is a plausible possible candidate to be an initiator. This assumption has a series of attractive features that fulfill the desiderata aimed by the SDCs-starting hypothesis. First, it fulfills the desiderata i') because of its influential position and role mentioned above, i.e., it gives rise to the cosmic structure. Furthermore, the decay of the inflaton in the so-called reheating stage is often hypothesized to have given rise to, at least, ordinary matter. Second, it fulfills the desiderata ii') given that the inflaton field doesn't seem to manifest itself in our present universe, we can, in principle, build arguments that consider that this field in the later stages of the universe didn't give rise anymore to SDCs or negligibly so. For instance, in the so-called reheating phase, it is standardly considered that the inflaton field reached the absolute minimum of its potential $V(\phi)$ and stayed there (and has been staying there).\footnote{At least in our universe.}  Let's assume that such minimum corresponds to the point where the field is zero or approximately zero (see \cite{Martin2014ThePlanck} for some empirically supported potentials by the Planck satellite that fulfill this requirement). Let's also consider that the coupling of the inflaton field to all other fields/systems in the Lagrangian density that describes or governs our universe depends on the value of the inflaton field in such a way that the interaction terms are zero (or approximately so) when the field zero. Given these two assumptions, we can consider that the inflaton field in the stages of the evolution of the universe after the reheating phase (which includes the phase where we are now) will at least rarely interact with other fields/systems. So, if we assume initiators of the kind a), it will (at least) rarely give rise to SDCs in these later stages.\footnote{One may worry that in other stages of the evolution of the universe (to put it in very rough and intuitive terms), there may be the creation of virtual particles-antiparticles pairs from the vacuum occupied by the inflaton field. These particles may give rise to SDCs. This isn't, in principle, a problem. In many inflaton models (see, e.g., \cite{Binetruy1996D-termInflation, Halyo1996HybridD-terms, Lyth1999ParticlePerturbation, McDonald2000ReheatingInflation}, the inflaton is considered to be very massive/energetic, and so those particles will be too short-lived (see \cite{Roberts2020Time-energyParticles} for a rigorous explanation for why this is the case) to be able to give rise to SDCs at least significantly. Note that the particles that arise from the vacuum also have to be able to decohere other quantum systems to give rise to SDCs.} 

Let's represent the Lagrangian density of our universe obeying the above desiderata, and whose initiator is the inflaton field, as $\mathcal{L}_{SDC}$. Since it fulfills the desiderata i') and ii'), I will call this hypothesis that appeals to inflation to explain the beginning of SDCs and the inflaton as the initiator, the $\textit{inflationary-starting hypothesis}$. This hypothesis could be stated in the following way,\\

Our universe is described/governed by the Lagrangian density $\mathcal{L}_{SDC}$.\\ 

\noindent This hypothesis is one concrete example of an SDCs-starting hypothesis.

How can we understand what are these initiators more concretely? This is largely an open question, and there are multiple possibilities. It is important that these possibilities concern local interactions. For example, \cite{Kiefer:2008kub} list various possible environmental systems that decohere the fluctuations of the inflaton field, such as some other quantum fields\footnote{Possibly coming from string theory.} or $parts$ of the fluctuations themselves.\footnote{E.g., an environment involving modes of the field with different momentum $\textbf{k}$.} If we adopt the inflaton and its fluctuations as initiators, they will have the DC concerning any system and give rise to other systems with determinate values by interacting with (i.e., decohering) other fluctuations and so on, forming SDCs. Note that if we consider instead that other fields have the DC, being these fields the initiators, we would need to explain why these other fields aren't currently still giving rise to SDCs in agreement with the SDCs-starting hypothesis.

The description of the inflaton field and the rest of the fields interacting with it needs a quantum field theoretical treatment. In this article, I haven't shown how EnDQT can be understood in the context of quantum field theory, but in principle, it won't be problematic to provide such treatment.\footnote{\cite{PipaToyEnDQT}.} Briefly, in this case, since we are interested in local interactions, we only consider quantum fields in single localized bounded spacetime regions. These are the fundamental quantum systems that transmit the DC. Briefly, in one possible approach to understanding these systems, we associate to a quantum field $\phi(x,t)$ in a finite spacetime region, such as the inflaton field in the spatial region $x$ at $t$, a wavefunctional $\Psi[\phi,t]$. The latter assigns a complex amplitude to each possible configuration of classical fields in spacetime regions, leading to a superposition of these configurations. A quantum field, which concerns a localized finite region, and the determinate values that may arise in that region will also be represented by the observables that act on the wavefunctional in that region. So, the DC is transmitted between quantum fields occupying a single localized region via local interactions. They are local because the observables that represent these systems concern local spacetime regions and interactions.\footnote{We guarantee this via conditions such as the so-called microcausality conditions, clustering decomposition (\cite{Weinberg1995TheFoundations}), and the Hamiltonian being local.} Thus, it’s expected that systems have determinate values and spread the DC in local regions of spacetime. The inflaton field is the initiator, but it transmits the DC via local interactions.

How can we understand what are these initiators more concretely? This is largely an open question, and there are multiple possibilities. It is important that these possibilities concern local interactions. For example, \cite{Kiefer:2008kub} list various possible environmental systems that decohere the fluctuations of the inflaton field, such as some other quantum fields\footnote{Possibly coming from string theory.} or $parts$ of the fluctuations themselves.\footnote{E.g., an environment involving modes of the same field as the target system, but with different momentum $\textbf{k}$.} If we adopt the inflaton (with its fluctuations) as initiators, they give rise to other systems with determinate values by interacting with them and so on, forming SDCs.\footnote{Note that if we adopt initiators of the kind a) and consider instead that other fields have the DC, being these fields the initiators, we would need to explain why these other fields aren't currently still giving rise to SDCs in agreement with the SDCs-starting hypothesis. Instead of adopting the simple initiators I have been talking about, an alternative possibility involves postulating another kind of initiator, which I will call reactive initiators. Contrary to non-initiator systems, reactive initiators S are systems that, upon specific interactions with other systems E that don't have the DC, acquire the DC either concerning an appropriate system in agreement with CDC1) or any system, and a determinate value during interactions. These interactions can be inferred via the pragmatic irreversible decoherence of $S$ by $E$.\\

In this case, the inflaton field system, with its fluctuations, would be a reactive initiator. The interactions with these perturbations, where the environmental systems would be other fields, would give rise to the fluctuations in a region having the DC.}

As we can see, the SDCs-starting hypothesis offers resources to establish what initiators may be. With the inflationary-starting hypothesis, I have pointed out that we already have models that agree with the above hypotheses and that simply appeal to the dominant paradigm in cosmology.

Furthermore, appealing to the inflaton as the initiator provides various advantages to EnDQT and possible scientific and philosophical payoffs. First, it potentially shows that EnDQT is a more parsimonious theory than other QTs.\footnote{Of course, it becomes less parsimonious if we adopt the initiator of the kind a).} On top of their ontological or mathematical additions to physics and quantum theory, other QTs very likely need to postulate the inflaton field (or some analog to this field) as belonging to the initial conditions of the universe and interpret it to address the problems that inflation is meant to address or as an instance of the so-called past hypothesis to explain the temporal asymmetries.\footnote{\cite{Albert2000TimeChance}.} This hypothesis constrains the state of the early universe so that the dynamics of the physical states in the future have the required temporal asymmetries, postulating a time-asymmetric boundary condition. Following \cite{Wallace2023TheHypothesis}, we consider that ``[t]he Past-Hypothesis, (...) is that the world came into being (or at least coalesced out of Planck-scale physics) in the local quantum vacuum state for a homogeneous, isotropic, inflationary spacetime.'' This vacuum state is also called the Bunch-Davies vacuum,\footnote{\cite{Bunch1997QuantumPoint-splitting}.} which is considered to be the initial state of the fluctuations of the inflaton field.\footnote{More concretely, it is the minimum energy eigenstate of the Hamiltonian for the primordial fluctuations infinitely back in the past.} However, EnDQT doesn't need those additional mathematical or ontological postulates. They are already a part of how EnDQT regards the inflaton as an initiator and the interactions that arise from there, having a fundamental role in the theory.\footnote{Even if one rejects inflation or the initial state of the inflaton as the past hypothesis, one needs to provide a substitute to these hypotheses that provide the same explanations, which is a tall order.} 

Second,  it may a) show that  EnDQT is also a more explanatory theory than other interpretations of quantum theory, or b) it at least offers a better scientific reductionist approach toward the above physical phenomena, which is often seen as valuable. More concretely, given the fundamental role of the inflaton mentioned above, EnDQT offers the possibility of explaining, or at least reducing the features of the initial conditions and the phenomena they aim to explain, together with the laws of physics, in something more fundamental. Buying the EnDQT picture, the story would go roughly like this: fundamental features of quantum phenomena involving initiators and their behavior are arguably more fundamental than phenomena described by classical cosmology, particle physics, thermodynamics, and statistical mechanics. The consequences of inflation (or any physical entity with a similar role), which is believed to involve the homogeneity of (relativistically speaking) causally separated different regions of space, perhaps the different temporal asymmetries, the geometry of the universe to be nearly flat, the seeds of structure formation, etc., and perhaps the state given by the Past Hypotheses, would be seen as fundamentally arising from an initiator with certain features plus the laws of physics that describe/govern its behavior and interactions. These features allowed for SDCs to spread throughout the universe in a specific way but also led the initiator to not manifest itself anymore, not currently giving rise to SDCs (being in the absolute minimum of its potential). 

Note that there is a sense in which this could be an explanation for the initial conditions of the universe because non-fundamental $special$ facts about the initial conditions of the universe can be grounded on the fundamental $special$ facts about QT. This is because initiators as special entities, and the phenomena that they give rise to, are fundamental for EnDQT and QT. There is much more to say about this. Whether one should regard the initiators and their features as explaining some features of the initial conditions and their consequences, or at least successfully reducing this to a more fundamental mystery, is a topic for future work. However, for now, note that it is plausible to consider that EnDQT offers this interesting possibility. 

So, even though EnDQT possibly ties a solution to the measurement problem in some features in the early universe, and that might be considered a problematic approach, any interpretation still has to postulate such initial conditions or events in the early universe. Furthermore, EnDQT can allow for other important advantages when it ties these initial conditions to initiators, which is a benefit of this approach.

Third, EnDQT diminishes the at-first-sight ad hocness of the inflaton field by providing it with a more fundamental role. This role involves solving problems with the more fundamental theory, quantum theory, rather than just solving some less fundamental problems.

Fourth, the inflationary-SDCs hypothesis offers predictions, which might allow us to constrain the very permissive models of inflation further.\footnote{\cite{Ijjas2013InflationaryPlanck2013, Dawid2023TestabilityScientific}.} For instance, as was mentioned, one of the possible constraints on inflationary models that EnDQT imposes is that the inflationary potential should have an absolute minimum when its value is zero in such a way that the Lagrangian \textit{turns off} the interactions of the inflaton field with other fields/systems once the inflationary potential reaches its absolute minimum. This can be regarded as a prediction of EnDQT.

Fifth, as I have mentioned, instead of explaining how primordial fluctuations arose from quantum fluctuations of the inflaton field by appealing to spontaneous collapse theories \cite{Perez_2006} that modify the equations of QT, or MWI and decoherence, which leads to the problem of quantum probabilities, EnDQT provides a conservative solution that doesn't suffer from the above potential issues. 

One might object that I didn't provide details about how these inflation models work to form the SDCs, and so the above hypothesis is not well-supported. This is very much a topic for future work. However, notice that as long as one believes in inflation, this shouldn't be a worry. This is because as long as inflationary models are well-developed and allow for local interactions, working out these models shouldn't be an issue (I have mentioned some decoherence models above that may be adapted to an SDC scenario like in the Appendix A). If it is an issue, it is also an issue to the inflationary paradigm and not just a problem for EnDQT. One might object that EnDQT, with its initiators, is built on top of speculative hypotheses. However, note that the above hypotheses are as speculative as any other hypotheses postulated by the currently popular interpretations of QT, and they turn out to be conservative in the sense that they don't involve any modification of standard QT.\footnote{MWI supporters might claim they make no speculative hypotheses, but a realist attitude to whatever lies beneath the multiplicity of worlds is itself a speculative hypothesis.} Furthermore, given the benefits of achieving UT), LC), and NR) (as we will see), it is worth taking the above hypotheses of EnDQT seriously. Also, I have shown that the inflationary-starting hypothesis provides a series of philosophical and scientific payoffs. So, overall, EnDQT yields worthy payoffs. On top of that, the beginning of SDCs occurs in regions of spacetime where we expect special events to occur. Finally, I should emphasize that every approach to QT so far needs to appeal to special initial conditions for one reason or another. However, EnDQT does that while providing additional benefits.  

One might also object that the inflationary paradigm has its problems, which places EnDQT in problematic foundations.\footnote{See, e.g., \cite{Ijjas2013InflationaryPlanck2013, Dawid2023TestabilityScientific} and references therein.} However, note that the SDCs-starting-hypothesis is a placeholder for current and future physics. Furthermore, whatever theory substitutes inflation, it should deal with the problems it pertains to solving.\footnote{Note that EnDQT could, in principle, incorporate a cyclic cosmological picture. The early stages refer to the early stages of this universe.} To solve those problems, it is plausible to consider that some features shared with the initiators will likely arise. This is because it is plausible to expect that in a future theory, it is likely that we will also have a rapid expansion of the universe in the early stages that doesn’t occur anymore. Such rapid expansion is likely due to some set of entities E with a considerable influence that don’t manifest themselves anymore or so much at least in this way. Furthermore, the SDCs-starting hypothesis can also be supplemented with another hypothesis if it turns out that there are phenomena that need to be explained via widespread initiators. So, EnDQT, conceived more broadly, is a view whose correctness doesn't just depend on inflation.\footnote{It is thus conceivable that there are other alternative mechanisms beyond inflation.} 

 In other words, the SDCs-decoherence hypothesis states that the SDCs in our world are such that the open environments involved in irreversible decoherence pragmatic models will give rise to many interactions between the target systems and systems that have the DC, allowing us to reliably approximate fundamental decoherence models using these pragmatic irreversible decoherence models. From now on, I will just assume that processes represented by pragmatic irreversible decoherence models occur in open environments. Under the plausible assumption that irreversible pragmatic decoherence models represent how the classical domain arises from quantum systems and that this classical domain is widespread, notice that the SDCs-decoherence hypothesis hypothesizes that systems tend to develop interactions that obey CDC1)-CDC4).

The fact that we have, on the one hand, fundamental decoherence models and, on the other hand, pragmatic decoherence models shouldn't be something objectionable if we accept other QTs because it isn't, in a sense, a feature exclusive to EnDQT. In the MWI and Bohmian mechanics, we also have, on one hand, the universal wavefunction and, on the other hand, the effective wavefunction.\footnote{See, e.g., \cite{Goldstein2021BohmianMechanics}.} Local observers tend to use the effective wavefunction locally and pragmatically, but its predictive success is grounded on the universal wavefunction. Analogously, we use pragmatic decoherence models, but their success is grounded in the phenomena represented by fundamental decoherence models.

One may also worry that there seems to be no guarantee that CDC1)-CDC4) are obeyed by enough systems so that SDCs tend to be formed. As I have said, under the plausible assumption that irreversible pragmatic decoherence models represent phenomena that are widespread, notice that the SDCs-decoherence hypothesis hypothesizes that this is the case. This is plausible because irreversible pragmatic decoherence models represent classical phenomena, and we have evidence that this process is widespread at certain scales. More concretely, we have evidence for irreversible decoherence being widespread because of our evidence concerning entanglement with large macroscopic objects being something widespread.\footnote{In other words, systems measured by macroscopic devices can be represented unitarily via this process.}

However, one may need more detailed evidence. Note that it is generically true in universes like ours that decoherence qua entanglement tends to increase over time, except under special conditions. Therefore, since CDCs are based on decoherence,  the above worry shouldn't be a problem because we have grounds to believe that more systems will follow CDC1)-CDC4) as time evolves.\footnote{Furthermore, there is a related plausible hypothesis that can be regarded at least in part as a consequence of a particular specification of the SDCs-starting hypothesis and the SDCs-decoherence hypothesis, which I think also defeats this skepticism. I will outline this hypothesis here and the argument that appeals to it and leave its details for future work. It has been argued by many, such as one of the pioneers of the decoherence program (\cite{Zeh1989TheTime}, see also \cite{sep-qm-decoherence} and references therein), that decoherence gives rise to its own arrow of time. Roughly, due to the irreversibility of decohering interactions, quantum states of systems in the universe tend to be more mixed in the future than in the past. Although EnDQT doesn't reify quantum states, given the SDCs-decoherence hypothesis, it interprets irreversible pragmatic decoherence as standing for a particular physical phenomenon, and thus, an analogous process should occur for EnDQT. Since CDCs are based on decoherence, we can instead consider that there is a tendency for more systems to follow the CDCs in the future than in the past, forming more SDCs in the future than in the past. More concretely, we can consider an SDCs-starting hypothesis that establishes that there were no SDCs and just initiators in the past, unentangled with other systems (as it is standardly assumed by decoherence models). Then, as the systems evolve under physically plausible Hamiltonians, there will be a tendency for more systems to interact, get entangled, give rise to decoherence, and form interactions that obey the CDCs, giving rise to SDCs. If we adopt the inflationary-starting hypothesis, we may end up considering that this progressively increasing number of systems that will tend to belong to SDCs at some point is accompanying the beginning and evolution of the universe. Thus, what I will call \textit{SDCs-time's arrow hypothesis} should be seen as a consequence of at least a specification of the SDCs-starting hypothesis and the SDCs-decoherence hypothesis. As we can see, this argument doesn't assume anything about decoherence being widespread. Thus, if we also endorse this hypothesis, the above worry shouldn't be a problem, and in principle, we have grounds to believe that more systems will follow CDC1)-CDC4) as time evolves.}

One may also object to the following: suppose agent $M$ interacts with $S$ such that $S$ acquires a determinate value for some binary observable. Then agent $M$ sends this value to $M^{*}$ using, say, the z-spin of a single particle as a signal. There is a period during which the signal particle isn't interacting with anything, and thus, that system $S$ doesn’t belong to an SDC. Therefore, it has an indeterminate value of the observable whose value represents the information $M$ wants to communicate to $M^{*}$.  Given this, it is hard to conceive how $S$ can carry reliable information about the result that $M$ got without itself having a determinate value.

As a reply, note that a system $S$ can carry reliable information if the complex amplitude $\alpha$ that is associated with the quantum state that concerns the information that we want to transmit is such that $|\alpha|^2>>1$. When we approach EnDQT, we need to deidealize quantum mechanics. Systems, after measurements upon their self-Hamiltonian, tend to evolve out of the quantum state that was measured, where this quantum state could encode information that aims to be sent somewhere else.
As mentioned before, the assumption that systems have indeterminate values of any observable by default except during certain interactions comes from the perspective that the so-called links (like the EEL) that connect state assignments to assignments of determinate values are idealizations that never occur in practice.

I have mentioned above some predictions that EnDQT provides, such as the dissolution of SDCs and those that are a consequence of the SDCs-inflationary hypothesis. Another prediction is the following: as we have seen with the example above, adopting the CDCs generates constraints on how SDCs are formed and new predictions. Decoherence timescales roughly serve as an indicator for the timescale it takes for environments of a system to decohere that system on average, where that system ends up having specific determinate values (that are observed in the lab). Given the SDCs-decoherence hypothesis, the CDCs predict that the decoherence timescale that we empirically observe of a kind of system $Z$ by a kind of system $Y$ should be superior or of the same order as the decoherence timescale of $Y$ by a kind of system $X$, where $Y$ is typically decohered by $X$ before $Y$ decoheres $Z$, and where the interaction between $X$ and $Y$ starts first. Otherwise, contrary to what is assumed by the CDCs, we can have situations where $Z$ will have a determinate value first (due to $Y$), then $Y$ will have a determinate value due to $X$. Since the decoherence timescales can be empirically determined, a further analysis of the current empirically determined decoherence timescales is needed to see if they agree with the predictions of the CDCs.

The predictions of the CDCs are supported in the case that $Y$ is a macrosystem (e.g., measurement devices), and $Z$ is a microsystem. This is because macroscopic systems have decoherence timescales much shorter than the microscopic systems that they can decohere.\footnote{The cross-section for larger systems is larger than the one for smaller systems. Moreover, the decoherence rate of a quantum system, which is the inverse of the decoherence timescale, is proportional to their cross-section, as well as the flux of systems of the environment. See the collisional models of decoherence in, e.g., \cite{Joos1985TheEnvironment, Kiefer1999Decoherence:Beyond, Schlosshauer2007DecoherenceTransition}, and references therein.} Furthermore, the conditions for a quantum system to be considered a classical controller of another quantum system support the CDCs, since our evidence for measurement-like interactions are based on these situations.\footnote{ \cite{Milburn2012DecoherenceSystems} provided two examples of interacting quantum systems where one system serves as a classical controller for the other. The conditions necessary for this to occur are as follows: First, the controller must be open to the environment to establish a pointer basis for the controller coupled with the target system. Second, the dynamics of the controller, as an open system, must ensure that the approach to that pointer basis is much faster than the timescales of the system being controlled. All these conditions support the CDCs.} So, so far, the CDCs seem to be favored. It would be interesting if we find further evidence for or against them.

Before ending this article, I would like to mention two additional features of this view. First, EnDQT provides a new interpretation of Born probabilities, viewing them as predictors of how SDCs evolve upon specific interactions.\footnote{If one is committed to objective probabilities qua chances, they track those chances that arise under these interactions} Second, given the above CDCs, we can adopt different strategies to build models of SDCs. One strategy is what I will call the \textit{recursive heuristic}: as we know, target systems of decoherence models can be environmental systems of other decoherence models. So, given this heuristic, we should consider that (pragmatic irreversible) decoherence models don't only model measurement-like interactions of the target system, but also how that target system can constitute an environment that gives rise to further measurement-like interactions. So, like in the above simple example involving $A$, $B$, and $C$ that makes certain assumptions, we can then build models of the behavior of SDCs piecewise. If we had the following SDC, $X \rightarrow Y \rightarrow Z$, we should consider at least a decoherence model where $X$ decoheres $Y$ and another one where $Y$ decoheres $Z$ to describe this process. Note that given the CDC4), SDCs have a certain structure. I have represented a toy example of such a structure in Figs. \ref{fig:example1} and \ref{fig:example2}. In Appendix A, I show in more detail how this can be done. Given the recursive heuristic and the fact that EnDQT doesn't modify the basic equations of QT, it should be possible to develop more realistic models of the evolution of SDCs.\footnote{Note also that given the CDCs and EnDQT perspective on quantum states, we cannot infer directly from a system whose quantum state is in an eigenstate of some observable that it has a determinate value of that observable. For instance, when Alice measures her system and assigns it an eigenstate of some observable, she cannot infer that the target system of Bob is in an eigenstate of some observable since Bob might not have interacted with the local SDC. Also, as I have mentioned above, we might assign out of convenience and idealization to a system an eigenstate of some observable, where the latter doesn't belong to an SDC, but that doesn't imply that it has a determinate value of that observable. This just implies that if it interacted with an SDC, it could have a determinate value with $100 \%$ of probability. Moreover, even upon a measurement of a local system ´´in an eigenstate of some observable,'' the system shortly after evolves into a superposition (modulo quantum Zeno-like measurements, which increase the probability of the system being found in the same quantum state in repeated measurements). That is also why I have been using decoherence to model measurement-like interactions in general. Furthermore, we cannot always infer from a system that it is not in an eigenstate of some observable, that it hasn't a determinate value of such observable. For instance, entangled states that arise in decoherence don't correspond to eigenstates of some observable. These inferences based on EnDQT are in tension with both directions of the Eigenstate-Eigenvalue link because this link neglects the SDCs to make such inferences.}

Given the CDC1)-CDC4), and the above two hypotheses, EnDQT has the important benefit of achieving UT), providing criteria for absolute determinate values to arise in a single world without modifying the fundamental equations of QT. It only uses decoherence to assign determinate values to a system and the SDCs, whose description appeals to such equations. Furthermore, arbitrary systems can, in principle, be placed in a superposition for an arbitrary duration concerning any observable as long as they don't interact with members of an SDC. Of course, given the SDCs-decoherence hypothesis, in principle, doesn’t mean in practice. Our pragmatic models of decoherence tell us that it’s very difficult to place large macroscopic systems in a superposition. Also, we have seen that EnDQT provides a series of other benefits and predictions, making it a QT worth taking seriously.

\section{Why is EnDQT local?}
In this section, I will argue that EnDQT achieves LC) and NR) by showing how it deals with Bell's theorem and provides a local common cause explanation of quantum/Bell-type correlations without adopting non-local/action-at-a-distance, relational, or superdeterministic/retrocausal strategies. In the EPR-Bell scenario, space-like separated Alice and Bob share a pair of quantum systems in an entangled state and randomly perform measurements on those systems.

First, like in standard QT, the Hamiltonians of interaction, representing the interactions between the agents and their systems, should represent local interactions. Second, EnDQT doesn't modify the equations of QT, and so, in principle, its laws can be rendered Lorentz covariant and generally covariant, and thus, it can be rendered compatible with relativity and be local in this sense. More precisely, if QT allows laws to be Lorentz and generally covariant (which seems to be the case at least given Quantum Field Theory in flat and curved spacetime),\footnote{See, e.g., \cite{Wald1994QuantumThermodynamics}.} then since EnDQT doesn't modify standard QT, it should uphold these relativistic symmetries.\footnote{More on this in Appendix A. I am bracketing issues with relativistic symmetries that may arise if we aim for a quantum theory of gravity, but this goes beyond the desideratum LC).} I will assume this here; however, future work will show this explicitly. Assuming that EnDQT achieves these two senses of locality, I will further argue that EnDQT addresses the EPR-Bell scenarios without violating relativistic causality, in the sense that it doesn't require us to assume that the causes of the events involved in those correlations are not within their past lightcone, and without invoking superdeterministic or retrocausal explanations. Furthermore, it provides a local common cause explanation of quantum correlations. Let's see how.
I will focus first on the 1976 widely influential version of Bell's theorem because it is generally considered applicable to indeterministic theories.
Since EnDQT is an indeterministic theory, this version is more relevant.\footnote{Another widely influential version is considered to rule out the existence of local deterministic hidden variables \cite{Bell1964OnParadox}. I will go over this version further below.} This version assumes the so-called statistical independence or no-superdeterminism assumption. This assumption states that any events on a space-like hypersurface are uncorrelated with any set of interventions subsequent to this hypersurface. It also assumes that there are single observed outcomes and not, for example, multiple outcomes that correspond to multiple worlds or perspectives.\footnote{Another way of putting this assumption is that there is a joint probability distribution involving the outcomes of Alice and Bob. See the absolutness of observed events assumption further below.} This is the assumption denied, for example, by the Many-Worlds Interpretation. The assumption more relevant for EnDQT is the factorizability condition. According to this condition,

\begin{equation}
P(A B\mid X Y \boldsymbol{\Lambda})=P(A \mid X \boldsymbol{\Lambda}) P(B \mid Y \boldsymbol{\Lambda}) . 
\end{equation}

\begin{figure}[ht]
    \centering
    \includegraphics[max width=\textwidth]{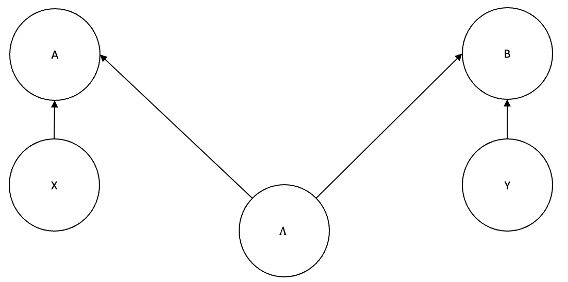}
    \caption{DAG of the common cause structure of Bell correlations, which respects relativity. This causal structure respects relativistic causality because $X$ or $A$ doesn't influence $Y$ or $B$, and vice-versa, where these events are spacelike separated. Moreover, no other variables influence the variables $A$, $B$, $X$, or $Y$, or they don't influence anything else. So, there are no retrocausal or superdeterministic causal relations.}
    \label{fig:example3}
\end{figure}

The variables $A$, $B$, $\boldsymbol{\Lambda}$, $X$, and $Y$ represent events embedded in a Minkowski spacetime. $A$ and $B$ represent the different measurement results of Alice and Bob, while $X$ and $Y$ are the different possible choices of measurement settings for Alice and Bob. $\boldsymbol{\Lambda}$ represents some set of (classical) ``hidden'' variables in the past lightcone of $A$ and $B$ (see also Figure \ref{fig:example3}), representing the common causes of the correlations between $X$ and $Y$.

This condition arises as a consequence of two assumptions:\footnote{\cite{Bell1976TheBeables, Bell2004LaCuisine}. See also, e.g., \cite{sep-bell-theorem} and references therein.}
\begin{itemize}
\item The causes of events are always in their past lightcone,\footnote{By always, we can also mean in any relativistic reference frame.}
\item The Classical Reichenbach Common Cause Principle (CRCCP).
\end{itemize}

I will be concerned here with a version of the CRCCP that is expressed in terms of variables whose different values represent different values of certain quantities or physical features. This is because this is the most appropriate notion to represent features of at least classical physical systems (i.e., systems represented via classical mechanics).\footnote{There are other versions that use propositions and their negation to state this principle (i.e., ``events''). See, .e.g., \cite{sep-physics-Rpcc}.} Briefly, this version of the CRCCP states that if variables $A$ and $B$ are correlated, then either $A$ causes $B$, or $B$ causes $A$, or both $A$ and $B$ have common cause variable $\boldsymbol{\Lambda}$, where conditioning on $\boldsymbol{\Lambda}$, $A$ and $B$ are decorrelated, i.e., $P(A, B \mid \boldsymbol{\Lambda})=P(A \mid \boldsymbol{\Lambda}) P(B \mid \boldsymbol{\Lambda})$. However, it is unclear whether we should accept these probabilistic relations given by the CRCCP as generally representing a causal structure involving quantum systems, given the exotic features of quantum systems. The above version of the CRCCP can be derived from the Classical Markov Condition (CMC), assumed by the so-called Classical Causal Models (CCMs).\footnote{See Appendix E for this derivation. See also \cite{sep-physics-Rpcc}. Note that the version of the CRCCP mentioned in the previous footnote cannot be derived from the CMC.}


The CMC connects the causal structure provided by some theory, and which is represented by a Directed Acyclic Graphs (DAGs), i.e., a directed graph with no cycles, with probabilistic statements. Note that the sense ``causation'' will be understood merely in terms of influence between quantum system, represented via QT (more on this below). The CMC says the following,\\

 \noindent \textit{Let's assume we have a DAG G, representing a causal structure over the variables $\boldsymbol{V}=$ $\left\{X_{1}, \ldots, X_{n}\right\}$. A joint probability distribution $P\left(X_{1}, \ldots, X_{n}\right)$ is classical Markov with respect to $\mathrm{G}$ if and only if it satisfies the following condition: for all distinct variables in $\boldsymbol{V}, P$ over these variables factorizes as $P\left(X_{1}, \ldots, X_{n}\right)=\prod_{j} P\left(X_{j} \mid P a\left(X_{j}\right)\right)$, where $P a\left(X_{j}\right)$ are the ``parent nodes'' of $X_{j}$, i.e., the nodes whose arrows point to $X_{j}$.}\\

The CMC for the above DAG, which respects relativity, allows us to derive the following equation (I will denote regions of spacetime, the related nodes, and variables whose values may be instantiated in those regions using the same letters),

\begin{equation}
P(A B \mid X Y)=\sum_{\Lambda} P(\Lambda) \mathrm{P}(A \mid X \Lambda) \mathrm{P}(B \mid Y \Lambda).
\label{causalMarkov}
\end{equation}

\begin{figure}[ht]
    \centering
    \includegraphics[max width=\textwidth]{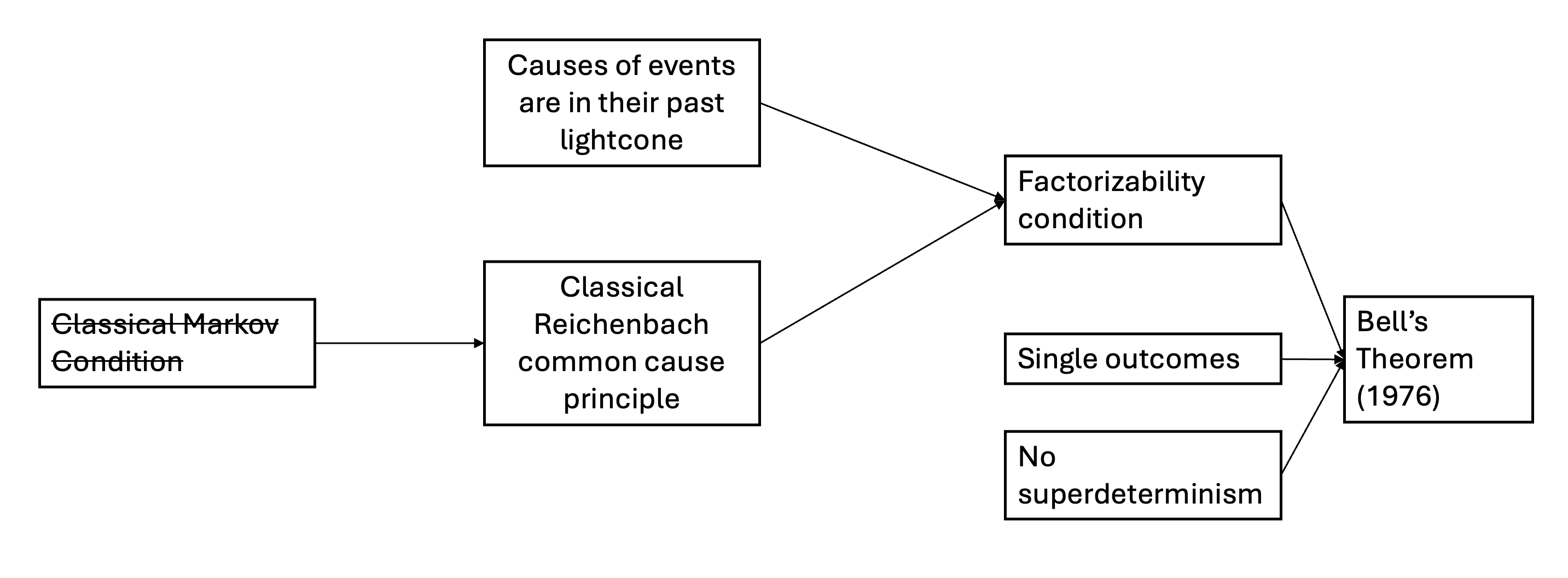}
    \caption{Diagram that helps to understand how EnDQT deals with Bell's 1976 theorem, which is considered to be applicable to indeterministic theories.  It also helps in understanding how EnDQT deals with it. It involves the rejection of the applicability of the Causal Markov Condition and Classical Causal Models to, in general, accurately represent causal relations between quantum systems. If a box with the claim $X$ is connected by one or more arrows, means that the one or more boxes with claims $Y$ that point to $X$ jointly lead to the derivation of $X$.}
    \label{fig:diagram1}
\end{figure}

Given the widespread empirical success of the application of the CMC via CCMs (e.g., \cite{Pearl2009Causality}), which can be used to derive the CRCCP, I will consider that the empirical success of the CRCCP in physics is supported by the empirical success of the application of the CMC via CCMs rather than the other way around (e.g., \cite{Pearl2009Causality}).\footnote{There is also a way of deriving the factorizability condition, as well as the no-superdeterminism condition directly from CCMs and the CMC. See \cite{Khanna2023ClassifyingInequalities}.} EnDQT responds to Bell's theorem by rejecting that the CMC, and hence the CCMs, can be applied in general to accurately represent causal relations between quantum systems, and hence, it rejects the applicability of the CRCCP and the factorizability condition to make such an accurate representation.\footnote{Therefore, note that EnDQT also rejects outcome independence and parameter independence \cite{Jarrett1984OnArguments}, which can be used to derive the factorizability condition by rejecting their applicability to represent causal relations between quantum systems.} This argument is schematized in \ref{fig:diagram1}.

There are at least two complementary arguments one can give to justify the rejection of the CMC, which aren't necessarily mutually exclusive. One argument looks directly at (to my knowledge) the most precise derivation of the CMC via structural equations to conclude that the CMC and the CCMs are inappropriate to represent causal relations between quantum systems. These equations involve relationships between endogenous variables $V_{j}$ (i.e., variables whose values are determined by other variables in the model) that depend on their endogenous parent variables $\operatorname{Pa}\left(V_{j}\right)$ plus exogenous variables $U_{j}$ (i.e., variables whose values are determined from outside the model) establishing a directed deterministic relationship $V_{j}=f\left(P a\left(V_{j}\right), U_{j}\right)$. \cite{Pearl1995ACausation} proved that if we have a DAG G' representing the causal structure on $V_{j}$ (i.e., a ``causal DAG''),\footnote{This DAG is often called a Bayesian network.} the probability distribution $P\left(V_{j}\right)$ that results from the marginalization of the noise sources if $U_{i}$ are probabilistically independent in P, will respect the $\mathrm{CMC}$ concerning $\mathrm{G}$.\footnote{See, e.g., \cite{Pearl2009Causality, Pearl1995ACausation,sep-physics-Rpcc}.}
The above derivation assumes features rejected by EnDQT. First, the origin of the probabilities of the CMC is in the ignorance about some underlying determinate values. Furthermore, note that these systems that travel to each wing, which, according to EnDQT, have indeterminate values, don't even have a probabilistic model independently of the measurements of Alice or Bob.\footnote{Note that we could assign a determinate value to the whole state $|\Psi\rangle$ of the entangled systems that would correspond to the eigenvalues of the observable that this state is an eigenstate of. However, Alice and Bob rather act on the subsystems of these systems. So, we should consider that it is not the whole state $|\Psi\rangle$ that determines the outcomes but its subsystems. Each subsystem of this entangled state influences locally the outcomes of Alice and Bob, and there is no way to assign a determinate value to each subsystem.} So, we cannot have a probability over the common causes independently of their interactions, as it is assumed by this proof. Third, the above causal relations between systems aren't described by QT. More precisely, they don't involve unitary evolutions, decoherence, and quantum indeterministic processes. Further below, I will show how EnDQT, by allowing indeterminate values, also permits a local explanation of Bell correlations. Given that this precise derivation of the CMC makes assumptions that are rejected by EnDQT, I, therefore, conclude that the CMC and CCMs are inadequate to represent causal relations between quantum systems according to EnDQT.

There is at least one possible objection to this argument for rejecting the adequacy of CMC. This justification makes it unclear whether $some$ causal explanation of quantum correlations can be provided, and this deficiency could press us to reject other assumptions of Bell's theorem instead of the CRCCP. In the end, one might think that CCMs are good inferential tools, and the problem is elsewhere.
Taking into account this objection, we can develop a better argument for why the CMC and CCMs are inappropriate to give a causal account of Bell correlations according to EnDQT. I will call it, \textit{the argument for locality}. This argument has the benefit of not necessarily appealing to the proof of Pearl and Verma, and we can use other considerations as well. The argument is based on the observation that a way of finding the limitations of the domain of applicability of the CCMs is by examining the more general models that putatively represent causal relations in the quantum domain, i.e., Quantum Causal Models (QCMs).\footnote{\cite{Costa_2016, Allen2017QuantumModels, Allen2017QuantumModels}.} Furthermore, we will analyze the limitations of CCMs by comparison to QCMs as interpreted by QT. More concretely, I will analyze how QCMs make some assumptions that CCMs don't make and that these assumptions concern the quantum domain, according to EnDQT. QCMs are, in principle, more general because they reduce to classical ones in a ``classical limit.'' Like we found what is wrong with classical mechanics when we examined the more general theory, QT, which reduces to classical mechanics in some limit, we will find what is wrong with the CCMs when we adopt QCMs interpreted via EnDQT.

As I will explain, QCMs will have the role of showing how EnDQT provides a local causal explanation of Bell-type/quantum correlations and how we can infer those relations. Note that QCMs currently are only formulated for finite-dimensional Hilbert spaces. However, this isn't as far as we can tell, in principle, a fundamental limitation.\footnote{See \cite{paunkovic2023challenges} for an overview of possible challenges that need to be addressed in order to extend QCMs to the infinite-dimensional case.} I will thus pose the following argument that I will present via three core premises.

Let's start with the first core premise of the argument:\\

P1) A causal model accurately represents causal relations between quantum systems for EnDQT if and only if it can be understood as explicitly considering that systems that participate in those causal relations i) only assume determinate values with a certain probability given by the Born rule when they interact with an SDC, and ii) where those relations are described via QT, but without reifying the quantum states like EnDQT assumes. i) and ii) are what we will call EnDQT-appropriate assumptions.\\

Given CDC1)-CDC4), this premise is plausible. Systems have indeterminate values of any observable by default, only having determinate values with a certain probability given by the Born when they interact with members of an SDC, hence i). Furthermore, obviously, the causal model should describe those relations explicitly via QT as interpreted by EnDQT, so that it is clear that what is representing is what EnDQT aims to represent. This also requires adopting the EnDQT perspective on quantum states. Note that causal models, both classical and quantum, don't wear their ontology on their sleeves. They require some interpretation. Their interpretation is theory-dependent to a certain degree, and so a theory-dependent argument has to be made as to whether they appropriately represent causal relations between quantum systems or not. If a causal model can be understood as providing i) and ii) explicitly, it should be seen as accurately representing causal relations according to EnDQT.

Let's turn to turn to the second premise:\\

P2) If QCMs consider that i) systems only assume determinate values with a given Born probability when they interact with an SDC, and ii) where those relations are described via QT, but without reifying the quantum states like EnDQT assumes, then QCMs provide a local non-relational, non-superdeterministic, and non-retrocausal explanation of quantum correlations.\\

Note that this local explanation includes a common cause local explanation of Bell correlations. Showing that P2) is true will require an interpretation of QCMs in agreement with EnDQT. We will turn to that now.

First, I will provide some minimal technical background. QCMs consider that each node in the causal DAG concerns a possible locus of interventions on the properties of a system. More concretely, each node is associated with a set of CP (completely positive) maps $\tau_{A_{1}}^{k_{A_{1}} \mid x_{A_{1}}} \otimes \ldots \otimes \tau_{A_{\mathrm{n}}}^{k_{A_{\mathrm{n}}} \mid x_{A_{\mathrm{n}}}},$\footnote{A quantum channel is a linear map $\varepsilon$ that is a completely positive trace preserving (CPTP) map. A map is a CPTP map if: a) it is trace-preserving, i.e., $\operatorname{Tr}(\rho)=\operatorname{Tr}(\varepsilon(\rho))$ for all density operators $\rho$, b) positive, i.e., $\varepsilon(\rho) \geq 0$ whenever the density operator $\rho \geq 0$, and c) completely positive. When only b) and c) are fulfilled, we have a completely positive (CP) map rather than a CPTP map. A CP-map can be associated with a positive operator-valued measure (POVM). See \cite{10.5555/1972505}.} also called quantum instruments, instead of random variables as in the CCMs case. This set gives the ``possibility space'' that can be associated with the different ways the properties of a system with its quantum state can change under local interventions $x$, which correspond to the preparation of quantum systems, transformations, measurements on them, etc., each leading to different outcomes $k$.

The QMC is defined through a causal DAG where the edges of the DAG are associated with quantum channels/completely positive trace-preserving (CPTP) maps.\footnote{See the previous footnote.} Examples of a quantum channel are unitary maps, evolution of the system with noise, etc.\footnote{Each (quantum) node $A_{i}$ is associated with an input Hilbert space $\mathcal{H}_{A_{i}^{\text{input}}}$, written as $A_{i}^{\text{input}}$, and an output Hilbert space $\mathcal{H}_{A_{i}^{\text{output}}}$, written as $A_{i}^{\text{output}}$. Each edge is associated with the output Hilbert space of one node and the input Hilbert space of another node. When written $\rho_{B \mid DA} \rho_{C \mid AE}$, it means $\rho_{B \mid DA} \rho_{C \mid AE} = \rho_{B \mid DA} \otimes \rho_{C \mid AE} = (\rho_{B \mid DA} \otimes I_{E^{\text{output}}} \otimes I_{C^{\text{input}}}) (\rho_{C \mid AE} \otimes I_{B^{\text{input}}} \otimes I_{D^{\text{output}}})$, where $X^{\text{input}}$ and $X^{\text{output}}$ are the inputs and outputs of node $X$. Moreover, $\operatorname{Tr}_{A} \rho_{AB \mid C} = \rho_{B \mid C}$ and $\operatorname{Tr}_{B} \rho_{AB \mid C} = \rho_{A \mid C}$.} Both CP and CPTP maps are written as positive semi-definite operators via the Choi-Jamiolkowski (CJ)-isomorphism.\footnote{See, e.g., \cite{Barrett2019QuantumCM}.}

The QMC representing a causal structure held fixed is written via the process operator $\sigma$, which is a CPTP map and factorizes analogously to the CMC. More precisely, a process operator $\sigma_{A_{1}, \ldots, A_{n}}$ is compatible with a DAG G with nodes $A_{1}, \ldots, A_{n}$, if and only if it obeys the Quantum Markov Condition (QMC), \cite{Barrett2019QuantumCM}). This condition says for all $i$, $l$ in the DAG G there are quantum channels such that $\left[\rho_{A_{i} \mid P a\left(A_{i}\right)}, \rho_{A_{l} \mid P a\left(A_{l}\right)}\right]=0$, and

\begin{equation}
\sigma_{A_{1}, \ldots, A_{n}}=\prod_{i} \rho_{A_{i} \mid P a\left(A_{i}\right)} . 
\end{equation}

We need to have $\left[\rho_{A_{i} \mid P a\left(A_{i}\right)}, \rho_{A_{l} \mid P a\left(A_{l}\right)}\right]=0$ because the product of two positive operators is positive if and only if they commute. $\sigma_{A_{1}, \ldots, A_{n}}$ factorize, which leads them to be analogous to the conditional probabilities in the CMC.

A version of the Born rule allows us to represent the overall causal structure, which also involves certain measurements on the nodes $A_{1}, \ldots, A_{n}$ with outcomes $k_{A_{1}}, \ldots, k_{A_{n}}$, given interventions $x_{A_{1}}, \ldots, x_{A_{n}}$,

\begin{equation}
\begin{split}
&P\left(k_{A_{1}}, \ldots, k_{A_{n}} \mid x_{A_{1}}, \ldots, x_{A_{n}}\right)\\
&=\operatorname{Tr}_{A_{1}, \ldots, A_{n}}\left[\sigma_{A_{1}, \ldots, A_{n}} \tau_{A_{1}}^{k_{A_{1}} \mid x_{A_{1}} S D C} \otimes \ldots \otimes \tau_{A_{n}}^{k_{A_{n}} \mid x_{A_{n}} S D C}\right]. 
\end{split}
\label{generalizedBornRule}
\end{equation}

\begin{figure}[htbp]
    \centering
    \includegraphics[max width=\textwidth]{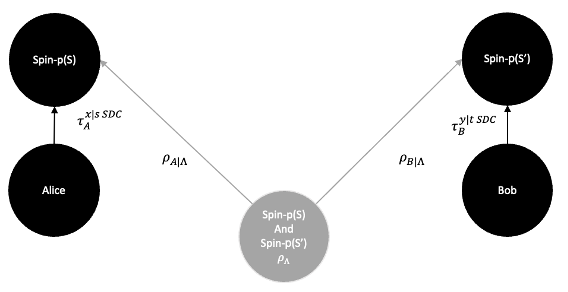}
    \caption{EnDQT-causal-DAG of the common cause structure of Bell correlations, which respects relativity being local, non-retrocausal, and non-superdeterministic, and is adopted by quantum causal models as interpreted by EnDQT. Furthermore, the outcomes are absolute/non-relative.}
    \label{fig:example4}
\end{figure}

An obstacle that one must face to provide a local causal explanation of Bell correlations via QCMs is to deal with their operationalism. Causal influences are typically understood by the possibility of ``signaling'' from one node to another.\footnote{When all the relevant systems participating in causal relations are included \cite{Barrett2019QuantumCM}.} The causal structure represented by QCMs represents the constraints on these signaling relations. So, node $X$ cannot signal to node $Y$ if and only if node $X$ doesn't precede node $Y$ in the DAG (e.g., see Figure \ref{fig:example4}, more on this below). Signaling between node $X$ and $Y$ can be understood as occurring when a variation in the choice of certain instruments/interventions performed at node $X$ can vary the probabilities of an outcome $\mathrm{k}$ concerning measurements performed at node $Y$.

However, one might worry that, as in other QTs such as Bohmian mechanics,\footnote{See, e.g., \cite{Goldstein2021BohmianMechanics}.} although there is no signaling, there could still be non-local influences, and QCMs might be hiding such influences. If we adopt EnDQT, which doesn't consider that there are hidden non-local influences that cannot be used for signaling, we don't need to have this worry because systems involved in QCMs, according to EnDQT, only have determinate values when they interact with members of SDCs. So, SDCs are necessarily involved in these influences that give rise to determinate values, and they concern local interactions between systems (see Section 2).

Furthermore, since EnDQT does not require agents at the fundamental level, using the concept of signaling and an operationalist language is unnecessary for understanding what QCMs fundamentally are about. Moreover, we don't need to adopt an account where signaling or causation is irreducible. We can rather consider that systems in a region influence the determinate value of certain systems in another region, where such influences are modally described/governed by QT, and QCMs allow us to represent and infer those influences.

Given this background, let's now see how if QCMs consider that systems i) only assume determinate values with a certain Born probability when they interact with an SDC, and ii) where those relations are described via QT but without reifying the quantum states like EnDQT assumes, then QCMs provide a local non-relational, non-superdeterministic, and non-retrocausal explanation of quantum correlations. To do this, we will show how the truth of the antecedent leads to the truth of the consequent. Thus, we will have to interpret QCMs according to EnDQT and in agreement with i) and ii).

Now, $A$, $B$, and $\Lambda$, represent spacetime regions, instead of classical variables. Consider below how, via the QMC and a version of the Born rule, we can represent the local common cause structure that explains Bell correlations (Figure \ref{fig:example4}),

\begin{equation}
P(x, y \mid s, t)=\operatorname{Tr}_{\Lambda \mathrm{AB}}\left(\rho_{\Lambda} \rho_{A \mid \Lambda} \rho_{B \mid \Lambda} \tau_{A}^{x \mid s\ S D C} \otimes \tau_{B}^{y \mid t\ S D C}\right).
\label{quantumMarkov}
\end{equation}

Note that eq. \eqref{quantumMarkov} is analogous to eq.\eqref{causalMarkov}. 



 According to EnDQT, systems prepared at the source act as common causes for Bell correlations, having indeterminate values until each system interacts with Alice and Bob's measurement devices, giving rise to the correlated outcomes/determinate values. $\rho_{\Lambda}$ via its subsystems represents the systems prepared at the source.

To see this more concretely, let's focus on the example of systems that have indeterminate values of spin-p (where p ranges over all possible directions of spin) in a Bell scenario with two parties sharing an entangled state. This example could, in principle, be extended to any finite-dimensional case. We use $\rho_{\Lambda}$ to represent each system in the different regions separately by keeping track of the labels $A$ and $B$ and the channels $\rho_{B \mid \Lambda}$ and $\rho_{A \mid \Lambda}$. Each system evolves locally to region $A / B$, where Alice/Bob influences the outcomes that arise in $A$/$B$. This influence is represented via the quantum channel $\rho_{A \mid \Lambda}$ in the case of $A$, and $\rho_{B \mid \Lambda}$ in the case of $B$. $\rho_{A \mid \Lambda}$ and $\rho_{B \mid \Lambda}$ are identity channels that acting on the density operator $\rho_{\Lambda}$ representing the systems in region $\Lambda$, evolve them to regions $A$ and $B$, respectively.\footnote{Note that an identity channel leave the quantum state unchanged.} The influence that gives rise to the outcomes/determinate values is also represented via the POVMs $\tau_{A}^{x \mid s\ SDC}$ in the case of Alice, where $s$ is her random measurement choice, and $x$ is her outcome/the determinate value of S, and analogously via $\tau_{B}^{y \mid t\ SDC}$ in the case of Bob. The superscript SDC means that these are interventions that give rise to a determinate value, connecting each one of the systems with an SDC, and correspond to other kinds of edges in the DAG in Figure \ref{fig:example4}. Alice and Bob, due to their measurements, will lead the systems to become part of an SDC because they also belong to SDCs. Importantly, the relations of influence represented via QCMs are represented via QT but without reifying the quantum states like EnDQT assumes. More concretely, by adopting EnDQT's view of quantum states, it isn't considered that the (local) measurement of Alice on the system affects the system of Bob and Bob, and vice-versa because we aren't reifying quantum states, viewing them as representing by themselves certain causal relations. Other elements of EnDQT and QT that we are seeing here, with the $auxiliary$ support of quantum states, provide such representation. The Born rule in (\ref{quantumMarkov}) is only applicable when we take into account the interventions that lead the target systems to belong to the SDCs in each lab.

Let's see how this works diagrammatically. We can represent the Bell situation via the following (what I will call) EnDQT-causal-DAG (Fig. \ref{fig:example4}), where the nodes in grey represent the systems that $don't$ belong to an SDC, and the arrows in grey represent their evolution and influences on the values of the systems these arrows point to. The nodes in black represent the systems that belong to an SDC (Alice/Bob). The arrows in black represent their interactions with other systems that give rise to the latter having determinate values, pointing to these systems. These arrows and interactions in grey are mathematically represented by POVMs. EnDQT-causal-DAGs aim to highlight the fact that measurement-like interventions in QCMs involve systems that are locally connected with SDCs.\footnote{Note that this DAG is different from the ones above involving the propagation of the DC.}

Thus, we can see that the DAG in Figure \ref{fig:example4} leads to a local common cause explanation of Bell correlations,\footnote{See \cite{Wood2015TheFine-tuning} for examples of non-local, superdeterministic, and retrocausal causal structures. They differ from the one in Figure \ref{fig:example4}.} which in principle could be extended to any quantum correlations (even the more tricky ones to explain, see below). The local interactions at each wing are mediated by the SDCs, and these interactions, plus the prepared systems at the source, provide a local non-relational, non-superdeterministic, non-retrocausal common cause explanation of quantum correlations, including Bell correlations. Therefore, EnDQT allows QCMs to be explicitly local and non-operational, and according to EnDQT, QCMs provide a local non-relational, non-superdeterministic, and non-retrocausal explanation of quantum correlations.

To be clear, note that EnDQT's approach to quantum causation is not restricted to Bell scenarios but also has the benefit of being applicable to scenarios where it is difficult to see how to apply QCMs coherently, such as in the popular extended Wigner's friend-like scenarios.\footnote{See, e.g., \cite{Bong2020AParadox, Brukner2020FactsRelative, Frauchiger2018QuantumItself,Ormrod2023WhichProblem, Myrvold2002ModalRelativity, schmid2023review}.} Suppose we have two friends/agents in isolated space-like separated labs in each wing,\footnote{I will focus on the scenario from \cite{Bong2020AParadox, Brukner_2018}.} and one Wigner/agent next to each lab, where the friends share an entangled pair prepared at the source, as in the Bell scenario. It is also assumed that the lab is isolated in such a way that Wigner can perform arbitrary unitary operations on the contents of the lab. Here, we have the case explained in Section 2, which concerns a process represented by pragmatic reversible decoherence models involving the target system and the friend or their measurement device. The friend is a macroscopic system that (supposedly) makes a measurement/decoheres her target system; however, Wigner can reverse this process.

So, if the isolation of the friend plus their system from the SDCs is achieved (see Section 2), the Wigners can unitarily manipulate the friend plus their system, possibly reversing their state. We would then treat each friend and their target systems at each wing as being in an entangled superposition of states. Thus, the evolution of each system of the entangled pair to each wing and the ``measurements'' of each friend would be treated via a unitary channel that entangles each friend and their target system, where these channels would also represent the causal structure of this situation.\footnote{In a toy scenario, we could represent the quantum channels that entangle each friend in each wing by a CNOT gate in the CJ-form, and each friend as a being initially in the state $|0\rangle$. The target systems would be in an entangled state in the computational basis, (i.e., a state belonging to a Hilbert space spanned by the basis $|00\rangle, |11\rangle, |10\rangle$  and $|01\rangle$).} Then, as I have mentioned, the Wigners in each wing can unitarily manipulate or measure these entangled states. We could then calculate the probabilities for these measurement outcomes for the different measurement settings of the Wigners using a version of the Born rule like in eq. \eqref{generalizedBornRule}, providing a local common cause explanation for this situation.

Notice that, in the (unlikely or perhaps even impossible) case that the above isolation from the SDCs is successful, contrary to what is assumed by the theorem underlying the scenario mentioned above, there is no joint probability distribution for the outcomes of the friends and Wigner. This is because the friends inside their labs don't obtain any outcomes since they don't interact with SDCs, which allows the Wigners to manipulate them unitarily. So, it rejects the so-called \textit{absoluteness of observed events} assumption of this theorem, not because events aren't absolute like relationalist views claim, but rather because the events concerning the outcomes of the friends don't occur.\footnote{This assumption is typically explained in the following way: ``[a]ny observed event is an absolute single event, and not relative to anything or anyone.'' \cite{Bong2020AParadox}. More precisely, there is a theoretical probability distribution \( P \) of theory $\theta$ that can give rise to the empirical probability distribution $p$, where

\[
p(a,b|x,y) = \sum_{c,d} P_{\theta}(a,b,c,d|x,y), \quad \forall a,b,x,y
\]
\noindent where \( a \), \( b \), \( c \), and \( d \) are the outcomes of Alice, Bob (which are the two Wigners, each in each wing), Charlie, and Debbie (which are the two friends, each within the isolated lab), respectively, and \( x \) and \( y \) are measurement choices of Alice and Bob. EnDQT is a theory that doesn't consider that there is such theoretical probability distribution \( P \) in the case the lab is isolated from the SDCs because Alice and Bob don't obtain any outcomes.} This theorem also assumes the locality assumption (explained in more detail below), which says that the outcomes of the system of one of the Wigners are statistically independent of the measurements of the other Wigner on their system and vice-versa. Furthermore, it assumes the no-superdeterminism assumption.

Given its appeal to indeterminacy, one may wonder about what, according to EnDQT, the friend experiences when it is in a superposition. In other words, in the unlikely possibility that we manage to isolate such a macroscopic system from SDCs, what's going on with their mental content (e.g., their thoughts, desires, etc.), according to EnDQT? More concretely, in the ``local friendliness theorem'' mentioned above,\footnote{\cite{Bong2020AParadox}.} Wigner sometimes opens the door and asks the friend which outcome they obtained; what's happening with the mental content of the friend? There are different possibilities that I don't have space to go in-depth here: one could consider that the friend lacks mental content. However, this position might seem unsatisfactory since it is hard to conceive what it is for a friend-like system to lack mental content. So, this possibility can be deemed as incoherent. Instead of adopting this position, I think that a more satisfactory possibility is to consider that friend-like systems, when isolated from SDCs, have indeterminate mental content, where this content depends on the indeterminate physical properties of their brain. When the lab is open, their indeterminate mental content becomes determinate.\footnote{See, e.g., \cite{Barrett2001TheWorlds} for a discussion of the relation between quantum theory and mental states.} This possibility has the advantage of not being foreign to the philosophy of mind. Externalism about mental content roughly consists of the thesis that mental content depends on the external environment of the subject that has that mental content.\footnote{\cite{Putnam1975Themeaning}.} The friend $having$ determinate mental content is dependent on the SDCs that render that content determinate.

Due to its oddness, one may still object to this consequence of EnDQT, but I will now briefly argue that in actually experimentally attainable extended Wigner's friend-like scenarios, many other respectable interpretations of QT will actually lead up to similar consequences. Extended Wigner's friend scenarios might never be experimentally feasible because, arguably, we can never isolate macroscopic systems that are capable of measuring systems as the friend does in this scenario. However, recently \cite{Wiseman2023ASuit} proposed a variation of this scenario where the Friend is an AI agent running on a quantum computer. Wigner is just a user of the quantum computer that is capable of manipulating the friend. Before Wigner erases the memory of the friend by unitarily manipulating this system, he can talk with the friend. This scenario makes extended Wigner's friend-$like$ scenarios experimentally feasible in the future. They call it the local friendliness theorem. Granting that such a system is like us in some ways and, therefore, has mental content, one can similarly also consider that before the friend delivers their output to the exterior, their mental content is indeterminate. Thus, in this version, we would deny the ``Friendliness'' assumption used to derive this no-go theorem. Various influential interpretations of QT, such as the MWI and spontaneous collapse theory, would also deny this assumption. This is because for quantum computers to be functional, their internal states will need to be maintained in a coherent superposition in order to perform quantum computations. So, no collapse or branching should occur inside a quantum computer.  Furthermore, in relationalist single-world theories, such as relational quantum mechanics,\footnote{\cite{Rovelli1996RelationalMechanics, DiBiagio2021StableFacts}.} the friend would be considered to have indeterminate mental content.

Therefore, although one might still object to this feature of EnDQT, it is something that, in actually experimentally feasible Wigner's friend scenarios, applies to other respectable interpretations of QT, such as collapse theories, MWI, and relational quantum mechanics. Thus, if we accept these other interpretations, we shouldn't see this feature as a problem in our noisy worlds, where isolating macroscopic systems might never be feasible. In Appendix B, I will go into more detail about this argument.


As a side note, it doesn't seem that any of the current QTs that don't modify the fundamental equations of QT, can use QCMs in this local, non-relationalist, and non-operational way to give a local common cause explanation of quantum correlations like those in the Bell and extended Wigner's friend scenarios. Thus, EnDQT currently appears to be the only theory capable of providing such an explanation in this manner. Note that relationalist theories are, along with EnDQT, the only non-operational, non-hidden variable theories that don't modify the fundamental equations of QT and consider it a universal theory. So, they are the only ones who could also consider that QCMs, which use standard QT, provide the whole causal story. Furthermore, spontaneous and gravity collapse theories will necessarily impose fundamental limitations on these macroscopic superpositions. As I have said, in the case of EnDQT, it all depends on the details of the histories of the SDCs (which includes not being subject to human ingenuity). However, typically, in relationalist theories, the shared correlations of the friends or Wigners only arise when they meet (if they ever do). Thus, there is no common cause explanation in the above sense. Moreover, QCMs in the single-world cases (at least) should be modified or adapted to account for these multiple varying perspectives since they don't consider that variation. So, contrary to the suggestions of others, EnDQT considers that QCMs don't need to be modified or adapted to a relationalist approach for them to explain the correlations that arise in the extended Wigner's friend scenarios.\footnote{\cite{e23080925, schmid2023review, ying2023relating}. See \cite{ormrod2024quantum} for a recently proposed relationalist adaptation.} Adopting EnDQT, we don't need to adopt this more complex approach to QCMs, which can be regarded as another benefit of this view.

Therefore, I have shown that if QCMs consider that i) systems only assume determinate values with a certain Born probability when they interact with an SDC, and ii) where those relations are described via QT, but without reifying the quantum states like EnDQT assumes, then QCMs provide a local non-relational, non-superdeterministic, and non-retrocausal explanation of quantum correlations.

Finally, let's turn to my last core premise:\\

P3) If we interpret Classical Causal Models according to EnDQT, we see that in comparison to QCMs i) they don't assume that systems only have determinate values when they interact with an SDC and ii) that the relations of influence aren't described via QT. The relations of influence that they represent only arise in a special limit from QCMs.\\


 QCMs consider that common causes can have indeterminate values represented via QT, i.e., via subsystems of an entangled state, and probabilities explicitly don't arise from the ignorance of underlying determinate values. Furthermore, in agreement with EnDQT, for QCMs, common causes, represented by the subsystems of the entangled state, don't have determinate values and a Born probabilistic model independently of the interactions with Alice or Bob. This is contrary to CCMs that assume that common causes and causes in general are represented via classical variables, which can always be assigned probabilities. Therefore, CCMs don't assume i).
 
 Also, CCMs don't assume ii) because the relations of influence aren't explicitly represented via QT, i.e., via CPTP maps when systems don't interact with members of an SDC and by CP-maps/POVMs when systems interact with members of SDCs. Only in a limit where we can consider the systems as having determinate values, the QMC reduces to the CMC.\footnote{The details about how to obtain this limit precisely are too evolved to be presented here. Basically, the classical limit should involve a process operator $\sigma_{A_{1} \ldots A_{n}}$, where there is an orthonormal basis at each node (that is, an orthonormal basis for $\mathcal{H}_{A_{i}^{\text {in }}}$, along with the basis for $\mathcal{H}_{A_{i}^{\text {out }}}$ ), such that $\sigma_{A_{1} \ldots A_{n}}$ is diagonal with respect to the product of these bases. This corresponds, for example, to the situation where the systems at the source are prepared in a product state.}

Thus, if we interpret classical Causal Models according to EnDQT, we see that i) systems only assume determinate values with a certain Born probability when they interact with members of an SDC and ii) the relations of influence that they represent aren't described via QT. The relations of influence that they represent only arise in a special limit from QCMs.

Therefore, given that\\

P4) QCMs, as interpreted by EnDQT, consider that i) systems only assume determinate values with a given Born probability when they interact with an SDC, and ii) where those relations are described via QT, but without reifying the quantum states like EnDQT assumes;\\

\noindent and given P1) and P2), QCMs accurately represent causal relations between quantum systems for EnDQT and where those relations of influence are local. Furthermore, given P1) and P3), contrary to QCMs, CCMs with their CMC for EnDQT don't accurately represent relations of influence between quantum systems. Contrary to the previous argument, notice that this argument doesn't necessarily appeal to the derivation by \cite{Pearl1995ACausation} of the CMC to reject the applicability of the CMC to infer causal relations between quantum systems (although it may $also$ appeal to this proof to help to compare both models). Rather, it appeals in general to the interpretation and comparison of CCMs with QCMs according to the EnDQT. Also, it shows that we can have a local explanation of Bell correlation via QCMs, and so it prompts us to reject the CCMs and the CMC.

Coming back to Bell's theorem, given the above arguments, such as the argument for locality, we reject CCMs and the CMC as accurately representing the relations of influence between quantum systems. Therefore, we reject the Classical Reichenbach Common Cause Principle and, therefore, the factorizability condition as accurately representing the relations of influence between quantum systems. EnDQT rather assumes a local and more accurate way of accounting for those correlations via QCMs. Therefore, via these reasons and arguments, EnDQT deals with Bell's theorem quantum correlations while ensuring that the theory does not conflict with relativity by favoring any specific reference frame, thus avoiding action-at-a-distance phenomena as seen in Bohmian mechanics. Also, it does not introduce hidden variables that result in retrocausality or superdeterminism. Furthermore, it deals with this theorem without adopting a relationalist interpretation of QT. Therefore, it fulfills desiderata LC) and NR).

One may object that according to EnDQT, systems can also have an indeterminate value of position because systems can be in a superposition of positions or be in an entangled state of positions. Thus, it seems that some sort of non-locality is built into EnDQT because systems won’t have a location in spacetime in the standard sense. As a reply, I should note that at the level of fundamental physics, i.e., at the level of quantum fields and quantum field theory, time and position aren’t observables in the standard sense. They are rather parameters that serve to parametrize the different configurations of quantum fields. So, since non-relativistic quantum theory arises as a limit of relativistic quantum theory (i.e., quantum field theory), we shouldn’t consider position and time per se as observables of quantum systems. So, in particular, we shouldn't consider a system in a superposition of positions as literally spread out over space at a time. Thus, the above objection is not worrying.

Sometimes, it is argued that QT is non-local and that the EPR argument \cite{Einstein1935CanComplete} ruled out the existence of local indeterministic theories (e.g., see \cite{Maudlin_2014} for an influential argument), and so one might worry that there is something wrong with my argument above. However, this argument concerning the non-locality of QT in principle should not be correct because EnDQT, as an indeterministic local theory, is a counterexample to its claims.

The EPR argument can be posed in the following way, let's assume QT and that experiments performed at arbitrary distances from each other, don't disturb the outcomes of each other. Furthermore, let's assume the following so-called EPR criterion of reality:\footnote{\cite{Einstein1935CanComplete}.} 

\begin{quote}
[i]f, without in any way disturbing a system, we can predict with certainty (i.e., with probability equal to unity) the value of a physical quantity, then there exists an element of reality corresponding to that quantity.    
\end{quote}

 Let's consider Alice and Bob in distantly separated labs, performing measurements in their systems in the same direction. Every time Alice gets spin up, Bob gets spin down, and vice-versa. If Alice/Bob, without disturbing the outcome, can predict with certainty the value of the observable of the system that Bob/Alice measures, then there should exist an element of physical reality that corresponds to these quantities that Bob/Alice measures prior to their measurement. However, QT is silent about those elements for all physical magnitudes in these situations hence it is incomplete.

Now comes another assumption for the above argument for non-locality:
\begin{quote}
The further conclusion that a final and complete physical theory must be deterministic \textit{at least with respect to these particular phenomena} just comes as an additional bonus. If the world is EPR-local, and there is no spooky action-at-a-distance, then not only must the quantum mechanical description of a system leave out some elements of reality, but the elements that it leaves out must be sufficient, in these circumstances, to completely predetermine the outcome of the ``measurement'' operation. For, as Bell remarks in the passage cited above, ``any residual undeterminism could only spoil the perfect correlation''. This further conclusion of predetermination obviously requires that the relevant correlations be perfect, which is also what is required here to apply the EPR criterion (``we can predict with certainty (i.e. with probability equal to unity)''). (\cite{Maudlin_2014}, pp. 10)   
\end{quote}



Thus, this completion should be a deterministic completion. Bell's 1964 theorem\footnote{\cite{Bell1964OnParadox}.} formalizes this conclusion by an assumption, which we may call outcome determinism or counterfactual definiteness or predeterminism.
Let's consider that $c$ is the variable that designates the preparation at the source of entangled particles, $\lambda$ represents the hidden variables of the particles that may exist between the source and the measurement of Alice and Bob. $A$ and $B$ are the measurement outcomes of Alice and Bob, and $a$ and $b$ are their choices of measurement settings. $P_{\theta}$ designates a probability model given by a theory $\theta$. Then, a theory $\theta$ obeys predeterminism if assumes variables $\lambda$ that determine the outcome A and B in such a way that:\footnote{See \cite{Wiseman2017}.}

\begin{equation}
    \forall A, a, B, b, \lambda,\ P_\theta(A, B \mid a, b, c, \lambda) \in \{0, 1\}.\label{predeterminism}
\end{equation}

We consider that these variables fully determine Alice and Bob's outcomes. In addition to this assumption and a no-superdeterminism assumption, this theorem also assumes the locality condition, also known as ``parameter independence''\footnote{\cite{Jarrett1984OnArguments}}, which says that the outcome for the quantum system in one of the wings in the Bell scenario is statistically independent of the measurement performed in the other wing and vice-versa, i.e., $P_\theta(B \mid a, b, c, \lambda) = P_\theta(B \mid b, c, \lambda)$ and similarly in the other case. The argument for non-locality proceeds by claiming that given that we should accept the EPR argument (as Bell supposedly did), we should already accept the predeterminism assumption. \footnote{As Maudlin (2014, pp. 10) argues,
\begin{quote}
   That
argument
is
one
line:
the
very
“element
of
reality”
that
the
EPR
argument
proves
to
exist—given
EPR‐locality—is
an
element
of
reality
defined
just
as
whatever
physical
characteristic
of
the
system
it
is
that
ensures
how
it
would
react
to
the
measurement
in
question.
So
any
system
that
has
that
element
of
reality
has
a
physical
characteristic
that
determines
how
it
would
react
to
the
measurement.
But
that
just
is
determinism
with
respect
to
that
particular
“measurement
operation”.
And
the
EPR
argument
can
be
repeated
for
any
“measurements”
for
which
quantum
theory
predicts
perfect
correlations
between
the
outcomes
and
that
can
be
made
arbitrarily
far
apart
in
space.
Hence,
in
an
EPR-
local
theory
both
the
reactions
to
a
“position
measurement”
and
to
a
“momentum
measurement”
must
be
predetermined
by
some
element
of
reality
in
the
system,
and
in
the
Bohm
spin
example
the
reactions
to
every
possible
“spin
measurement”
must
be
predetermined.
That
is
enough
to
get
Bell’s
1964
argument
off
the
ground.
Not
by
assuming
determinism,
but
by
assuming
EPR-locality
and
deriving
determinism.
\end{quote}}

\begin{figure}[ht]
    \centering
    \includegraphics[max width=\textwidth]{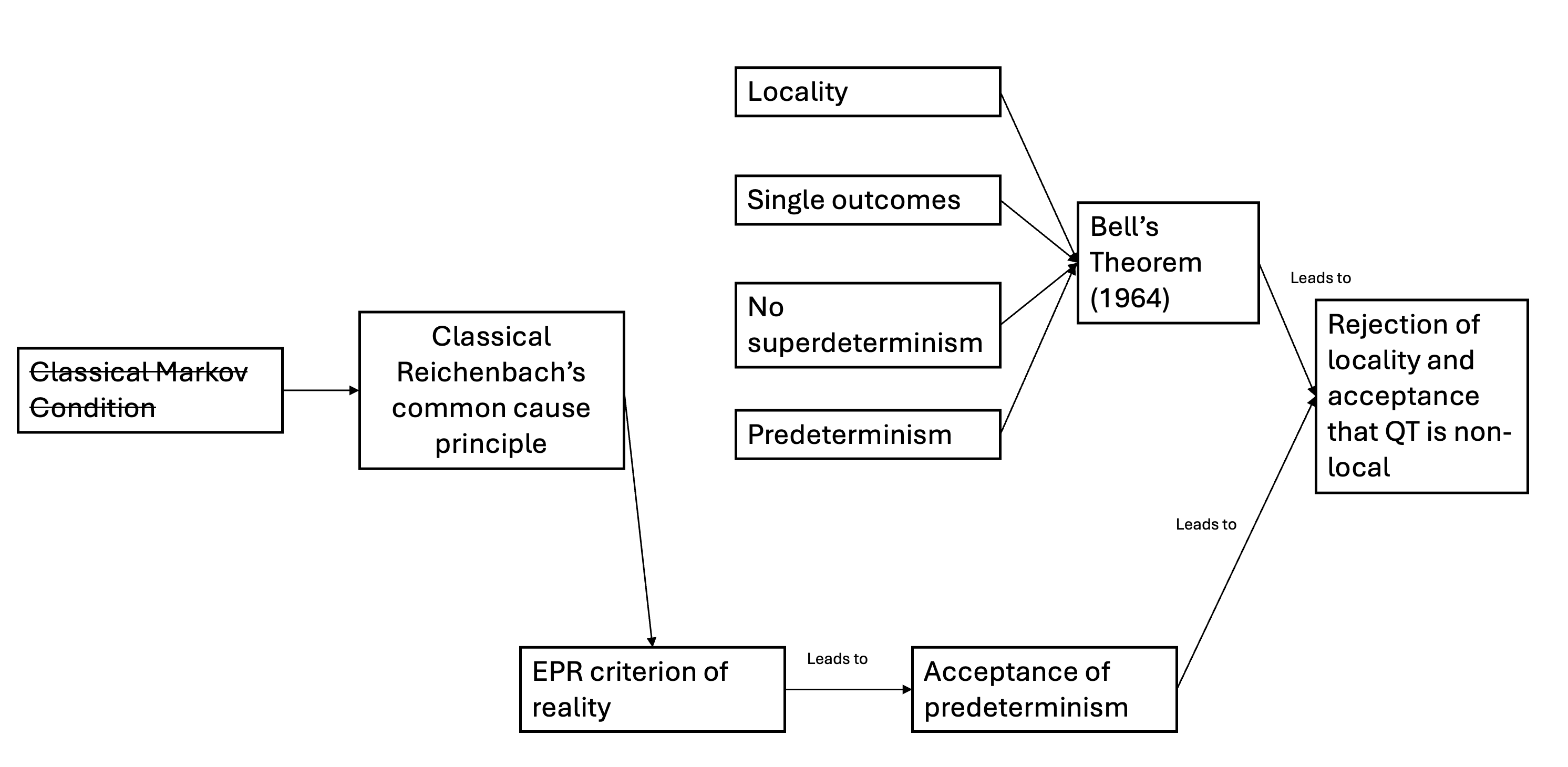}
    \caption{Diagram that helps understand the argument for non-locality based on Bell's 1964 theorem works and how EnDQT deals with it. It again involves the rejection of the applicability of the Causal Markov Condition and Classical Causal Models to, in general, accurately represent causal relations between quantum systems. If a box with the claim $X$ is connected by one or more arrows without a legend, means that the one or more boxes with claims $Y$ that point to $X$ jointly lead to the derivation of $X$. If an arrow has the legend ``lead to,'' it means that one claim leads to another, as explained in the text.}
    \label{fig:diagram2}
\end{figure}

Thus, since denying the ``no-superdeterminism'' and the ``single outcomes'' assumption also assumed by this no-go theorem is problematic,\footnote{\cite{Maudlin_2014}.} we should rather reject the locality assumption. Hence, quantum theory is non-local, i.e., it leads to action at a distance. This argument for non-locality is summarized in Figure \ref{fig:diagram2}.


There are some problems with this argument.

First, Bell's 1964 theorem is restricted to perfect correlations that occur in certain circumstances in the Bell scenario when Alice and Bob measure their entangled systems on the same basis. Predeterminism (equation (\ref{predeterminism})) assumed by this theorem is that there is a theory that allows us to predict with certainty the outcomes of Alice and Bob by specifying certain variables in the past of these outcomes plus the measurement choices of Alice and Bob. These variables are any information that represents entities that determine that outcome. 

The EPR criterion of reality is about the postulation of an inference from certain predictions to something that determines those predictions. Predeterminism isn't that assumption, and an indeterministic theory like EnDQT doesn't obey it.\footnote{It is also unclear if Alice can ever predict with $100\%$ of probability the outcome of Bob and vice-versa in a world represented by an indeterministic theory. The particles that constitute Bob obey indeterministic dynamics, and thus, it is impossible to be certain whether Bob will measure his particle on the same basis as Alice, as it is assumed by the EPR criterion. So, one could argue that the EPR criterion assumes physically unrealistic circumstances for an indeterministic theory.}

Second, it is not even clear to what extent this assumption follows from the EPR criterion of reality (CR) since it is unclear what this principle means. To my knowledge, the best precisification of this principle views it as a consequence of the classical Reichenbach common cause principle \cite{Gomori2021OnCriterion}, which, as I have mentioned, is a special case of the more general CMC applied in realistic physical situations \cite{sep-physics-Rpcc}. 


First of all, instead of variables, \cite{Gomori2021OnCriterion} express the CRCCP using events described using propositions and their negation. However, in the case of their proof, in principle, we can think of these propositions in terms of binary variables. Now, some terminology: there exists a \textit{directed path} from vertex \(X\) to vertex \(Y\), denoted by \(X \rightsquigarrow Y\), if there is a direct arrow from \(X\) to \(Y\) (\(X \rightarrow Y\)) or there is an intermediate node \(Z\) such that \(X \rightsquigarrow Z\) and \(Z \rightarrow Y\). Furthermore, we write $X \nrightarrow Y$ when event $X$ doesn't influence event $Y$.

Suppose A is the outcome of a measurement $a$ of a certain physical quantity measured by Alice. Suppose that B denotes the event that involves the prediction that the quantity in question has a certain value, which is made by Bob. Furthermore, suppose event B doesn’t influence event A, as it should, because the act of prediction doesn’t influence what is being predicted. Also, suppose that event B doesn’t influence event A. In other words, suppose that Bob doesn’t influence the outcome of Alice, i.e.,“without in any way disturbing a system” the outcome of Alice. Also, suppose that the prediction involved in event B is certain, and by certain, it’s meant that whenever B happens and a is performed, the measurement results in outcome A, i.e., $p(A|a \land B) = 1$. Then, there is an element of reality $\alpha$ that will cause the outcome A to occur when the measurement of a is performed. In other words, $p(A|a \land \alpha) = 1$. Notice that the notion of determination of an outcome by elements of physical reality is clearer here. We can summarise the above description in the following way (\cite{Gomori2021OnCriterion}, pp. 13459-13460):
\begin{quote}
\textbf{Reality Criterion} Suppose that \( A \) is an outcome of measurement \( a \). Assume that there is an event \( B \) such that
\begin{equation}
   A \nrightarrow B \text{ and } B \nrightarrow A
   \tag{24}
\end{equation}
\begin{equation}
   p(A|a \land B) = 1
   \tag{25}
\end{equation}

Then there exists a further event \( \alpha \) such that
\begin{equation}
    \alpha \rightarrow A
    \tag{26}
\end{equation}
\begin{equation}
 p(A|a \land \alpha) = 1
 \tag{27}
\end{equation}
\end{quote}

How can we make sense of the above reality criterion claim as concerning a causal claim? We can make sense of this claim as concerning a causal claim because it can be shown that the reality criterion follows from the CRCCP (and ultimately from the CMC, the CMC, and its semantics). The proof would take us too far afield,\footnote{\cite{Gomori2021OnCriterion}.} but to give an intuition about how this goes, we can see that the predictor of an element of reality is the predictor of a special kind of classical common cause. To see this, consider the following formulation of the CRCCP by Gömöri and Hofer-Szabó (\cite{Gomori2021OnCriterion}, pp. 13460-13461):\footnote{$X\rightarrow Y$, means that $X$ influences $Y$. $X \nrightarrow Y$, means that $X$ doesn't influence $Y$.}\\

\begin{quote}
\textbf{Common Cause Principle} Suppose that $A$ is an outcome of measurement $a$. Assume that there is an event $B$ such that

\begin{equation}
A \nrightarrow B \text{ and } B \nrightarrow A 
\tag{33}
\end{equation}
\begin{equation}
    p(A \land B | a) \neq p(A | a) p(B | a)
    \tag{34}
\end{equation}

Then, there exists a further event $C$ such that
\begin{equation}
C, \neg C \rightarrow A, B
\tag{35}
\end{equation}
\begin{equation}
p(A \land B | a \land C) = p(A | a \land C) p(B | a \land C)
\tag{36}
\end{equation}
\begin{equation}
p(A \land B | a \land \neg C) = p(A | a \land \neg C) p(B | a \land \neg C)
\tag{37}
\end{equation}
\end{quote}

$C$ in the equation above would be the element of reality. If we take into account this definition of the CRCCP and the above one of the RC, it can be shown that the reality criterion is just a special case of the CRCCP. More concretely, it can be shown that the reality criterion is an application of the CRCCP to the case where the prediction is certain and if the act of prediction does not disturb the predicted phenomenon. 

 So, even if we grant the above connection between the EPR criterion of reality and predeterminism since EnDQT rejects the CRCCP's applicability to accurately represent causal relations between quantum systems, it also rejects the CR's applicability to represent such causal relations (since one follows from the other). Thus, (as expected) it should also reject the predeterminism condition as accurately representing such relations.

Maudlin\footnote{\cite{Maudlin_2014}.} called the reality criterion analytic, but as Lewis puts it\\

\begin{quote}
I doubt that Einstein’s criterion of reality is really analytic [...] It seems perfectly
conceivable that an event could be predicted with certainty even when there is
nothing physical that brings that event about. [...] Indeed, Maudlin is perfectly
sanguine about fundamentally probabilistic laws (e.g. in spontaneous collapse
theories), according to which there is in general no physical reason why this
result is obtained (as opposed to that result) when the probabilities differ from
zero and one. Why should things be different when the probabilities are zero
and one? (\cite{Lewis2019BellsLocality}, pp. 38–39)
\end{quote}

 Gömöri and Hofer-Szabó reply: 
 
 \begin{quote}
[t]he answer to this is now simple and clean: the reason why probability-1 predictions
are distinguished is because those predictions mean perfect correlation between the
act of prediction and the predicted outcome, and a perfect correlation can only be
explained, in accord with the CCP’s requirement [i.e., the CRCCP], by a deterministic common cause;
hence the outcome must be predetermined. (\cite{Gomori2021OnCriterion}, pp. 13463). 
 \end{quote}

If the CRCCP and the CMC, more generally, are based on determinism, and this seems to be the case given what we saw above with one justification of the CMC, this is unsurprising. Moreover, as my arguments above indicate, quantum causal models show how the CRCCP leads to wrong inferences about how systems influence each other. They also show us how Bell's theorems, more generally, can be considered as telling us how quantum indeterminacy and the determinacy that arises from it behave and that we need new inferential tools to understand this process. 
More concretely, the inferences we make based on classical causal models (and their assumptions) are wrong for EnDQT, even when perfect correlations arise between measured quantum systems. When Alice, at one wing of a Bell experiment, makes her measurement, she can't make claims about the physical state of Bob's system a la elements of reality/hidden variables, which will determine his result. She is applying the wrong causal model to make these inferences. The right theories to use to make inferences are quantum causal models and EnDQT in this view. If Alice used them, she would see that not only her target system but also Bob is an indeterministic system connected with SDCs that determine his behavior. Furthermore, Bob, as well as his target system, are described via quantum states, observables, as well as SDCs and QCMs. She would also notice that there is no ``element of physical reality'' with the above features in the quantum domain, and nothing is predetermined in Bell's theorem sense.

Thus, for EnDQT, the above argument doesn't work to show that QT is non-local, even if we grant that we can go from this argument to accepting predeterminism (which we can also deny on independent grounds, as well as the predeterminism assumption, as I have explained above).

One may object to the claim that EnDQT provides a local view of the world and argue that there seems to be some non-locality going on in this view. Suppose that quantum systems have some kind of irreducible and intrinsic dispositions that are manifested with a given probability under some measurement-like interactions. Also, quantum states work as predictors for those manifestations. Let's consider the case where the two entangled spin-1/2 particles end up being anti-correlated in the Bell scenario upon measurements of Alice and Bob on the same basis. For instance, it seems that when Bob measures his system and obtains spin-up, the probabilistic disposition of the particle of Alice changes instantaneously at a distance. More concretely,  it changes to  $100\%$ being spin-down under a measurement of Alice on the same basis as Bob. 
However, note that it doesn't seem that any of the theorems or arguments we have examined and assumptions that we have dealt with oblige us to infer that there is some non-local influence between the dispositions of both particles. Furthermore, the view taken here on these theorems and arguments doesn't lead to the commitment of some sort of irreducible and intrinsic dispositions.

Relatedly, notice that,\footnote{Thanks to Peter Lewis for pressing me to clarify this.} even if we have theories where like EnDQT systems don't have determinate values (described by hidden variables) until they reach Alice or Bob, EnDQT via its theoretical posits and tools (such as SDCs and QCMs) shows that we don't need to pose those influences and adopt these theories. Let's consider the Ghirardi–Rimini–Weber (GRW) theory with flashes (determinate values of position) that arise from the collapse of the wavefunction randomly in spacetime. Flashy GRW doesn't (at least necessarily) adopt a realist position about the wave function -- the wave function can be regarded as a bookkeeping device that allows us to keep track of the probabilities of flashes at various locations (i.e., of dispositions). Suppose Alice in a Bell scenario measures the spin of her particle. That triggers a series of flashes at Alice's location, and those flashes instantly and non-locally change the probabilities for flashes associated with Bob's particle. For example, if Alice measures z-spin and gets spin-up, the probabilities at Bob's location instantly change to those associated with a z-spin-down particle. But flashy GRW doesn't postulate any hidden variables (in the sense of determinate values) underlying the probabilities -- they can appeal just to brute dispositions to account for these correlations. 

However, note that EnDQT provides a theory that shows how we can also dispense with posing these non-local dispositions while explaining Bell correlations. According to EnDQT, systems represented via QCMs evolve locally and only have determinate values when they interact with members of SDCs. So, SDCs are necessarily involved in these influences that give rise to determinate values, and they concern local interactions between systems, not non-local ones. Alice doesn't influence non-locally the probabilities of the system of Bob or vice-versa. Thus, EnDQT provides a local $causal$ pathway that dispenses with the adoption of non-local dispositions.


\section{Conclusion and future directions}

I have proposed EnDQT and argued that contrary to the other well-known QTs, it has the great benefit of being a local, non-relational, non-superdeterministic, and non-retrocausal quantum theory. Systems have determinate values only while interacting with other systems of SDCs. On top of that, EnDQT has the benefit of being conservative, not modifying the fundamental equations of QT, and, in principle, arbitrary systems can be placed in a coherent superposition for an arbitrary amount of time. Also, EnDQT is able to give a local causal explanation of quantum correlations.




There are many future directions. For instance, one should develop realistic quantum field theoretic models of SDCs,\footnote{\cite{PipaToyEnDQT}.} explore the explanatory potential further, seek to develop techniques to model and map SDCs, and test and extract further predictions from EnDQT, which might distinguish it empirically from other QTs.

\section*{Acknowledgments}
I want to thank Peter Lewis and John Symons for valuable feedback on multiple earlier drafts. I also want to thank Harvey Brown, Claudio Calosi, Ricardo Z. Ferreira, Sam Fletcher, Richard Healey, Gerard Milburn, Kim Soland, Noel Swanson, Chris Timpson, and the members of the Caltech Philosophy of Physics reading group, Philosophy of Physics graduate seminar and the Foundations of Physics workshop at the University of Oxford for helpful discussions and/or valuable feedback on earlier drafts.

\section*{Appendix A: Toy-model of an SDC}
In this section, I will present a non-relativistic toy model of an SDC. The notation of this model may look complicated, but I will essentially be modeling a chain of simple decohering interactions. Thus, I will also provide a simple and pedagogical example of decoherence.

Since I am considering non-relativistic quantum theory, it cannot be guaranteed that the laws will be Lorentz and, more generally, general covariant. However, I will provide indications that these symmetries will be respected in a relativistic model. These indications will come from showing that given CDC1)-CDC4), all successive events where systems have determinate values will be time-like separated from each other. So, the order of events, which is important for EnDQT to maintain (to explain how determinacy arises), won't vary according to the reference frame, and the events that arise from SDCs arise locally (although the duration of interactions can be described according to different reference frames). Also, even if systems are entangled with each other, since I am not reifying the wavefunction if a system has a determinate value (i.e., ``collapses''), it doesn't affect any other system that was entangled with. I will get back to these points at the end of this section.

 I will represent each system in the model by $s^{k}_{ij}$, and each system will occupy a fixed spatial point that I will specify further below in a way that will emulate a quantum field in a spatial point (modulo the difference in degrees of freedom). This will facilitate the transition to quantum field theory in future work. $i$ in $s^{k}_{ij}$ will represent a layer in the graph in which each system is. The higher the value of $i$, the later will be the events that the system $s^{k}_{ij}$ will give rise to (see Figure 1 with an example of a graph with three layers with systems with their indices). Thus, $i$ represents a temporal order. $k$ will represent groups of systems in each layer $i$. Each group of systems $k$ in a layer $i$ will serve as an $environment$ that will decohere a single system $s^{l}_{ij}$ in the layer $i+1$, where in $i+1$ that system will have an index $j=k$. So, the upper script $k$ will help us designate which groups of systems in a layer $i$ decohere which systems in a layer $i+1$. When I omit specific indices, I will be implicitly referring to all systems that have the indices that weren't omitted. So, when I write $S_0 \rightarrow S_1 \rightarrow S_2$, I will be referring to all the interactions between all systems with $S_{i=0}$ (in the layer $i=0$) and $S_{i=1}$ (in the layer $i=1$), as well as the interactions between all systems with $S_{i=1}$ and systems with $S_{i=2}$, where these interactions obey the conditions CDC1)-CDC4).

 Importantly, I will assume that the temporal order of interactions in the graphs runs from the top to the bottom: systems $S_0$ first interact with $S_1$, and then (in agreement with the CDCs), systems $S_1$ interact with $S_2$. As I have said, this can also be written as $S_0 \rightarrow S_1 \rightarrow S_2$, where the temporal order of interactions runs from left to right. The interaction between systems $S_0$ and $S_1$ gives rise to the set of spatiotemporal events that I will designate as $E_{S_0 \rightarrow S_1}$, which corresponds to systems $S_0$ and $S_1$ having determinate values. The interaction between systems $S_1$ and $S_2$ gives rise to the set of events $E_{S_1 \rightarrow S_2}$. Note that given CDC1), events $E_{S_0 \rightarrow S_1}$ are time-like separated from the event $E_{S_1 \rightarrow S_2}$, although some events in the set $E_{S_0 \rightarrow S_1}$ may be space-like separated from each other. This gives rise to the following ``causal'' order of events that I will write like this $E_{S_0 \rightarrow S_1}\rightarrow E_{S_1 \rightarrow S_2}$, where the temporal order of events runs from the left to the right. So, each edge in these graphs corresponds to one event, which involves both relata of interactions having a determinate value.

\begin{figure}[h]
    \centering
    \includegraphics[width=1.07\linewidth]{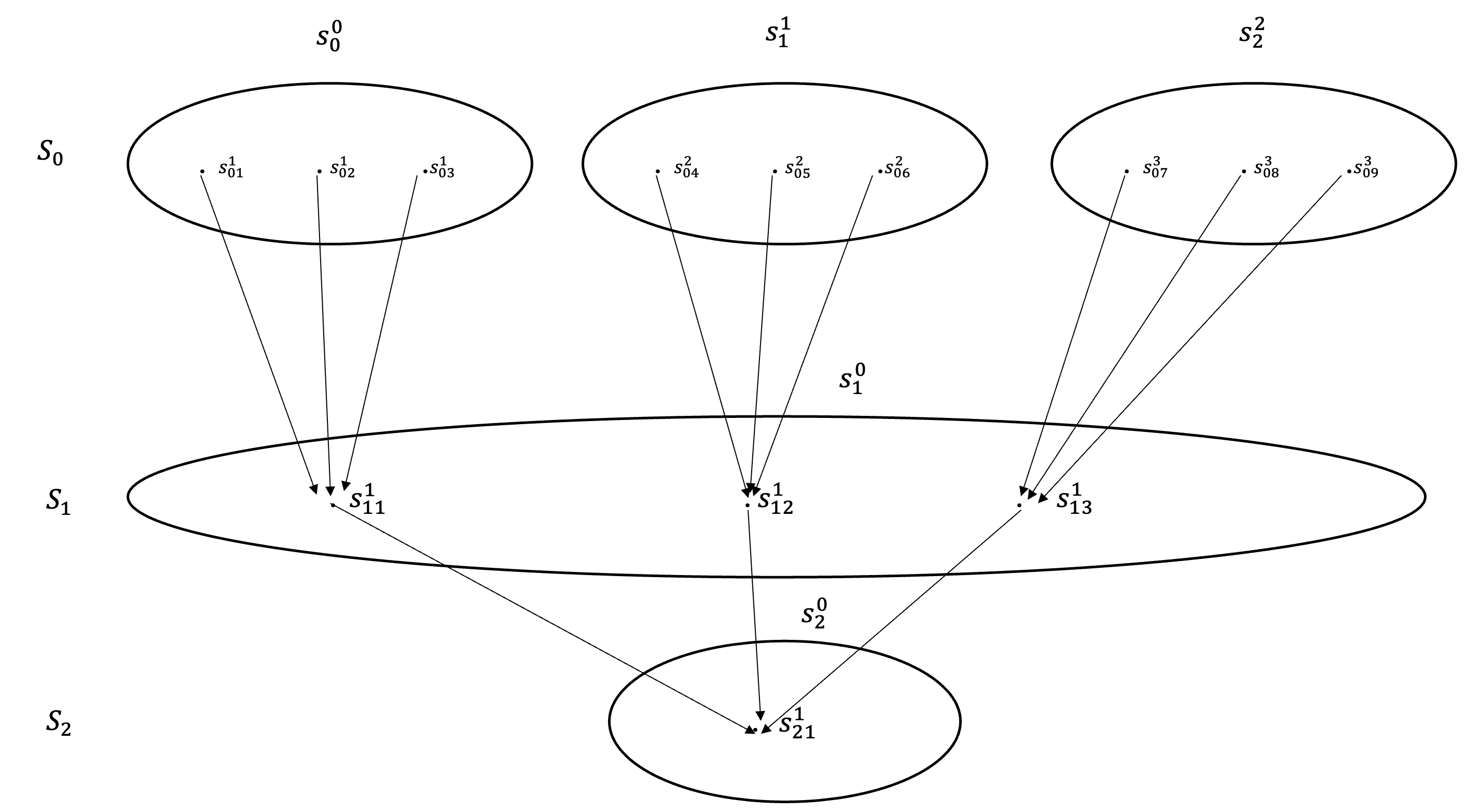}
    \caption{Representation of an SDC with the the structure $S_0 \rightarrow S_1 \rightarrow S_2$ and with three labels. Temporal order runs from the top to the bottom. Each edge corresponds to one event, which involves both relata of interactions having a determinate value. In this graph, we have 12 interactions depicted, which corresponds to 12 events.}
    \label{Figure1}
\end{figure}

In agreement with postulate CDC1*-ii*), let's consider that in this SDC, systems $S_i$ start interacting with systems $S_{i-1}$  while $S_{i-1}$  are $already$ interacting with systems $S_{i-2}$ (where $S_{i-2}$ have the DC-$S_{i-1}$) so that $S_{i-1}$  has the DC-$S_i$, and $S_{i-1}$  can end up transmitting the DC to $S_i$  concerning some other system that $S_i$ might end up interacting with. However, when $S_{i}$  and $S_{i-1}$  begin interacting, let’s assume that we can neglect the evolution of the quantum states of $S_{i-1}$ while $S_{i-2}$  and $S_{i-1}$  interact, such that we can consider that systems $S_{i-1}$  and $S_i$  start interacting only when the interaction between $S_{i-2}$  and $S_{i-1}$  ends. In a more complicated model, this assumption will be translated in terms of different timings in which the coupling constants $g$ in the Hamiltonians specified above are non-zero.\footnote{As I have explained in Section 2, there is a way of making this assumption more precise via the commutativity criterion.}

For simplicity, I assume that each interaction between a pair of systems, $S_i \rightarrow S_{i+1}$, constitutes a step in the model. The first step involves the interaction between systems $S_0$ and $S_1$, the second step will involve systems $S_1$ and $S_2$, and so on. At each step, I implement a state update, which corresponds to one of the possible determinate values of the relata of interactions that indeterministically arise with a certain probability. The probability for these determinate values is predicted via the reduced density operator for each interaction (more on this below).

Let's turn to the Hamiltonian for the decoherence models, which will represent the interactions between the systems $S_i$ and $S_{i+1}$. I will start with the total Hamiltonian for this chain. I omit the indices that represent the systems and the groups of systems within each layer. I will have \(N+1\) number layers, where \(S_0 \rightarrow S_1 \rightarrow \ldots \rightarrow S_{N}\), where $N$ is the value of the index of the systems in the last layer, and where I start counting the layers from $i=0$. I will adopt a certain reference frame to describe these interactions. For simplicity, I will assume that each pair of systems $S_i$ and $S_{i+1}$ interact in a given frame during a fixed duration $\Delta$ that corresponds to the duration of each step. To express this interaction succinctly via the Hamiltonian of the SDC, let's define the function $\chi(t_i+i\Delta, \Delta)$ given by the following difference between Heaviside functions:\footnote{\[
\theta(x) =
\begin{cases} 
0 & \text{if } x < 0 \\
1 & \text{if } x \geq 0
\end{cases}
\]}
\begin{equation}
    \chi(t,i\Delta, \Delta)= \theta(t-i\Delta)-\theta(t-(i\Delta+\Delta)).
\end{equation}

Now, I can express the sequential interaction between systems that obey the laws given by CDC1)-CDC4), where their interactions start at $t=0$,

\begin{equation}
\hat{H} =\sum^{N}_{i=0} \hat{H}_{S_{i} \rightarrow S_{i+1}} \chi(t, i\Delta, \Delta).
\end{equation}

$\hat{H}_{S_{i}} \rightarrow S_{i+1}$ designates the Hamiltonians of interaction describing the interaction between systems $S_{i}$ and $S_{i+1}$. As we will see, due to certain approximations, I will neglect the self-Hamiltonians of all systems. \(S_0\) and its subsystems will be the initiators, which means that they will have the DC by default. Thus, they are able to transmit the DC to any other systems without having to interact with other systems that have the DC, and, therefore, they are able to start an SDC.

Consider the following interactions between systems that belong to two layers, given by  \(S_{i} \rightarrow S_{i+1}\). As it was mentioned above, I am assuming that if we have a system \(S_{i+1}\) with \(a\) number of elementary subsystems, we will have a system \(S_{i}\) with \(N_{i}=a\) number of groups of systems, which will constitute the environment that will decohere each elementary subsystem of \(S_{i+1}\). Thus, in a toy model with the chain \(S_0 \rightarrow S_1 \rightarrow \ldots \rightarrow S_{N}\), the higher the index \(i\) of  \(S_i\), the lower the number of elementary subsystems each \(S_i\)  needs to have for this model to be adequate in describing processes of decoherence.

I will examine interactions modeled by the well-known spin-spin decoherence model \cite{Zurek1982Environment-inducedRules, Cucchietti2005DecoherenceEnvironments} because it is the most straightforward non-relativistic model. So, I will not be concerned with solving the model in general since it has been solved elsewhere. Rather, I will solve it for the simplest case to show how we can model SDCs via this decoherence model.

The initiator \(S_0\) is constituted by sets of (initiator) two-level/spin-$1/2$ systems. Each set of systems interacts with one two-level/spin-\(\frac{1}{2}\) system that is a subsystem of a larger system \(S_1\). Furthermore, the spin-spin model will represent the successive interactions between the non-initiator systems, \(S_1 \rightarrow S_2 \rightarrow \ldots \rightarrow S_{N}\). I will consider that in these interactions, the timescales in which systems are affected by their interactions with other systems are much shorter than their intrinsic evolution, and thus I will ignore their self-Hamiltonians. For reasons that will become clearer, I will divide the Hamiltonian in eq. (\ref{centralHamiltonian}) into two Hamiltonians,
\begin{equation}
\begin{split}
\hat{H} &=  \sum_{i=0, \text{i even}}^{i=N} \hat{H}_{S_{i} \rightarrow S_{i+1}} \chi(t, i\Delta, \Delta) \\
& \quad + \sum_{i=1, \text{i odd}}^{i=N} \hat{H}_{S_{i} \rightarrow S_{i+1}} \chi(t, i\Delta, \Delta).
\end{split}
\label{centralHamiltonian}
\end{equation}
Let's consider $N^k_i$ as a variable whose values concern the numbers assigned to the last member of the group $k$ in a layer $i$. Let's consider $N^{k'}_i$ as a variable whose values concern the numbers assigned to the first members of the group $k$ in a layer $i$. The Hamiltonian of the kind presented below will be applied to all the interactions between subsystems of $S_{i}$ and $S_{i+1}$, i.e., \(S_{i} \rightarrow S_{i+1}\), where $i$ is odd. So, they will correspond to a specific \(i\), \(k\), and \(l\) where, as a reminder, $k$ will be the group number in the layer $i$ whose value corresponds to an elementary system in the layer $i+1$. $l$ is the group number of systems in the layer $i+1$:
\begin{equation}
\begin{aligned}
\hat{H}_{s^k_{i} \rightarrow s^l_{(i+1)k}} &= \frac{1}{2} \hat{\sigma}_{x s^l_{(i+1)k}} \otimes \sum_{j=N^{k'}_{i}}^{j=N^k_{i}} g_{s_{ij}^k \to s_{(i+1)k}^l} \hat{\sigma}_{x s^k_{i j}} \\
&= \frac{1}{2} \hat{\sigma}_{x s^l_{(i+1)k}} \otimes \hat{E}_{s^k_{i}}.
\end{aligned}
\label{HamiltonianSpin-Spinx}
\end{equation}

This Hamiltonian obeys the commutativity criterion of the following kind, 

\begin{equation}
[\hat{H}, \sigma_x] = 0,
\label{commutativityx}
\end{equation}
and thus $\sigma_x$ will be the pointer observable selected by the interactions represented via this Hamiltonian (Section 2).

On the other hand, the Hamiltonian below will apply to specific \(i\), \(k\), and \(l\) where \(S_{i} \rightarrow S_{i+1}\) and $i$ is even:

\begin{equation}
\begin{aligned}
\hat{H}_{s^k_{i} \rightarrow s^l_{(i+1)k}} &= \frac{1}{2} \hat{\sigma}_{z s^l_{(i+1)k}} \otimes \sum_{j=N^{k'}_{i}}^{j=N^k_{i}} g_{s_{i j}^k \to s_{(i+1)k}^l} \hat{\sigma}_{z s^k_{i j}} \\
&= \frac{1}{2} \hat{\sigma}_{z s^l_{(i+1)k}} \otimes \hat{E}_{s^k_{i}}.
\end{aligned}
\label{HamiltonianSpin-Spinz}
\end{equation}

In this case, this Hamiltonian obeys the commutativity criterion of the following kind, 

\begin{equation}
[\hat{H}, \sigma_z] = 0,
\end{equation}
and thus $\sigma_z$ will be the pointer observable selected by the interactions represented via this Hamiltonian.

 Let $k^{tot}_i$ be a variable that designates the total number of groups within each layer $i$. I am thus assuming that the number $k^{tot}_i$  of groups of systems of each \(S_i\) will be equal to the total number of systems that constitute \(S_{i+1}\), where \(S_i \rightarrow S_{i+1}\).

I will assume the following initial states of all the systems that will give rise to an SDC,

\begin{equation}
\begin{aligned}
|\Psi(t=0) \rangle &= |\psi \rangle_{S_{0}} \bigotimes^{i=N}_{i=1, \text{i odd}} |\psi \rangle_{S_{i}} \bigotimes^{i=N}_{i=2, \text{i even}} |\psi \rangle_{S_{i}},
\end{aligned}
\end{equation}

Then, we have that the initial states of the initiators are

\begin{equation}
    |\psi \rangle_{S_{0}} = \bigotimes_{j=N^{1'}_{1}}^{N^{1}_{1}} \vert + \rangle_{s^{1}_{1j}} \bigotimes_{j=N^{k'}_{1}}^{N^{k}_{1}} \ldots \bigotimes_{j=N^{k'^{\text{tot}}_{1}}}^{N^{k^{\text{tot}}_1}_{1}} \vert + \rangle_{s^{k^{\text{tot}}_1}_{1j}}.
    \label{InitialStateInitiators}
\end{equation}

For the non-initiator, we have for $i$ odd,

\begin{equation}
|\psi \rangle_{S_{i}} = \bigotimes_{j=N^{1'}_{i}}^{N^{1}_{i}} \vert + \rangle_{s^{1}_{ij}} \bigotimes_{j=N^{k'}_{i}}^{N^{k}_{i}} \ldots \bigotimes_{j=N^{k'^{\text{tot}}_{i}}}^{N^{k^{\text{tot}}_i}_{i}} \vert + \rangle_{s^{k^{\text{tot}}_i}_{ij}}.
\label{InitialStateZ}
\end{equation}

For the non-initiators, we have for $i$ even with $i>0$,

\begin{equation}
|\psi \rangle_{S_{i}} = \bigotimes_{j=N^{1'}_{i}}^{N^{1}_{i}} \vert 0 \rangle_{s^{1}_{ij}} \bigotimes_{j=N^{k'}_{i}}^{N^{k}_{i}} \ldots \bigotimes_{j=N^{k'^{\text{tot}}_{i}}}^{N^{k^{\text{tot}}_i}_{i}} \vert 0 \rangle_{s^{k^{\text{tot}}_i}_{ij}}.
\label{InitialStateX}
\end{equation}

These initial states are set up in such a way that, given the above Hamiltonians of interaction, there will be decoherence. These are the initial states of our toy universe, and I will consider that it is a brute fact why they are the way they are. So, I am postulating a kind of ``past hypothesis'' \cite{Albert2000TimeChance} that assumes that these are special initial states that are uncorrelated with each other.\footnote{More on this in Section 2.}

Let's turn to a more detailed analysis of the spin-spin interactions. I will focus on the case of the interactions $S_i\rightarrow S_{i+1}$ with $i$ even. The cases where $i$ is odd will be analogous and give rise to the same results, except that systems will have determinate values $+$ or $-$. First of all, it is important to notice that we can assume that the ``environments'' composed of non-initiator systems along the SDCs have initial states randomly distributed due to their previous interactions with other systems of an SDC. This is important because it allows us to consider that decoherence will end up occurring since the initial states of the environment need to be randomly distributed. Furthermore, we can assume that we have randomly distributed coupling constants since the strength of those interactions will be random, given the random initial states of environmental systems of non-initiator systems.\footnote{To see why, note that when it comes to overlap terms, we are summing over the different phases of the off-diagonal terms of the reduced density matrix of the target quantum system, and in order for the phases to cancel each other, they need to be randomly distributed. See, e.g., \cite{Schlosshauer2007DecoherenceTransition} for more details.}  The initial states of the non-initiator target systems and the initiator systems will have, as a brute fact, randomly distributed amplitudes.  Thus, we have for specific values of $k$, $l$ and $i$, the following initial state:
\begin{equation}
| \psi (0) \rangle_{s^k_{i} \rightarrow s^l_{(i+1)k}} =
a | 0 \rangle_{s^l_{(i+1)k}} + b | 1 \rangle_{s^l_{(i+1)k}} 
\prod_{j=N^{k'}_{i}}^{j=N^k_{i}} (\alpha_{s^k_{ij}} | 0 \rangle_{s^k_{ij}}+ \beta_{s^k_{ij}}| 1 \rangle_{s^k_{ij}}).
\end{equation}

Afterward, we have that under the evolution driven by $\hat{H}_{s^k_{i} \rightarrow s^l_{(i+1)k}}$ in eq. (\ref{HamiltonianSpin-Spinz}),
\begin{equation}
\begin{aligned}
| \psi (t) \rangle_{s^k_{i} \rightarrow s^l_{(i+1)k}} = 
& \; a | 0 \rangle_{s^l_{(i+1)k}} 
\prod_{j=N^{k'}_{i}}^{j=N^k_{i}} 
\left(
\alpha_{s^k_{ij}} \exp(ig_{s^k_{ij} \rightarrow s^l_{(i+1)k} }t) | 0 \rangle_{s^k_{ij}} + \right. \\
& \quad \left. \beta_{s^k_{ij}} \exp(-ig_{s^k_{ij} \rightarrow s^l_{(i+1)k} }t) | 1 \rangle_{s^k_{ij}}
\right) \\
& + b | 1 \rangle_{s^l_{(i+1)k}} 
\prod_{j=N^{k'}_{i}}^{j=N^k_{i}} 
\left(
\alpha_{s^k_{ij}} \exp(-ig_{s^k_{ij} \rightarrow s^l_{(i+1)k}}t) | 0 \rangle_{s^k_{ij}} + \right. \\
& \quad \left. \beta_{s^k_{ij}} \exp(ig_{s^k_{ij} \rightarrow s^l_{(i+1)k} }t) | 1 \rangle_{s^k_{ij}}
\right).
\end{aligned}
\end{equation}

When we trace out the degrees of freedom of the environmental systems, we obtain the following equation,

\begin{equation}
\begin{aligned}
\text{Tr}_{s^k_{i}} |\psi(t)\rangle_{s^k_{i} \rightarrow s^l_{(i+1)k}} \langle \psi(t)| = 
& \; |a|^2 |0\rangle \langle 0| 
+ z(t)ab^* |0\rangle \langle 1| \\
& + z^*(t)a^*b |1\rangle \langle 0| 
+ |b|^2 |1\rangle \langle 1|,
\end{aligned}
\end{equation}
where it can be shown that\footnote{\cite{Zurek1982Environment-inducedRules}.}
\begin{equation}
    z(t) = \prod_{j=N^{k'}_{i}}^{j=N^k_{i}}  \left[ \cos 2g_{s^k_{ij} \rightarrow s^l_{(i+1)k}}t + i \left( |\alpha_{s^k_{ij}}|^2 - |\beta_{s^k_{ij}}|^2 \right) \sin 2g_{s^k_{ij} \rightarrow s^l_{(i+1)k}}t \right].
\end{equation}

We can make a simple analysis of the evolution of this expression for $|\alpha_{s^k_{ij}}|=|\beta_{s^k_{ij}}|$,
\begin{equation}
    z(t) = \prod_{j=N^{k'}_{i}}^{j=N^k_{i}}   \cos 2g_{s^k_{ij} \rightarrow s^l_{(i+1)k}}t.
\label{decoherencemodel}
\end{equation}

Then, we can analyze how many systems it takes for decoherence to occur for randomly distributed coupling constants $g_{s^k_{ij} \rightarrow s^l_{(i+1)k}}$, by analyzing when $z(t)$ goes quasi-irreversibly to zero over time.

\begin{figure}[H]
    \centering
    \begin{subfigure}[b]{0.32\textwidth}
        \centering
        \includegraphics[width=\textwidth]{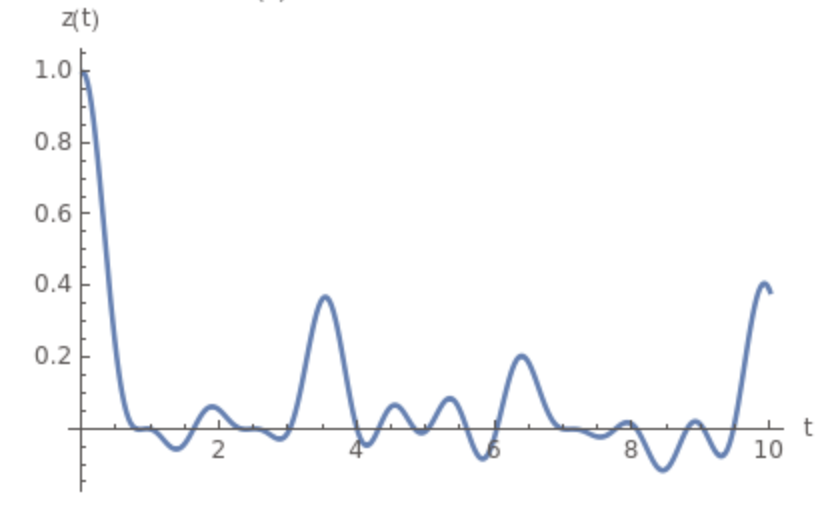}
        \caption{$G=6$}
        \label{fig:plot1}
    \end{subfigure}
    \hfill
    \begin{subfigure}[b]{0.32\textwidth}
        \centering
        \includegraphics[width=\textwidth]{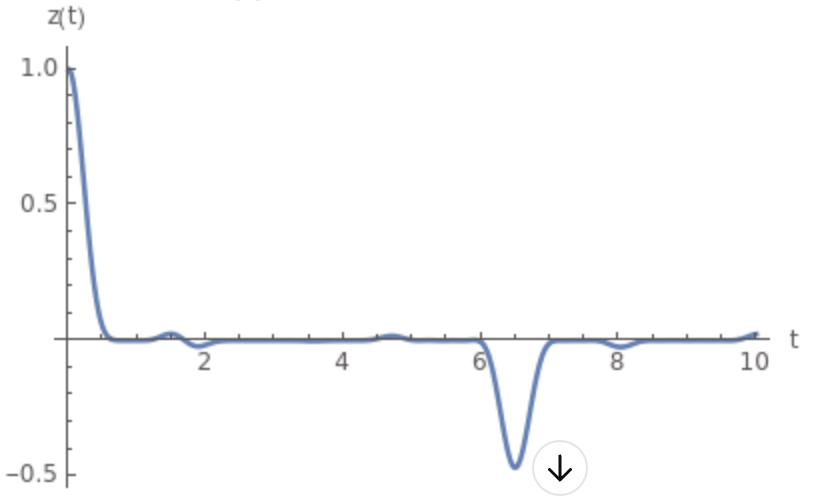}
        \caption{$G=12$}
        \label{fig:plot2}
    \end{subfigure}
    \hfill
    \begin{subfigure}[b]{0.32\textwidth}
        \centering
        \includegraphics[width=\textwidth]{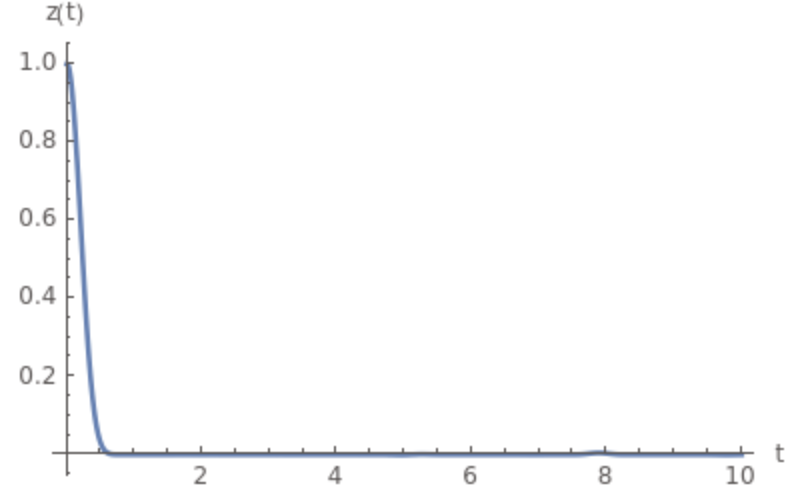}
        \caption{$G=17$}
        \label{fig:plot3}
    \end{subfigure}
    \caption{Plots for the decoherence factor $z(t)$ in eq. (\ref{decoherencemodel}) with randomly distributed $g_{s^k_{ij} \rightarrow s^l_{(i+1)k}}$, taking values between 0 and 1 for different numbers $G$ of environmental systems. It is necessary for them to be randomly distributed in order for decoherence to occur. As we can see, the higher the value of $G$, the lower the fluctuations of the decoherence factor $z(t)$. Thus, the time it takes decoherence to occur can be estimated, for example, by directly looking at plots like these ones and seeing at what time $z(t)$ goes to zero on average without any further future oscillations.}
    \label{fig:three_plots}
\end{figure}

\FloatBarrier

More generally,  \cite{Zurek1982Environment-inducedRules, Cucchietti2005DecoherenceEnvironments} have shown that for a broad kind of distribution of couplings \( g \),  and a sufficiently large amount of environmental systems,  the function $z(t)$  exhibits over time an approximately Gaussian decay,

\begin{equation}
  z(t) \approx e^{-\Gamma^2 t^2}. 
\end{equation}

The exact value of the decay constant $\Gamma^2$ depends on the initial state of the environment and the distribution of the coupling constants $g_i$.

Let's focus on the simple SDC model involving decoherence models that assume the simplification $|\alpha_{s^k_{ij}}|=|\beta_{s^k_{ij}}|$. This will be enough for our purposes of illustrating how an SDC can be modeled. Let's consider that in this model, we have randomly distributed coupling constants $g$. Let's designate the number of systems that constitute group $k$ in layer $i$ as $|N^{k}_{i}|$. To simplify, let's assume that the groups of systems that constitute each \(S_i\) will have the same number of subsystems, and that number will be equal to a constant \(G\). Thus, I will designate by $|N^{k}_{i}|=G$ for all layers $i$ and groups $k$ within each layer. Thus, given decoherence, in order for this model to work, we will need that \(G>>1\). I will make a more detailed analysis of this feature in Figure \ref{fig:three_plots}. 

Since the results in Figure \ref{fig:three_plots} hold across all the spin-spin models considered in this SDC model, we can see that systems will decohere each other, obeying CDC1)-CDC4), thus giving rise to determinate values and transmitting the DC. To see under which conditions this is obtained and how it is obtained in more detail, let's see how many systems we need to have in the rest of the layers in order for systems in the last layer $N$ to have determinate values, where $N+1$ is the total number of layers. This layer will be the layer of systems $S_{N}$ with their respective subsystems. Given our simplifications, the following formula allows us to calculate the number of systems in layer $i$ that are needed in order for decoherence to occur in layer $N$, where $G$ needs to be an adequate number (such as a number higher than 17):

\begin{equation}
\begin{aligned}
\textit{Number of systems in layer i} &= \textit{Number of subsystems of $S_N$} \\
& \quad \times G^{\textit{layer number i}}.
\end{aligned}
\end{equation}

Given this simple model, the value of the index $N$ of the system in the last layer, $G$, and the number of systems in each layer, we can calculate the number of events involving systems having determinate values that arise in this SDC,

\begin{equation}
\begin{aligned}
\text{Number of events in an SDC} &= \sum_{i=0}^{i=N-1} \text{Number of systems} \\
&\quad \text{in the layer } N \times G^i.
\end{aligned}
\label{layerSystems}
\end{equation}

Note that systems have determinate values at interactions, so I will assume that they have a determinate value at the spatial location of the target system. Thus, we associate to each target system $s$ a spatial point $(x, y, z)$, and the environmental systems $E$ of $s$  and $s$ will have determinate value in the spatiotemporal point $(t,x, y, z)$. Note that I am simplifying here, and it is more appropriate that systems have values in spacetime regions. If we have 5 systems in layer $N$ and 30 systems per group, which is a reasonable number to give rise to decoherence (Figure \ref{fig:three_plots}), and the total of $4$ layers (and this $N=3$), the SDC will give rise to $139655$ events involving systems having determinate values.


Now, we can see how the dynamics of this SDC is going to develop by implementing the following algorithm via simple decoherence models like the one above (Figure \ref{fig:three_plots}). I will explain this algorithm schematically. Before implementing it, we should use eq. (\ref{layerSystems}) to assign an appropriate number of systems to each layer, given the number of subsystems in $S_N$ that are assumed. Within each layer $i$, we should number each system with natural numbers $j$ with $j>1$, and group them in groups of size $G$. The last layer, $N$, just needs one group. As I have said, we make each group correspond to an individual system in the next layer by making the group number in the layer $i$ correspond to an individual system in the layer $i+1$ (or a subsystem of $S_{i+1}$). These numbers, $i$, $k$, and $l$ will individuate systems $s^l_{ik}$. At the end of each step, we update the states in agreement with the determinate values that will arise randomly with a given Born probability. Then, we have\\

-Even steps:\\

Step 0: We analyze a decoherence model with the initial states for the initiators given by eq. (\ref{InitialStateInitiators}), which will be the subsystems of $S_0$ and will be the environmental systems. The target systems will be the subsystems of $S_1$ and will be given by eq. (\ref{InitialStateX}). The Hamiltonian for this interaction will be given by  ($\ref{HamiltonianSpin-Spinz}$).\\ 

The rest of the even steps: We analyze decoherence models where the states $|+\rangle$ and $|-\rangle$ of the systems that arise from the odd steps will serve as initial states of the environment and the states in eq. (\ref{InitialStateZ}) will serve as the initial states of the target system. The Hamiltonian for this interaction will be given by eq. (\ref{HamiltonianSpin-Spinz}).\\

-Odd steps:\\ 

We analyze decoherence models where the states $|0\rangle$ and $|1\rangle$ of the systems that arise from the prior even step will serve as initial states of the environment and the states in eq. (\ref{InitialStateX}) will serve as the initial states of the target systems. The Hamiltonian that governs/describes these interactions will be given by eq. (\ref{HamiltonianSpin-Spinx}).\\

Since I have supposed that in a certain reference frame, the duration of the interaction $\Delta$ is the same in all the interactions, we get that the time in this frame for the systems in layer $N$ to have a determinate value is $\Delta \times N$, where we start counting time in our model at $t=0$. For simplicity, we can assume that $\Delta$ will be equal to the average decoherence time as measured in a rest frame.\footnote{And in a frame far from any strong gravitational field. Assuming these frames, allow us to consider that the decoherence timescale can only get longer under certain relativistic transformations.} In the model with four layers, when $G=30$, after an analysis involving diverse random coupling constants, the average decoherence time is approximately $0.6$ seconds. So, it will take approximately $1.8$ seconds for the systems $S_3$ to have determinate values. As we increase $G$, this time becomes shorter.

I will end this section by mentioning some important features that this model illustrates. First, notice that I am making some important simplifications. Via the coupling constant $\chi$, I am assuming that systems in each layer start interacting at the same time in a given frame. In a more realistic model where this assumption is not made, the coupling constants $g$ for each elementary system would depend on time, and therefore, we would just have one kind of coupling constants $g$, instead of $\chi$ and $g$. Although this seems to give rise to a very complex model, as we will see in future work, when the more fundamental quantum field theoric models are taken into account, in principle, we just need to consider one kind of coupling constant mediating the interaction between fields. So, in principle, we can describe the above dynamics of the SDC using any other reference frame. However, only when we have a quantum field theoretic model of SDCs this reference frame independence will become manifest.

Second, note also that the duration of decoherence may vary between reference frames. However, given quantum field theory in curved spacetime,\footnote{\cite{Wald1994QuantumThermodynamics}.}, this variation won't be a problem as long as the laws are generally covariant.

Third, given the CDC1)-CDC4), the events involving systems in the layer $0,...,N-1$ that lead to the events involving systems in the layer $N$ (which necessarily involve the systems in layer $N-1$) having a determinate value are time-like separated from each other. So, they will assume a fixed temporal order in all reference frames. To see this, note that CDC1)-CDC4) establishes that in order for systems $S_{N}$ in layer $N$ to have a determinate value due to systems $S_{N-1}$ in layer $N-1$ at $t$ (where systems have values in the same spacetime point), systems $S_{N-1}$ in layer $N-1$ need to have a determinate value due to systems $S_{N-2}$ in layer $N-2$ at $t'<t$ in a given reference frame, and so on for the rest of events involving interactions between systems in the other layers. 
Thus, given CDC1)-CDC4), all successive events where systems have determinate values will be time-like separated from each other. So, the order of events, which is important for EnDQT to maintain (since the criteria for systems to have determinate values depend on this order), won't vary according to the reference frame.

Fourth, it should be kept in mind that since there is no literal collapse of the wavefunction but only a state update, even if the initial states of each system in each layer were initially entangled with all the other systems in that layer, they don't cause the other systems to collapse. So, there is no action at a distance or the adoption of some preferred reference frame.

Fifth, the above model involves the simple spin-spin decoherence models, but in principle, any other decoherence model that involves local interactions can be used to model SDCs.\\

\section*{Appendix B: The Wigner's friend experiences}
As I have mentioned in Section 3, one might object that in some extended Wigner's friend theorems,\footnote{See \cite{Bong2020AParadox}.} it is plausible to consider that the friend Alice inside her isolated lab sees a determinate outcome. In a sense, this theorem assumes that Wigner, without performing any operations on Alice and her lab and after her measurement, simply opens her lab door and asks her about what outcome she obtained. In the simple case where the friends share a spin-$1/2$ entangled particles, she will answer that she obtained spin-up or spin-down with $50 \%$ of probability each (i.e., if Wigner makes a projective measurement on the state of Alice after her measurement, without performing any other operation on the lab, he will obtain these outcomes). So, it seems that Alice sees a determinate outcome contrary to what EnDQT claims in (the highly idealized) situations where we manage to isolate the friend from interacting with SDCs. To put the objection more dramatically, the measurement problem can be regarded as the problem of accounting for the experiences of determinate outcomes of experimentalists upon measurements, despite QT predicting that measurement-like interactions can yield indeterminate outcomes. The friend inside the isolated lab seems to experience a determinate outcome, but EnDQT seems to give no account of what this agent is experiencing. Hence, EnDQT doesn't solve the measurement problem.

 First, note that in the case where we manage to isolate the lab's contents from the SDCs, according to EnDQT, Wigner opening the lab triggers a physical process that leads to Alice obtaining determinate outcomes and reporting them to Wigner. It is not necessarily the case that Alice sees a determinate outcome inside her lab before opening the door. Seeing a determinate outcome can arise due to the interactions with the SDCs when the door is opened.

Second, there are different positions one may adopt regarding the friend's experiences, and which one is the correct one depends on deep philosophical and empirical issues, which I don't have space to settle here. The main point that I want to make now is that the above objection is not worrying, and there are different ways of answering it. On top of that, I will argue that the above objection could also be applied to other more accepted interpretations of QT in certain circumstances (such as the MWI), and in so far, it is a legitimate worry, it could also be a worry applicable to these interpretations. Given how accepted these interpretations are, it shouldn't be a reason to reject EnDQT.

Regarding the different positions, as I have mentioned in the main text, a possible one is that a) the agent lacks mental content underlying its perception of the outcome: this is the \textit{absent experience hypothesis}. The claim would be that we shouldn't worry that EnDQT (and other interpretations, as we will see below) can lead us to friend-like agents without experiences. We shouldn't follow our intuitions in the extreme (and quite possibly unrealistic) environments of a completely isolated agent and think that that agent will be exactly like us. The problem with this possibility is that it is hard to make sense of an agent without mental content.

However, as I have also mentioned, EnDQT can even consider that the friend experiences something in the isolated lab via particular hypotheses, dissolving the above worry. We might consider that b) friend-like systems in isolated regions have indeterminate mental content, where this content depends on the indeterminate physical properties of their brain. When the lab is open, their indeterminate mental content becomes determinate. I will call this possibility the \textit{quantum experience hypothesis}. This hypothesis has the benefit of not being foreign to the philosophy of mind. Externalism about mental content is roughly the thesis that mental content depends on the external environment of the subject that has that mental content.\footnote{\cite{Putnam1975Themeaning}.} The friend $having$ determinate mental content is dependent on an environment that renders that content determinate.

One might object that it is conceivable that we have a situation where the friend is in a coherent superposition (we don't open the lab's door) and could send messages via a sheet of paper in a sealed box so that the paper maintains its superposition. The box is only opened much later and/or in a faraway location from where the friend is. Furthermore, if we open the box, the message seems to make perfect sense. So, it seems plausible that the friend has determinate mental content already inside their lab in the spatiotemporal location where the message was produced. 

To deal with these cases, together with b), we can adopt a version of the extended mind hypothesis of \cite{Clark1998TheMind}, which I will call the \textit{quantum extended mind hypothesis}. The idea is that the bearers of the friend's mental content would be the outputs of the friend to the exterior (i,e, the sheet of paper). At first, their mental content is indeterminate; then it becomes determinate when the box is opened. Like the most sophisticated technology is perhaps an extension of our mind, for an agent like the friend, its outputs that interact with the external environment are an extension of their mind. Note that there isn't any action at a distance here according to EnDQT. So, Wigner measuring the output doesn't influence the friend's body.\footnote{Note also that the extended quantum mind thesis differs from the traditional extended mind thesis by considering that even phenomenal content can have extended bearers. I don't see any problem with considering that. More concretely, some might justify the extended mind thesis via individuating mental content through its functional roles \cite{Clark1998TheMind}. However, some may reject the claim that phenomenal content can be individuated by its functional roles (e.g., \cite{Chalmers1996TheTheory}). It is unclear that my thesis requires a functionalist account of phenomenal content. I will leave the investigation of this topic for future work.}  

So, we have here familiar situations in the philosophy of mind, which involves externalism. Alice could, in fact, have experiences in these situations, and EnDQT can account for them. There is much more to say about this. Future work will go into more detail on a), b), and c). Note that a), b), and c) are options that may be adopted if we reject the absoluteness of observed events assumption in the way EnDQT did (Section 3).

It is important to notice that if we consider realist Wigner's friend scenarios, the position adopted by EnDQT regarding the friend's experiences and the adoption of the above hypotheses shouldn't be seen as something restricted to EnDQT. More concretely, if extended Wigner's friend scenarios become realizable one day, it will very likely be via quantum computers and quantum agents running on those quantum computers as friends instead of human friends (see \cite{Wiseman2023ASuit} for a proposal). Assuming the controversial position that such quantum agents have mental content, which is a requirement if we want this version to mimic the original extended Wigner's friend version, many realist interpretations of QT will be pressed to assume that quantum agents don't have internally determinate mental content. This is because, plausibly,\footnote{I am setting aside strong emergentist and dualist perspectives about such content here.} their experiences will depend on superpositions of qubits. As it is recognized by many MWI proponents,\footnote{See most prominently, (\cite{Wallace2012TheInterpretation}, Section 10.3).} we can have robust branching into worlds when there is decoherence, but inside some quantum computers, we shouldn't often have such branching because there isn't a lot of decoherence (at least ideally). Many proponents of interpretations such as the MWI \textit{won't consider} that, in many situations, there is enough robust branching inside the quantum computer so that we could have something like an agent with determinate mental content running on those circuits. Spontaneous collapse theories won't also consider that there is such an agent because they don't consider that collapses happen (at least frequently) in situations like these ones within a functional quantum computer.

\cite{Wiseman2023ASuit} basically acknowledge the above in the case of spontaneous-collapse theories, saying that the ``thoughts [of the artificial agent] are thus not real in the way that my thoughts as a human are real.'' This amounts to the rejection of the ``Friendliness'' assumption of the theorem of \cite{Wiseman2023ASuit}.  My claim is that EnDQT also rejects this assumption, as well as the (at least some influential versions of the) MWI.\footnote{Of course, accepting b) and c), one shouldn't talk in terms of the reality of the thoughts/mental content. Instead, we should talk in terms of how different they are from our thoughts because the quantum agents have thoughts, they are just different from what we typically conceive our thoughts to be.}

So, if we ever come up with a scenario where that replicates the original extended Wigner's friend scenario, EnDQT leads to the same account of the agent's experiences as (at least) these realist and consistent quantum theories, and so the above objection could also apply to them. Thus, these views are on an equal footing when it comes to realistic scenarios in terms of accounting for the agent's experiences, and they could also adopt one or more (i.e., b) and also c)) of the above hypotheses concerning the friend's experiences along with EnDQT.

Furthermore, although single-world relationalists can account for the relative friend's experiences and prima facie this is an advantage relative to EnDQT, there is a good case to be made that these experiences aren't absolute. A more careful inspection of single-world relationalist views, such as Relational Quantum Mechanics,\footnote{See, e.g., \cite{Rovelli1996RelationalMechanics, DiBiagio2021StableFacts}.} shows that relative to some systems, other systems' mental content can be indeterminate since relative to one system, the other system might be in a superposition of quantum states that the mental content depends on. So, views such as Relational Quantum Mechanics, in these circumstances, would be in a similar position as EnDQT and be subject to a version of the above worry.

\section*{Appendix C: The basics of an ontology of quantum properties}
One might object that EnDQT doesn't offer a clear ontology since an ontology that views the world in terms of systems, observables, and determinate or indeterminate values is unclear and not so satisfactory when we compare it with the richer ontologies where the wavefunction is reified. As I have said, EnDQT offers the possibility of different ontologies that reject the view that quantum states are entities in the world. I have also mentioned the alternative ontology of determinable and determinates in Section 1. So, the above objection has no force.

However, there is another alternative ontology where the world is filled with matters of fact even when systems are not interacting, and not just observables and flashes, for example. Also, contrary to the previously mentioned ontologies friendly to EnDQT, the changes modeled and inferred via the irreversible process of decoherence that give rise to determinate values become manifest via specific interactions. This is an ontology of quantum properties, where systems are collections of quantum properties. Quantum properties have a certain structure or features that impact the determinacy of the values that systems having them give rise to, which I will call the differentiation $D^{*}$ of quantum properties.

So, for example, we have spin in a given direction, which comes in terms of different degrees of differentiation. These features of quantum properties are represented through observables concerning P (e.g., where P could be energy, momentum, etc.) and quantum states that are eigenstates of those observables. Systems have, by default, quantum properties with the lowest degree of differentiation, i.e., undifferentiated. Certain interactions change the degree of differentiation of such properties.

At least in the simple cases of decoherence that I have been assuming, the degree of differentiation is measured via the non-diagonal terms of the reduced density operator of the system subject to decoherence by systems that have the DC, when we trace out the degrees of freedom of the environmental system that are interacting with the system of interest. The quantum state of some system $S$ with $\alpha, \beta \neq 0$,

\begin{equation}
\alpha\left|\uparrow_{z}\right\rangle_{S}+\beta\left|\downarrow_{z}\right\rangle_{S}, 
\end{equation}

\noindent and the observable $S_{z}$ that acts on the Hilbert space of $S$, represents the quantum property spin-z of $S$. The spin-z of $S$ has a degree of differentiation $D^{*}=0$ and we consider that the system has an undifferentiated spin-z. This is because this system is not interacting with any other systems (note that it would still have an undifferentiated spin if it interacted with systems that don't have the DC).

Let's consider a system $E$, constituted by many subsystems that are interacting with $S$. For instance, $S$ with quantum properties spin in different directions that interacts strongly with the many systems, also with spin in multiple directions, that constitute $E$. I will again put a subscript SDC in the systems that belong to an SDC. A system belonging to an SDC will have a stably differentiated quantum property represented via its quantum states when it interacts, in agreement with the CDCs (Section 2), with another system $S$, decohering it and thus giving rise to interactions belonging to an SDC. So, if $S$ is interacting with a system $E$ belonging to an SDC and having the DC-$S$ (Section 2), we have that

\begin{equation}
\alpha\left|\uparrow_{z}\right\rangle_{S}\left|E_{\uparrow}(t)\right\rangle_{\text{E SDC}}+\beta\left|\downarrow_{z}\right\rangle_{S}\left|E_{\downarrow}(t)\right\rangle_{\text{E SDC}} . 
\end{equation}

The degree of differentiation of a quantum property that systems end up with after their interaction can be inferred and calculated via the overlap terms that concern the distinguishability of the states of $E$ concerning $S$, such as $\left\langle E_{\uparrow}(t) \mid E_{\downarrow}(t)\right\rangle_{\text{E SDC}}$ and $\left\langle E_{\downarrow}(t) \mid E_{\uparrow}(t)\right\rangle_{\text{E SDC}}$. Generally, given

\begin{equation}
\hat{\rho}(t)=\sum_{i=1}^{N}\left|\alpha_{i}\right|^{2}\left|s_{i}\right\rangle_{S}\bra{s_{i}}+\sum_{i, j=1, i \neq j}^{N} \alpha_{i} \alpha_{j}^{*} \ket{s_{i}}_{S}\left\langle s_{i}\right|\left\langle E_{j}(t) \mid E_{i}(t)\right\rangle_{\text{E SDC}} 
\end{equation}

\noindent a measure of the degree of differentiation of the different $D^{\*}$-P of $S$ in spacetime region over time $t$ for the simple scenarios that we are considering will be given by the von Neumann entropy\footnote{Given a density operator $\rho_{S}$ for quantum system $S$, the von Neumann entropy is $S\left(\rho_{S}\right)=-\operatorname{tr}\left(\rho_{S} \ln \rho_{S}\right)$. $S\left(\hat{\rho}_{S}\right)$ is zero for pure states and equal to $\ln N$ for maximally mixed states in this finite-dimensional case.} $S\left(\hat{\rho}_{S}(t)\right)$ of $\hat{\rho}_{S}(t)$ over $\ln N$, where $N$ is the number of eigenvalues of $\hat{\rho}_{S}(t)$,

\begin{equation}
D^{*}(P, S, t)=\frac{S\left(\hat{\rho}_{S}(t)\right)}{\ln N} 
\end{equation}

Thus, we can measure and represent the degree of differentiation $\mathrm{D}^{*}$ of the quantum property D*-P that $S$ will have at the end of the interaction with $E$ at $\mathrm{t}$ with $0 \leq$ $D^{*}(P, S, t) \leq 1$, and the differentiation timescale (inferred from the decoherence timescale).

$S$ ends up having a stably (qua irreversibly) differentiated quantum property if $D^{*}(P, S, t)$ goes quasi-irreversibly to one over time (in the sense that the recurrence of this term back to significantly different from zero is astronomically large). We also consider that system $E$ decohered system $S$, and that both systems have undergone a so-called process of stable differentiation, which leads them to each have a determinate value. Upon knowing the actual result, we update the state of $S$ to one of the $\left|s_{i}\right\rangle_{S}$, and consider that the system has a determinate value, which is an eigenvalue of the observable that $\left|s_{i}\right\rangle_{S}$ is an eigenstate of. Similarly in the case of $E$ for $\left|E_{i}\right\rangle_{E}$.

A quantum property of $S$ might not be fully stably differentiated and just be stably differentiated to some $degree$ $D^{*}$ by $E$, and thus, it has a value with a degree of determinacy $D=D^{*}$. This can be inferred if the quantum states of an environment, which has the DC, have a non-zero overlap that is stable over time.

I will return below to the intuition for why we can consider that there are values with degrees of determinacy. For now, note that not all interactions with a system give rise to systems having a determinate value, although there is something that changes in the quantum properties of the systems under these interactions. As a toy example, consider the spin of a particle in different directions in a series of Stern-Gerlach devices without letting the particles hit a screen between each device. The inhomogeneous magnetic field leads the subsystem of the particle with a spin in a certain direction to interact with the subsystem with the quantum property position, leading to their entanglement. Something changes in the spin direction of the quantum systems when the particle goes from one magnet to another, but no determinate value arises. If there was, we would have an irreversible process, and thus, we wouldn't be able to reverse the result of the operations of the magnet via a Stern-Gerlach interferometer. So, the spin of the system that interacts with the other subsystem of the particle has an indeterminate value, although there is something that changes in the quantum property that corresponds to this indeterminate value.

As I have mentioned in Section 2, pragmatic reversible decoherence models allow us to infer and represent $E$ and $S$ interacting but having indeterminate values. This occurs when $E$ doesn't belong to an SDC, not having the DC. The Stern-Gerlach case above is a case appropriately modeled via a pragmatic irreversible decoherence model.\footnote{See, e.g., \cite{Oliveira2006CoherenceAE} for one such model.} I will call the interactions represented and inferred via the pragmatic reversible decoherence models, unstable differentiation interactions. During these interactions, both systems continue with their quantum properties undifferentiated.

As I have been assuming above, it is plausible to consider that some quantum properties can be stably differentiated to a certain degree, and this impacts the subsequent degree of determinacy of the value that arises from a quantum property. Let's look at the intuition for this. In the double-slit experiment, if the detectors at the slits interact with a quantum system weakly in such a way that we cannot fully distinguish in which slit it passed we get some disappearance of interference. These interactions will give rise to a low entanglement between the position and the degrees of freedom of the detector. Furthermore, the more these interactions distinguish the path of the system, the more entanglement we have between the position of the target system and the degrees of freedom of the detector, and the more the interference disappears until it disappears completely under maximal entanglement. So, I have considered that stable differentiation of a quantum property comes in degrees and the determinacy of the resultant values.

Differentiation and determinacy are related and this will allow us to provide an analysis of quantum indeterminacy and determinacy. This relation will establish that property $P^{*}$, in this case, a value property, is the property of having some other property P having specific features. I will thus consider that

For a system to have a value v of P (where P could be energy, position, etc.) with a non-minimal degree of determinacy D is to have stably differentiated quantum property D*-P to a non-minimal degree D* where D=D*. A system with a quantum property fully stably differentiated will have a determinate value of P. 

On the other hand, indeterminacy and undifferentiation are related,

For a system to have an indeterminate value of P is to have an undifferentiated quantum property.

Note that according to this relation, we have multiple quantum properties concerning P, represented by quantum states and observables,  that correspond to a non-maximally determinate value of P.\footnote{Note also that this relation doesn’t imply that undifferentiated quantum properties are more fundamental than indeterminate value properties.}

\begin{nodotdef}
\noindent For a system to have a value $v$ of $P$ (where $P$ could be energy, position, etc.) with a nonminimal degree of determinacy $D$ is to have stably differentiated quantum property $D^{*}-P$ with a non-minimal degree of differentiation $D^{*}$ where $D=D^{*}$. A system with a quantum property (fully) stably differentiated will have a determinate value of $P$.
\end{nodotdef}

On the other hand, indeterminacy and differentiation are related when the systems have a quantum property undifferentiated (which is the lowest degree of differentiation).

\begin{nodotdef}
\noindent For a system to have an indeterminate value of P is to have an undifferentiated quantum property.
\end{nodotdef}

\cite{Pipa2024AnTheory} enters into further details about this ontology. It may initially seem pedantic compared with the simpler ontology of flashes and observables. However, it captures more structure represented by quantum states (and decoherence) than the flashes. Systems don't only have determinate values under interactions (which would be analogous to the flashes), they have quantum properties with different degrees of differentiation that change over time.

\section*{Appendix D: Interference phenomena according to EnDQT}
In this appendix, I put into practice some of the above features of EnDQT to see how it can account for interference phenomena via a simple example.\footnote{This model is based on \cite{vonderLinde2021OpticalIssue}. See \cite{Cohen-TannoudjiClaude1997PhotonsElectrodynamics} for a more extensive discussion of this framework.}

The electromagnetic field can be quantized, where such quantization proceeds by associating to each radiation mode a quantum harmonic oscillator and the corresponding so-called creation and annihilation operators, allowing us to express the particle number operator $\widehat{N}_{C h}$. Each channel of the interferometer's beam splitters is associated with a number $C h$. Let's consider that the eigenvalues $\mathrm{Ch}$ of the operator $\widehat{N}_{C h}$ obtained from

\begin{equation}
\widehat{N}_{C h}|n\rangle=n_{C h}|n\rangle
\end{equation}

\noindent represents the number of photons (the particle number) in the channel $\mathrm{Ch}$, where each channel is associated with a radiation mode.

\begin{figure}[h]
    \centering
    \includegraphics[max width=\textwidth]{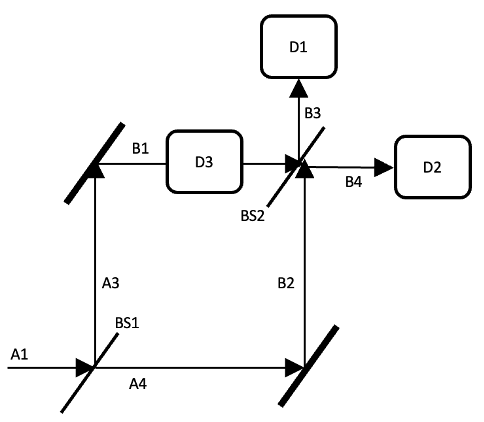}
    \caption{Mach-Zender interferometer}
    \label{fig:example5}
\end{figure}

Now, let's consider the following states of the channels whose numbers appear in Figure \ref{fig:example5}, $|1000\rangle=|1\rangle_{1} \otimes|0\rangle_{2} \otimes|0\rangle_{3} \otimes|0\rangle_{4}$, the same in the case of $|0100\rangle,|0010\rangle$, and $|0001\rangle$. The context will make clear whether, for example, 1 refers to A1 or B1, and so on. $|0\rangle$ is the vacuum state. Channels will allow us to represent the subsystems of the system under analysis, occupying different spatial regions of the interferometer.

Let's start with the case where detector D3 is not present and consider that the initial state of the quantum system inserted into channel A1 together with the other systems is given by

\begin{equation}
\ket{Input}=\ket{1000}. 
\end{equation}

This system has an indeterminate particle number since it won't be interacting with systems that belong to an SDC at least after being prepared. After the interaction with the first beam-splitter, we obtain two subsystems with an indeterminate particle number whose state is given by the following entangled state,

\begin{equation}
\begin{aligned}
\ket{Final}_{B S 1}=1 / \sqrt{2}|0010\rangle \\
+i / \sqrt{2}|0001\rangle .
\end{aligned}
\end{equation}

Afterward, these systems will pass by the beamsplitter BS2, which gives rise to a system with the following state:

\begin{equation}
\ket{Final}_{B S 2}=|0001\rangle . 
\end{equation}

After BS2, the system will interact with the detector D2, giving rise to a system having a 1 particle number determinate value during the interaction. Note that it is assumed that D2 is connected to an SDC.

Let's clarify how interactions leading to a determinate value work by examining what happens when detector D3 is placed at B1 (see Figure \ref{fig:example5}). This detector interacts with the quantum system, annihilating the above interference phenomenon. I am going to adopt the same SDC subscripts convention that I have adopted in the last sections. The interactions at time t' involving D3 (and omitting the interactions with D1 and D2) lead to the following state,

\begin{equation}
\begin{aligned}
\left|\operatorname{Final}\left(t^{\prime}\right)\right\rangle= & \frac{1}{\sqrt{2}}|1000\rangle\left|E_{1}(t)\right\rangle_{S D C}-\frac{1}{2}|0010\rangle\left|E_{0}(t)\right\rangle_{S D C}  \\
& +\frac{i}{2}|0001\rangle\left|E_{0}(t)\right\rangle_{S D C},
\end{aligned}
\end{equation},

\noindent where these interactions can be represented via decoherence models.\footnote{For simplicity, I will not analyze in detail decoherence in the Fock basis and assume that the Schrödinger picture is applicable. See \cite{Walls1995QuantumOptics, McClung2010DecoherenceState, Myatt2000DecoherenceReservoirs} for a detailed account. I will also assume that a notion of spatiotemporal localization of particles arises during these interactions. See \cite{Fraser2022ParticlesTheory} for a survey of different options that consider particles as non-fundamental, but emergent.} How does EnDQT interpret the above phenomenon? First, note that contrary to $\left|E_{1}(t)\right\rangle,\left|E_{0}(t)\right\rangle$ concerns the inexistence of the measurement signal. It will also mean that the measurement device interacted with a subsystem, giving rise locally to a 0 particle number determinate value (i.e., the vacuum).

So, upon the placement of $\mathrm{D} 3$, there is also the probability of $1 / 2$ of a photon arising at D3, and a 0 particle number arising at the other detectors. Furthermore, there is a $1 / 4$ probability of one of the systems with an indeterminate particle number interacting with D1 or D2 and having a 1 particle number. Also, the other system giving rise to 0 particle number at D3. As I have argued via the Bell scenario in Section 3, note that all these interactions are local and there isn't any non-local influence. Here, we have a similar situation, but with quantum systems that can also give rise to interference.

Let's now consider instead the situation where the detectors are isolated from interacting with elements of an SDC, not belonging to an SDC as well. The quantum state $|Final(t')\rangle$ isn't anymore applicable to correctly represent the situation inside the lab. We would rather have

\begin{equation}
\ket{Final(t)}=\frac{1}{\sqrt{2}}\ket{1000}\ket{E_{1}(t)}-\frac{1}{2}\ket{0010}\ket{E_{0}(t)}+\frac{i}{2}\ket{0001} \ket{E_{0}(t)}, 
\end{equation}

\noindent and no systems would have determinate values.

\section*{Appendix E: From the Classical Markov Condition to the Classical Reichenbach Common Cause Principle}
Let's start with some key definitions.\footnote{This content is mainly based on \cite{Wronski2014ReichenbachsCauses, Williamson2004BayesianFoundations}.} A directed graph \( G \) over a collection of nodes \( V \) is defined as the pair \( \langle V, E \rangle \), where \( E \) is a collection of directed edges. A \textit{path} between nodes \( X \) and \( Y \) consists of a sequence of nodes that begin at \( X \) and terminate at \( Y \), in such a way that for each pair of consecutive nodes in the sequence, there exists a directed edge connecting them (the orientation of the edge is irrelevant). A node \( X \) on a path is identified as a \textit{collider} if, jointly with its adjacent nodes \( Y \) and \( Z \) on the path, it forms an inverted fork: \( Y \rightarrow X \leftarrow Z \). Furthermore, nodes will be associated with random variables. Moreover, I will consider that there exists a directed path between nodes \( X \) and \( Y \), which I will denote as \( X \rightsquigarrow Y \), if there is an arrow from \( X \) to \( Y \) (\( X \rightarrow Y \)) or if there is some node \( Z \) such that \( X \rightsquigarrow Z \) and \( Z \rightarrow Y \).

\textit{Par}(X) will be the set of parents of a node \( X \), and it will include the nodes \( Z \) in such a way that \( Z \rightarrow X \). \textit{Childr}(X), the set of \( X \)'s ``children'', involves the nodes \( Z \) such that \( X \rightarrow Z \). Furthermore, I will establish the sets \textit{Anc}(X) (referred to as ``ancestors'') and \textit{Desc}(X) (referred to as ``descendants'') by substituting \( \rightarrow \) with \( \rightsquigarrow \) in the earlier definitions. It is important to note that a node is always considered its own ancestor and descendant, but it is never its own child or parent.

As discussed in Section 3, the DAG in CCM and the probability distributions over the variables of the DAG are linked by the Classical Markov Condition (CMC). The CMC states that any variable, given its parents, is probabilistically independent of all other variables except its descendants. This is denoted as \( R \perp\!\!\!\perp S \mid T \), meaning ‘\( R \) is probabilistically independent of \( S \) given \( T \)’, which implies \( p(R \mid ST) = p(R \mid T) \) provided \( p(ST) > 0 \). I will use the notation \( R \not\!\!\perp\!\!\!\perp S \mid T \) to indicate ‘\( R \) and \( S \) are probabilistically dependent given \( T \)’. Unconditional independence is denoted by \( R \perp\!\!\!\perp S \), and \( R \perp\!\!\!\perp S \mid \emptyset \) represents unconditional independence \( R \perp\!\!\!\perp S \). Similarly, \( R \not\!\!\perp\!\!\!\perp S \mid \emptyset \) means unconditional dependence \( R \not\!\!\perp\!\!\!\perp S \). Let \( ND_A = V \setminus (\{A\} \cup Des_A) \) represent the non-descendants of \( A \). Therefore, the CMC can be expressed as \( A \perp\!\!\!\perp ND_A \mid Par_A \) for each \( A \in V \).

\begin{figure}[ht]
    \centering
    \includegraphics[scale=0.5]{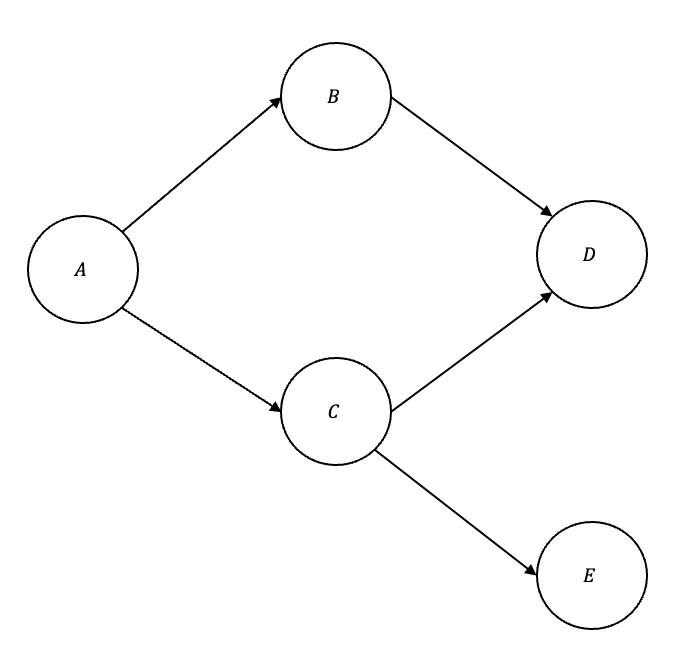}
    \caption{Example of a DAG}
    \label{fig:DAG}
\end{figure}
This definition can be shown to be equivalent to the one in the main text if it concerns a DAG. To see how it works, consider the DAG in the Figure \ref{fig:DAG}. The CMC determines the following conditional independencies:

\begin{equation}
B \perp \!\!\! \perp C, E \mid A
\end{equation}
\begin{equation}
C \perp \!\!\! \perp B \mid A
\end{equation}
\begin{equation}
D \perp \!\!\! \perp A, E \mid B, C
\end{equation}
\begin{equation}
E \perp \!\!\! \perp A, B, D \mid C
\end{equation}

Then, we have the following definition of the principle of common cause:\\

\noindent The Classical Reichenbach Common Cause Principle (CRCCP) holds for a directed acyclic graph \( G \) over \( V \) and a probability distribution \( P(V) \) if, whenever \( A \rightsquigarrow B \) or \( B \rightsquigarrow A \) or there is a \( U \subseteq V \) such that \( C \in U \) implies \( C \rightsquigarrow A \) and \( C \rightsquigarrow B \), and \( A \perp\!\!\!\perp B \mid U \).\\

It is typically considered that the shortest proof of the connection between the CMC and the CMC appeals to \( d \)-separation. d-separation is not an intuitive concept. I will use the one in \cite{sep-causal-models}. I have explained above the notion of a path. Now, consider \(X\), \(Y\), and \(Z\) as disjoint subsets of \(V\). \(Z\) \(d\)-separates \(X\) and \(Y\) if and only if every path \(\langle X_1, \ldots, X_k \rangle\) from a variable in \(X\) to a variable in \(Y\) includes at least one variable \(X_i\) such that either: (i) \(X_i\) is a collider, and no descendant of \(X_i\) (which includes \(X_i\) itself) is in \(Z\); or (ii) \(X_i\) is not a collider, and \(X_i\) is included in \(Z\). A path that fulfills these criteria is considered to be blocked by \(Z\). On the other hand, if \(Z\) does not \(d\)-separate \(X\) and \(Y\), then \(X\) and \(Y\) are \(d\)-connected by \(Z\). If \(X\) and \(Y\) are not d-separated by $Z$, then \(X\) and \(Y\) are \(d\)-connected by \(Z\). 

The \( d \)-separation is useful for our purposes due to the following feature (\cite{Williamson2004BayesianFoundations}, pp. 17):

\begin{quote}
 Given a directed acyclic graph \( G \) over \( V \) and \( R, S, T \subseteq V \), \( T \) \( d \)-separates \( R \) and \( S \) if and only if \( R \perp\!\!\!\perp S \mid T \) for all probability functions that satisfy the Markov Condition with respect to \( G \).   
\end{quote}

For instance, if we have knowledge that a certain DAG with a probability distribution fulfills the CMC, and we notice that different variables \( X \) and \( Y \) are \( d \)-separated by \( \emptyset \), we are able to make the inference that \( X \) and \( Y \) are not correlated.

 Let's now turn to the proof that the CMC leads to the CRCCP. Let's assume that the CMC holds for a DAG \( G \) over \( V \) and a \( P(V) \). Let \( A, B \in V \). Let's assume that it is not the case that \( A \rightsquigarrow B \) or \( B \rightsquigarrow A \) or there is a \( C \in V \) such that \( C \rightsquigarrow A \) and \( C \rightsquigarrow B \). Then, variables \( A \) and \( B \) are \( d \)-separated by \( \emptyset \), because a collider has to be included in any path between them. Thus, in this case \( A \perp\!\!\!\perp B \).

Now, let's assume that \( A  \not\!\!\perp\!\!\!\perp B \). From the above by contraposition we have that if it is not the case that \( A \rightsquigarrow B \) or \( B \rightsquigarrow A \), there should be at least one \( C \in V \) such that \( C \rightsquigarrow A \) and \( C \rightsquigarrow B \). Let \( U \) be the set of these \( C \)'s. Then, we have that \( U \) \( d \)-separates \( A \) and \( B \), and therefore \( A \perp\!\!\!\perp B \mid U \).

\bibliographystyle{plainnat} 
\bibliography{references} 

\end{document}